\documentclass{aa}  

\usepackage{graphicx}
\usepackage{txfonts}
\usepackage{xcolor}
\usepackage{ragged2e}   
\usepackage{caption}   
\usepackage{multirow}   
\usepackage{subcaption}  

\usepackage[colorlinks, citecolor = blue]{hyperref}
\begin{document}

   \title{PRUSSIC III -  ALMA and NOEMA survey of dense gas in high-redshift star-forming galaxies}

    \authorrunning{M. Rybak et al.}
    \titlerunning{PRUSSIC III - dense gas in high-redshift galaxies}

   \author{M. Rybak\inst{1,2,3},
   G. Sallaberry\inst{1},
   J. A. Hodge\inst{1},
   D. Riechers\inst{4},
   N. N. Geesink \inst{1,5},
   T. R. Greve  \inst{6,7,8},
   S. Viti\inst{1,9},
   F. Walter\inst{10},
   P. P. van der Werf\inst{1},
   C. Yang\inst{11}
   }
   \institute{Leiden Observatory, Leiden University, P.O. Box 9513, 2300 RA Leiden, The Netherlands  \\
         \email{mrybak@strw.leidenuniv.nl}
        \and Faculty of Electrical Engineering, Mathematics and Computer Science, Delft University of Technology, Mekelweg 4, 2628 CD Delft, The Netherlands
        \and SRON - Netherlands Institute for Space Research, Niels Bohrweg 4, 2333 CA Leiden, The Netherlands
        \and Institut f\"ur Astrophysik, Universit\"at zu K\"oln, Z\"ulpicher Stra{\ss}e 77, D-50937 K\"oln, Germany
        \and European Southern Observatory (ESO), Karl-Schwarzschild-Stra{\ss}e 2, Garching 85748, Germany 
        \and Cosmic Dawn Center (DAWN), K{\o}benhavn, Denmark
        \and DTU--Space, Technical University of Denmark, Elektrovej 327, 2800 Kgs. Lyngby, Denmark
        \and Department of Physics and Astronomy, University College London, Gower Street, London WC1E 6BT, United Kingdom
        \and Transdisciplinary Research Area (TRA) ´Matter´/Argelander-Institut f\"ur Astronomie,  University of Bonn
        \and  Max–Planck Institut f\"ur Astronomie, Königstuhl 17, 69117 Heidelberg, Germany
        \and Department of Space, Earth and Environment, Chalmers University of Technology, SE-412 96 Gothenburg, Sweden
             }
\date{Received 14th August 2025, Resubmitted 4th November 2025}

  \abstract{
  Characterising the relationship between dense gas and star formation is critical for understanding the assembly of galaxies throughout cosmic history. However, due to the faintness of standard dense-gas tracers -- HCN, HCO$^+$, and HNC -- dense gas in high-redshift galaxies remainslargely unexplored.
  
  We present ALMA and NOEMA observations targeting HCN/HCO$^+$/HNC (3--2) and (4--3) emission lines in eleven (mostly) gravitationally lensed dusty star-forming galaxies (DSFGs) at redshift $z=1.6-3.2$.
  We detect at least one line in 10 out of 11 galaxies. Altogether, we detect 34 dense-gas transitions, more than quadrupling the number of extant high-redshift detections. Additionally, in two targets, we detect lower-abundance CO isotopologues $^{13}$CO and C$^{18}$O, as well as CN emission.
  
  We derive excitation coefficients for HCN, HCO$^+$ and HNC in DSFGs, finding them to be systematically higher than those in nearby luminous infrared galaxies.
  Assuming a canonical dense-mass conversion factor ($\alpha_\mathrm{HCN}=10$), we find that DSFGs have shorter dense-gas depletion times (median 23~Myr) than nearby galaxies ($\approx60$~Myr), with a star-forming efficiency per free-fall time of 1--2\%, a factor of a few higher than in local galaxies. We find a wide range of dense-gas fractions, with HCN/CO ratios ranging between 0.01 and 0.15. Finally, we put the first constraints on the redshift evolution of the cosmic dense-gas density, which increases by a factor of $7\pm4$ between $z=0$ and $z=2.5$, consistent with the evolution of the cosmic molecular-gas density. 
  }


   \maketitle
%

\section{Introduction}

\subsection{Dense gas in high-redshift galaxies}

Understanding the process of star formation -- how the cold, molecular gas is converted into newborn stars -- is one of the key questions in astrophysics. Present-day galaxies, on average, convert gas into stars inefficiently, depleting their gas over billions of years \citep[e.g.][]{Tacconi2018, Tacconi2020}. However, in the early Universe, the star-forming activity was much higher, peaking between redshifts $z=1-4$ (e.g., \citealt{Madau2014, Zavala2021, Harikane2023}), during the so-called ``Cosmic Noon''. This intense star formation is dominated by massive, dusty star-forming galaxies (DSFGs\footnote{{In this work, we consider DSFGs to have continuum flux at 850-$\mu$m (observed-frame) $S_\mathrm{850~\mu m}\geq1$~mJy.}}), which were producing stars at rates up to 100$\times$ higher than present-day galaxies (e.g., \citealt{Magnelli2011,Dudzeviciute2020, Zavala2021}). 

This trend is qualitatively matched by the evolution of \emph{molecular} gas density, based on  surveys of low-J CO, [\ion{C}{ii}], [\ion{C}{i}], and cold dust emission, primarily with the NOrthern Extended Millimeter Array (NOEMA) and Atacama Large Millimeter / Sub-millimeter Array (ALMA) (see, e.g., \citealt{Riechers2019,Walter2020,hodge2020, Tacconi2020, Bollo2025} and references therein). These studies have shown that the elevated star-forming activity at the Cosmic Noon is primarily due to the increase in molecular gas content of galaxies, with only a mild increase in star-forming efficiency (by a factor of $\approx$3).

However, these large-volume average trends provide a potentially biased view of the actual star-forming processes in high-redshift galaxies. First, while conveniently bright, CO, [\ion{C}{ii}], [\ion{C}{i}], and dust continuum trace gas across a wide range of densities (down to $\approx$100 cm$^{-3}$ for [\ion{C}{ii}]), rather than dense gas from which stars actually form ($n\geq10^4$ cm$^{-3}$). Moreover, spatially resolved observations of [\ion{C}{ii}] (e.g., \citealt{Gullberg2018, Fujimoto2020, ginolfi2020, Rybak2019, Rybak2020b, Ikeda2025}) and CO(1--0) emission (e.g., \citealt{Ivison2011,Riechers2011,Stanley2023, rybak2025a}) indicate that the majority of cold gas (up to $\approx$80\%, \citealt{rybak2025a}) is in extended, diffuse halos and does not directly contribute to the observed star-forming activity. Finally, theoretical studies suggest that the bulk of cold, molecular gas resides in low-mass, sub-mm faint galaxies, rather than high-mass, sub-mm bright galaxies which dominate the star-forming activity at the Cosmic Noon (e.g., \citealt{Lagos2020}).

To properly understand the star-forming processes in high-redshift galaxies, we need to characterise the relationship between their \emph{dense} gas\footnote{While the definition of ``dense'' gas is arbitrary, following \citet{Gao2004a}, we consider gas with $n\geq3\times10^4$~cm$^{-3}$ as ``dense''.} and star formation. Tracing dense gas requires targeting spectral lines of molecules that have high critical densities -- particularly HCN, HCO$^+$, and HNC, whose ground-state rotational transitions have critical densities $n_\mathrm{crit}=13\times10^5$, $1.9\times10^5$, and $3\times10^5$~cm$^{-3}$ at a temperature of $T=20$~K, respectively. 

Indeed, surveys of local galaxies have established HCN(1--0) as the ``gold-standard'' tracer of dense gas (see, e.g., recent reviews by \citealt{Saintonge2022} and \citealt{Schinnerer2024}). The HCN(1--0) luminosity and star-formation rate correlate linearly over $\sim$8~orders of magnitude, from individual molecular clouds (e.g., \citealt{Wu2005, Kauffmann2017, Dame2023, Forbrich2023} to entire galaxies (e.g., \citealt{Nguyen1992, Gao2004b, Gracia2008, Garcia2012, Costagliola2011}) and sub-galactic scales (e.g., \citealt{Bigiel2015,  Gallagher2018a, Jimenez2019, Neumann2023, Stuber2025}). Compared HCN, HCO$^+$ and HNC are more sensitive to gas thermodynamics. For example, HCO$^+$ is sensitive to the free-electron abundance (e.g., \citealt{Papadopoulos2007}), while HNC abundance depends on gas temperature due to its temperature-sensitive formation and destruction pathways (e.g., \citealt{Hacar2020}).

Extending local studies of dense-gas tracers to high redshift remains challenging, as the HCN/HCO$^+$/HNC emission can be more than 10$\times$ fainter than CO. Additionally, at high redshift, the ground-state HCN/HCO$^+$/HNC lines move into the ``high-frequency'' bands of Karl G. Jansky Very Large Array (JVLA), which can be used only under good weather conditions. As a result, ''\textit{to date, there have been no systematic studies of HCN (or other high-dipole-moment molecules) at high z, but in principle these are now feasible with the capabilities of ALMA and NOEMA}'' \citep[][p.\,168]{Tacconi2020}. 
In fact, despite almost two decades of effort, there are only three $z\geq$1 DSFGs detected in the HCN(1--0) emission\footnote {{In addition to these primarily star-forming galaxies, several high-z QSOs have been detected in the HCN(1--0) \citep{Solomon2003,vdBout2004, Carilli2005} or HCO$^+$(1--0) (e.g., \citealt{Riechers2006}).}} \citep{Gao2007, Oteo2017, Rybak2022a}. Indeed, as shown by the recent JVLA survey of \citet{Rybak2022a}, DSFGs might have low dense-gas fractions, making HCN(1--0) emission even harder to detect.

An alternative to observing the ground-state transitions are the mid-$J$ rotational HCN/HCO$^+$/HNC lines. These are both intrinsically brighter than the ground-state transitions and at $z\geq1$ fall conveniently into the easily-accessible 2-mm and 3-mm atmospheric windows. The mid-J HCN lines have been proposed to be better tracers of dense gas than HCN(1--0) \citep{Viti2017}, but their interpretation is more complex, as they might be sensitive to, e.g., shock heating, and mid-IR pumping \citep{Aalto2007,Kazandjian2015}.

In local galaxies, the HCN/HCO$^+$/HNC (3--2) and (4--3) lines have been systematically targeted from the scales of entire galaxies \citep{Baan2008, Bussmann2008, Gracia2008, Krips2008, Zhang2014, Imanishi2018b, Israel2023} to sub-kpc scales \citep{Wilson2008, Tan2018,Nishimura2024, Butterworth2025} and pc-scales (e.g., \citealt{Imanishi2018, Impellizzeri2019}). In contrast, only a handful of $z\geq1$ DSFGs have been detected in the $J_\mathrm{upp}$=3, 4, 5 emission lines \citep{Danielson2013, Oteo2017, Bethermin2018, Canameras2021, Yang2023}, {alongside several quasars \citep{Wagg2005, Riechers2010}}. Even spectral stacking studies have yielded only a handful of HCN/HCO$^+$(4--3) detections \citep{Spilker2014, Reuter2022, Hagimoto2023}. However, stacking studies only cover $J_\mathrm{upp}\geq3$ lines and miss anchoring to low-$J$ emission lines.

Converting these mid-$J$ HCN detections to dense-gas masses requires knowing the HCN spectral line energy distribution. However, the excitation of dense-gas tracers is almost completely unconstrained. Even in local galaxies, HCN/HCO$^+$/HNC ladders have been largely unexplored, with only a handful of studies dedicated to the topic (e.g., \citealt{Krips2008, Papadopoulos2014, Israel2023}). At high redshift, just one DSFG  has been detected in both the ground-state and a mid-$J$ HCN transition: SDP.9 ($z=1.575$), which was detected in HCN(1--0) and HCN(3--2) by \citet{Oteo2017}. The stacking studies are not helpful in this regard, as they do not cover the ground-state transition. Moreover, different transitions -- e.g., HCN(3--2) and (4--3) lines -- arise from \emph{different} galaxies, which can hide the true scatter of the excitation ratios (see a recent work on CO excitation by \citealt{Frias2023}).

\subsection{The PRUSSIC survey}

\textsc{Prussic}\footnote{“Prussic acid” is an alternative name for HCN, which was first isolated from the Prussian blue pigment.} is a comprehensive census of dense-gas tracers in high-redshift star-forming galaxies, aiming to drastically expand the number of high-redshift galaxies detected in dense-gas tracers.

In the first \textsc{Prussic} paper, \citet[Paper~I]{Rybak2022a} have presented the JVLA observations of the $J_\mathrm{upp}=1$ HCN, HCO$^+$, and HNC emission in six $z\sim3$ lensed DSFGs, finding low dense-gas fractions and elevated dense-gas star-forming efficiencies. The second paper \citep[Paper~II]{Rybak2023} presented ALMA imaging of the HCN, HCO$^+$, and HNC $J_\mathrm{upp}=4$ lines in a $z\approx3$ lensed DSFG SDP.81, finding an unusually high HCO$^+$/HCN ratio, indicating a subsolar metallicity and low mechanical heating.

In this third paper of the \textsc{Prussic} series, we present observations of the mid-$J$ HCN/HCO$^+$/HNC emission in eleven redshift\footnote{Throughout this paper, we assume a flat $\Lambda$CDM cosmology, with $\Omega_m=0.315$ and $H_0=67.4$ km s$^{-1}$ Mpc$^{-1}$ \citep{Planck2016}, and the Salpeter initial stellar mass function \citep[IMF,][]{Salpeter1955}.
}
$z=1.6-3.2$ lensed star-forming galaxies, obtained with ALMA and NOEMA.

The paper is structured as follows. In Section~\ref{sec:obs}, we present the targeted galaxies, details of NOEMA and ALMA observations, and data reduction. Section~\ref{sec:results} presents the resulting continuum (\ref{subsec:continuum}) and line measurements (\ref{subsec:results_hcn}). In Section~\ref{sec:discussion}, we place our observations in the context of low- and high-redshift surveys of dense gas (\ref{subsec:scaling_rels}), discuss the HCN/HCO$^+$/HNC line ratios (\ref{subsec:line_ratios}), present the dense-gas excitation ladders (\ref{subsec:ladders}), derive the star-forming efficiencies (\ref{subsec:sfe}), place constraints on the redshift evolution of the dense-gas content of galaxies (\ref{subsec:z_evolution}), and discuss the underlying systematics (\ref{subsec:systematics}).

\begin{table*}[ht]
\caption{Target galaxies, ordered by increasing redshift. Individual columns list the source position, source and lens redshift ($z_S$, $z_L$), sky-plane (lensed) FIR and CO(1--0) luminosities, CO(1--0) FWHM and FIR continuum magnification. $L_\mathrm{FIR}$ is inferred from modified black-body fits integrated over 8-1000~$\mu$m. For G09v1.40 and G12v2.43, $z_L$ are still unknown. For completeness, we also include SDP.81 from Paper~II \citep{Rybak2023}.}
    \centering
    \begin{tabular}{l|ccccccc}
    \hline
      Source & RA \& DEC & $z_s$ & $z_L$ & $L_\mathrm{FIR}^\mathrm{sky}$ & $L{'}^\mathrm{sky}_\mathrm{CO(1-0)}$ & FWHM CO(1--0)& $\mu_\mathrm{FIR}$\\
     & [J2000] & & & [$10^{12}$ $L_\odot$] & [$10^{10}$ K km s$^{-1}$ pc$^{-2}$] & [km/s]& \\
     \hline
   SDP.9 & 09:07:40.0 --00:41:59.8 & 1.575 & 0.613 & $66\pm1^\mathrm{O17}$ & $13\pm3^\mathrm{O17}$ & 370$\pm$10$^\mathrm{d}$& $8.8\pm2.2^\mathrm{B13}$ \\ 
   SDP.11 & 09:10:43.1 --00:03:22.8 & 1.786 & 0.793 & $64\pm1^\mathrm{O17}$ & 13$\pm$1$^\mathrm{O17}$ & 500$\pm$100$^\mathrm{O17}$ & $10.9\pm1.3^\mathrm{B13}$ \\
     G09v1.40 & 08:53:58.9 +01:55:37.7 & 2.092 & --- & $66\pm25$ $^\mathrm{Y17}$& $30\pm8^\mathrm{Y17}$ & $310\pm90^\mathrm{Y17,a}$ & 15.3$\pm$3.5$^\mathrm{B13}$ \\
    SDP.17 & 09:03:03.0 --01:41:27.1 & 2.305 & 0.23 & $71\pm26^\mathrm{Y17}$ & $26\pm3^\mathrm{H12}$ & $1000\pm130^\mathrm{H12}$ & 4.9$\pm0.7^\mathrm{B13}$\\

   J1202 & 12:02:07.6  +53:34:39 & 2.442 & 0.21 &  $81\pm4^\mathrm{H21}$ & $142\pm50^\mathrm{H21}$ & 600$\pm$20$^\mathrm{e}$ & 25.0$^\mathrm{H21,b}$\\ 

        G15v2.235 & 14:13:52.1 -00:00:24.4 & 2.478 & 0.547 & $28\pm7^\mathrm{Y17}$ &$43\pm5^\mathrm{H12}$ & $500\pm40^\mathrm{H12}$ & $1.8\pm0.3^\mathrm{B13}$ \\

J0209 & 02:09:41.3  +00:15:59 & 2.553 & 0.20 & $133\pm4^\mathrm{H21}$  & $84\pm11^\mathrm{R}$ & 410$\pm$20$^d$ & 14.7$\pm$0.3 \\

     G09v1.326 & 09:18:40.9 +02:30:45.9 & 2.581 & --- & $40\pm5^\mathrm{H12}$ &$33\pm8^\mathrm{H12}$ & $680\pm140^\mathrm{H12}$ & 1 \\
     SDP.130 & 09:13:05.4 --00:53:41 & 2.625 & 0.22 & $32\pm3^\mathrm{B13}$ &$25\pm4^\mathrm{F11}$ & $360\pm40^\mathrm{F11}$ & $8.6\pm0.4^\mathrm{F18}$\\
     
     NAv1.195 & 13:26:30.2 +33:44:07.6 & 2.951 & 0.785 & $75\pm20^\mathrm{Y17}$ &$57\pm20^\mathrm{H21}$ & $266\pm19^\mathrm{H21}$ & $4.1\pm0.3^\mathrm{F18}$ \\
     SDP.81 & 09:03:11.6 +00:39:07 & 3.042 & 0.299 & $43\pm1^\mathrm{R20}$ & $54\pm9^\mathrm{V11}$ & $435\pm54^\mathrm{V11}$ & $18.2\pm1.2^\mathrm{R20}$\\
     
     G12v2.43 & 11:35:26.3 -01:46:06.6 & 3.128 & --- & $90\pm2^\mathrm{Y17}$ &$15\pm3^\mathrm{H12}$ & $350\pm80^\mathrm{H12}$ & $9.2\pm3.2^\mathrm{G23,c}$\\

     \hline
    \end{tabular}
    \\
   \justify
    References: B13 -  \citet{Bussmann2013}, F11 - \citet{Frayer2011}, F17 - \citet{Falgarone2017}, G23 - \citet{Giulietti2023}. H12 -  \citet{Harris2012}, H21 - \citet{Harrington2021},
    I12 - \citet{Iono2012},
    O17 - \citet{Oteo2017}, R - Riechers et al., in prep; R20 - \citet{Rybak2020b}, V11 - \citet{Valtchanov2011}, Y17 - \citet{Yang2017}.\\
    $^a$ - Derived from CO(2--1), assuming CO is thermalised.\\
    $^b$ - No lens model available.\\
    $^c$ - Mean magnification for ALMA Bands 6, 7, and 8 continuum \citep{Giulietti2023}.\\
    $^d$ - derived from the CO(3--2) lines from our NOEMA observations.\\
    $^e$ - derived from ALMA observations, Project ID \#2023.1.00432.S (PI: M. Rybak).\\

    \label{tab:sources}
\end{table*}

\section{Targets, observations and imaging}
\label{sec:obs}

\subsection{Target sample}
\label{subsec:targets}

Our sample comprises eleven dusty star-forming galaxies at $z=1.6 - 3.1$ (median $z=2.52$), i.e., 2.1 -- 4.1~Gyr after the Big Bang. The targets are drawn from the \textit{Herschel} H-ATLAS survey \citep{Negrello2010, Negrello2017}, and the \textit{Planck} sample of strongly lensed DSFGs \citep{Canameras2015, Harrington2021}.

The intrinsic (magnification-corrected) far-infrared luminosity spans almost 1~dex ($L_\mathrm{FIR}=2.5\times 10^{12}-2.2\times 10^{13}$~$L_\odot$; median: $1.0\times10^{13}$~$L_\odot$), which corresponds to a star-formation rate SFR = 400 -- 4000~$M_\odot$~yr$^{-1}$ (median SFR = 1800~$M_\odot$~yr$^{-1}$). The targeted DSFGs thus include both extremely star-forming systems, as well as more ``normal'' galaxies at these redshifts. We list the properties of individual galaxies in Table~\ref{tab:sources}.

Five galaxies from our sample were observed with NOEMA, while six additional sources were observed by ALMA. SDP.130, was observed by both NOEMA ($J_\mathrm{upp}=3$ lines) and ALMA ($J_\mathrm{upp}=4$ lines). Five galaxies (including SDP.130) were targeted in both $J_\mathrm{upp}=3$ and 4 transitions. Finally, for completeness, we also include ALMA observations of SDP.81, previously presented in Paper~II.

\subsection{NOEMA observations and calibration}

The observations were carried out under the NOEMA projects S21CB and S23CB (PI: M.\,Rybak). {In total, five sources were observed: SDP.9, SDP.11, J0209, J1202, and SDP.130}. The S21CB observations were carried out between June 2021 and January 2022. The S23CB observations were carried out in October 2023. The array consisted of 9 to 12 antennas with 15~m diameter deployed in either D (the most compact) or C configuration. All observations were taken in the standard wideband mode, with two 7.744-GHz sidebands and spectral resolution of 2~MHz. As the observations spanned both summer and winter semesters, the precipitable water vapour (pwv) varied considerably between individual scheduling blocks, from $\sim$2~mm to $\sim$12~mm. Details of the NOEMA observations are given in Appendix~\ref{app:noema_obs}. 

The NOEMA data were calibrated and reduced using the \textsc{Gildas} package\footnote{\texttt{https://www.iram.fr/IRAMFR/GILDAS/}}. We used the new baseline-based amplitude calibration where applicable. Tracks for J0209 and SDP.130 showed some shadowing at low elevations; we manually flagged the shadowed antennas. Additionally, we tried to further improve the data fidelity by using self-calibration; however, this did not yield an appreciable increase in S/N. The data presented below are thus not self-calibrated. The resulting beam size FWHM varied from 2.6''$\times$0.9'' to 7.5''$\times$5.0''.

\subsection{ALMA observations and calibration}

We supplement our NOEMA observations with archival ALMA data from project 2017.1.01694.S (PI: I. Oteo). This project targeted a total of eight lensed DSFGs discovered in \textit{Herschel} surveys. Seven sources were targeted in the $J_\mathrm{upp}=4$ transitions, while G09v1.40 (which is at a lower redshift than the rest of the sample) was observed in the $J_\mathrm{upp}=3$ transitions. Data for SDP.81 have been recently presented in \citet{Rybak2023}, here we present the remaining seven sources.

The observations were carried out between 2018 March 30 and 2018 May 1 using ALMA Band~3 receivers. The time on-source ranged between 0.5 and 1.7~hours. The baselines ranged between 15 and 500~m for all sources but NAv1.195, for which baselines out to 740~m were used. The spectral setup consisted of a single sideband with two spectral windows, each with 128 channels with a frequency resolution of 31.250~MHz. The other sideband was not used.

We process the data using the standard ALMA pipeline; the data quality was generally excellent. The resulting beam size FWHM varied from 2.1''$\times$1.6'' to 2.7''$\times$2.1''.

\subsection{Imaging procedure}
{We imaged the data using \textsc{Casa}'s \texttt{tclean} task. To maximise the sensitivity, we used natural weighting for all images. We used a pixel size of 0.1''$\times$0.1'', ensuring that the synthesised beam is properly sub-sampled (see Tab.~\ref{tab:obs_summary}).} 

For both NOEMA and ALMA data, we first identified potential line emission by examining dirty-image cubes. {The resulting cubes had a default channel spacing of 20~MHz; in case these did not show any signal, we re-imaged the data using a coarser resolution of 100~MHz}. We then created wide-band continuum images excluding channels with emission lines. For NOEMA, we created separate continuum images for the upper/lower sideband. 

\begin{figure*}[ht]
    \centering
            
    J1202 (z=2.442)\\
     \includegraphics[height=4.25cm]{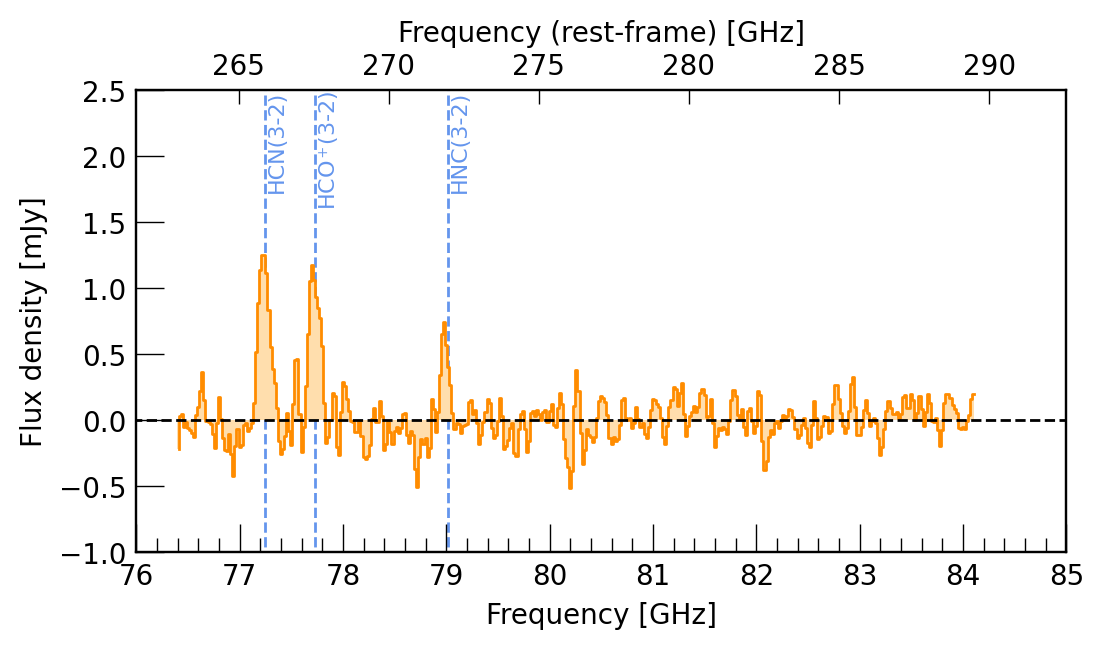}
    \includegraphics[height=4.25cm]{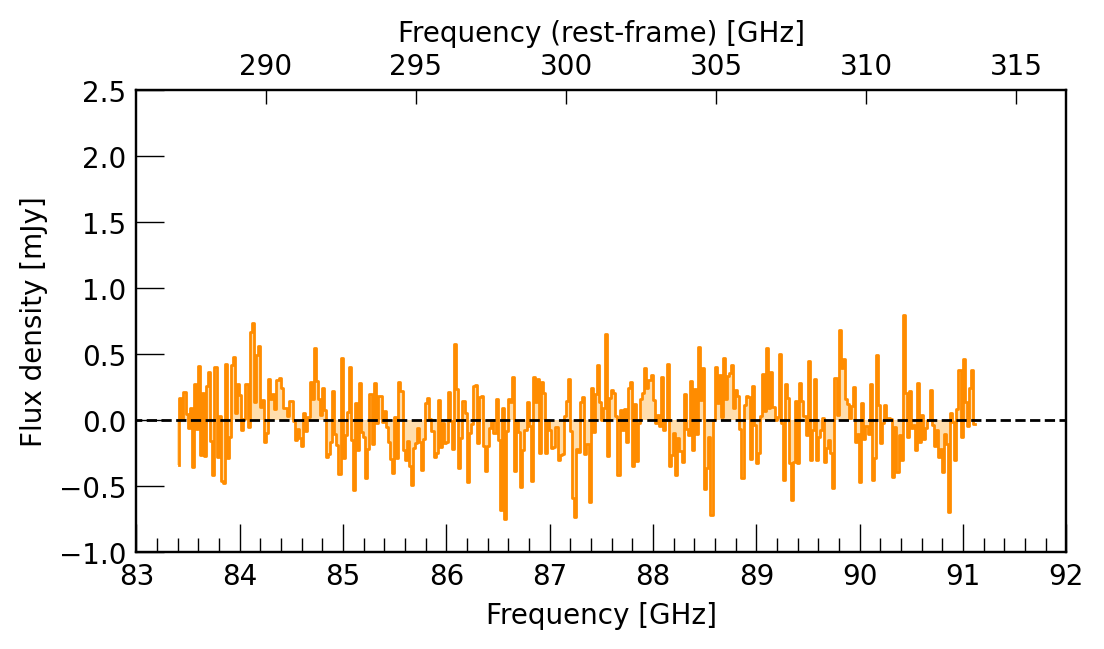}\\
    \includegraphics[height=4.25cm]{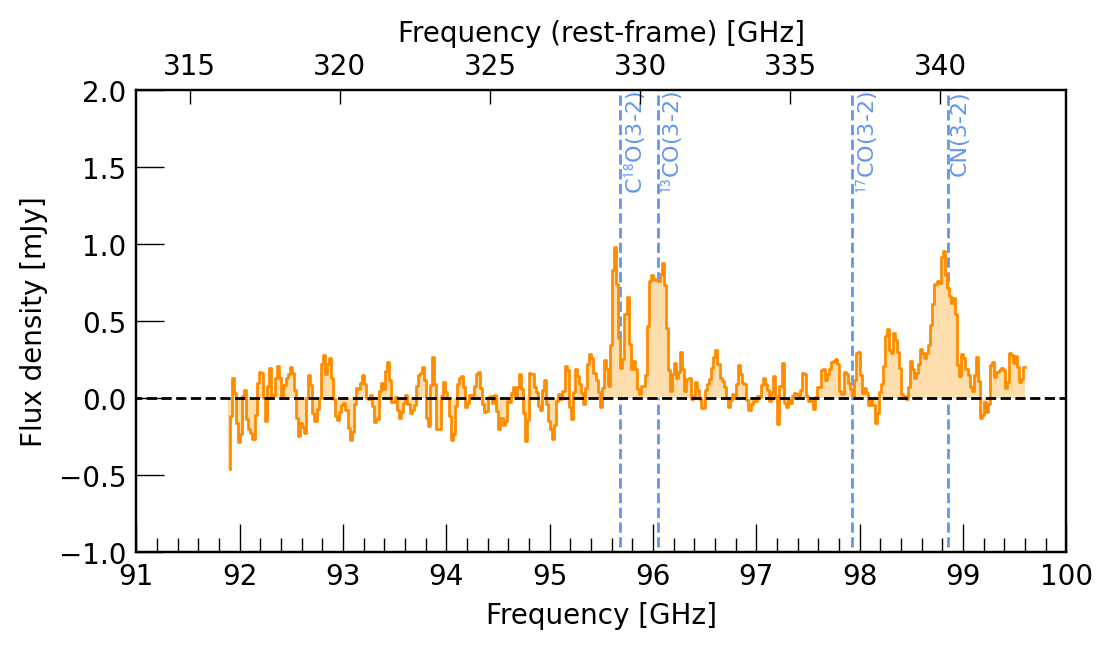}
    \includegraphics[height=4.25cm]{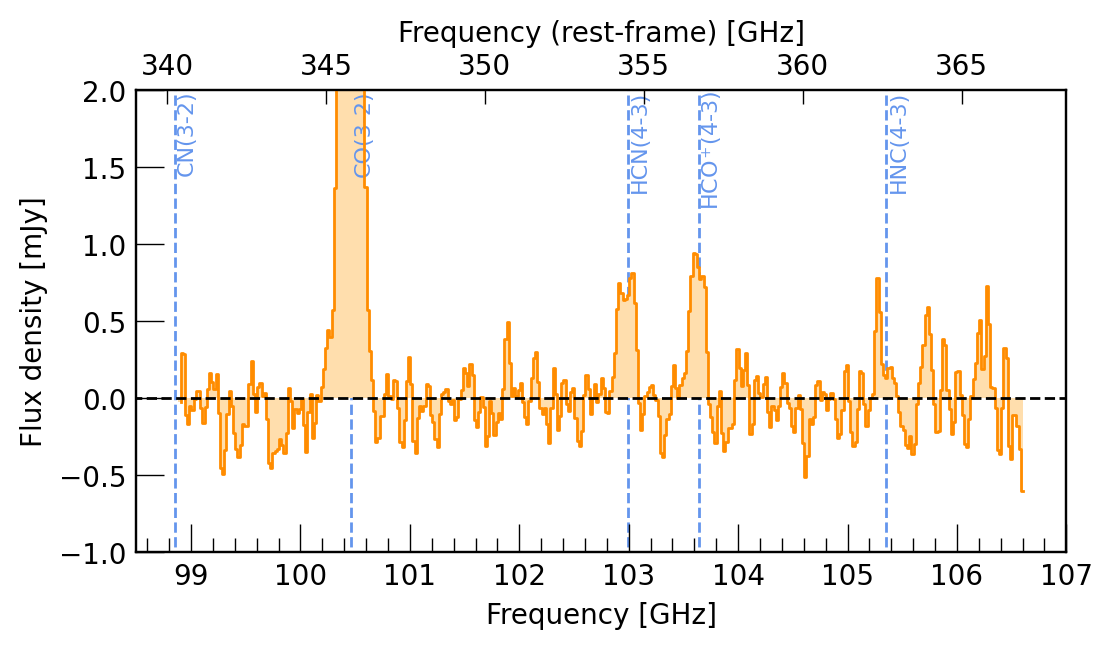}\\
       
    \caption{NOEMA Band-1/2 spectra for J1202. We detect the HCN, HCO$^+$, HNC (3--2) and (4--3) lines, alongside the CO(3--2) isotopologues and CN(3--2). The HNC(3--2) and (4--3) line profiles differ significantly from those of HCN and HCO$^+$. See Appendix~\ref{app:spectra} for the spectra of remaining targets.}
    \label{fig:noema_spectra}
\end{figure*}

In the next step, we subtracted the continuum signal by fitting a constant function to the line-free channels using CASA's \texttt{uvcontsub} task. We then extracted spectra for each source using hand-drawn apertures corresponding to the 2-$\sigma$ contour in continuum maps; these are presented in Fig.~\ref{fig:noema_spectra}. Finally, we created narrow-band images of each line by imaging the channels within $\pm$0.5 FWHM of the low-$J$ CO line (Tab.~\ref{tab:sources}). We performed the deconvolution manually, with a stopping threshold of 1.5$\sigma$.

\begin{figure*}
\centering

\includegraphics[height=4cm]{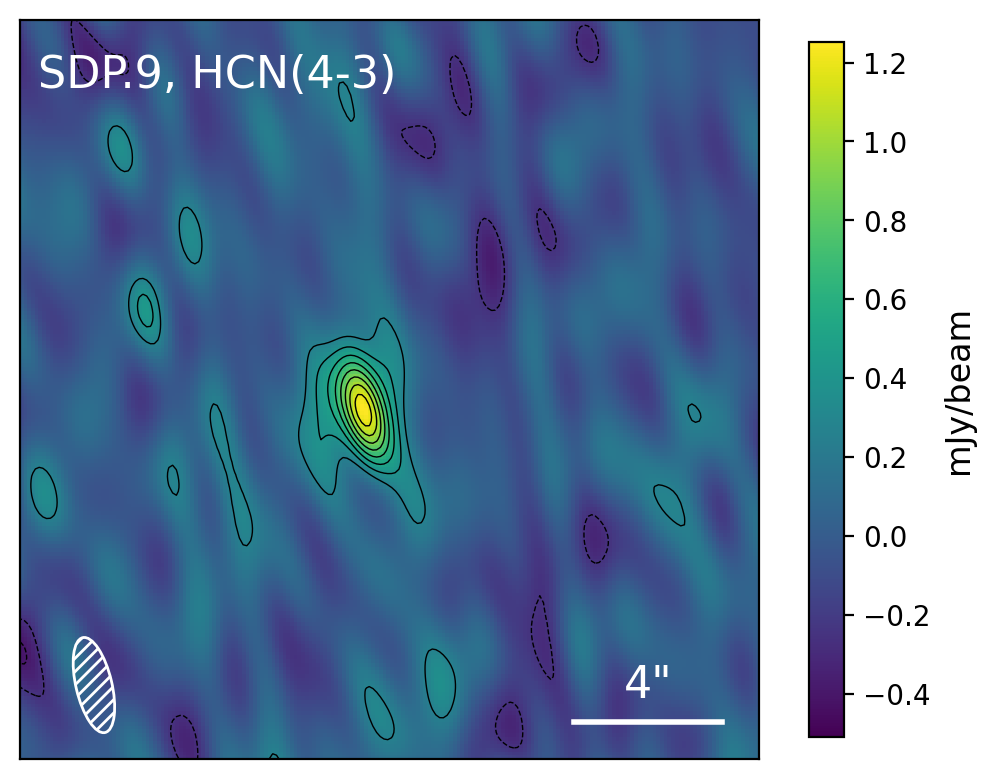}
    \includegraphics[height=4cm]{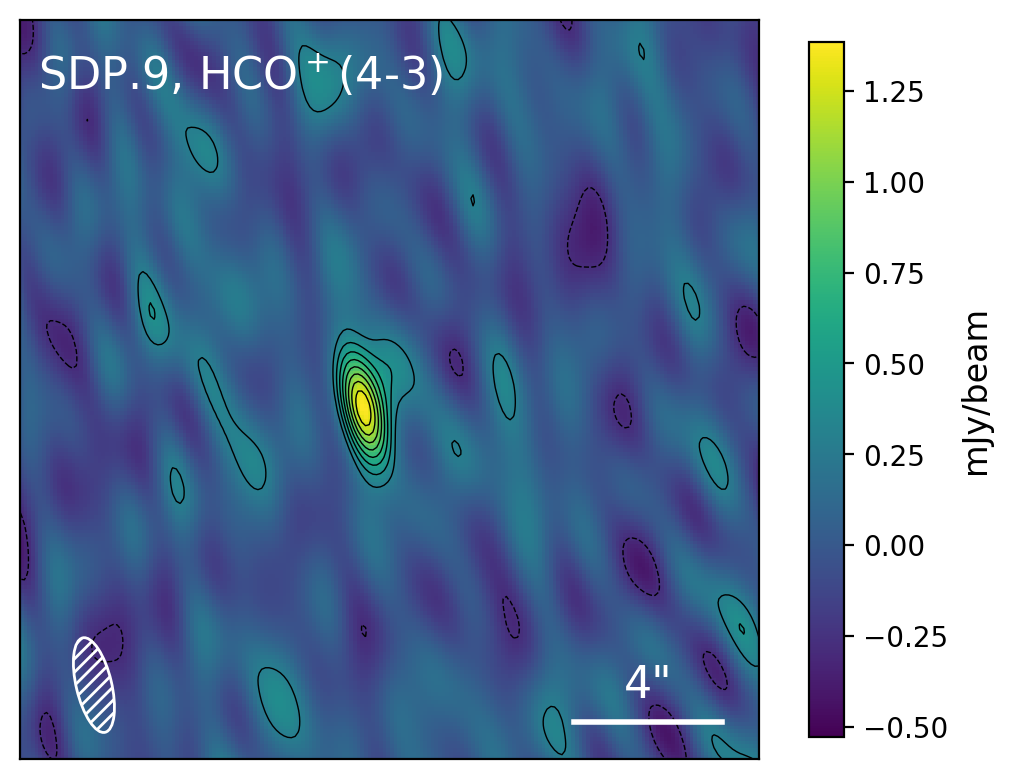}
    \includegraphics[height=4cm]{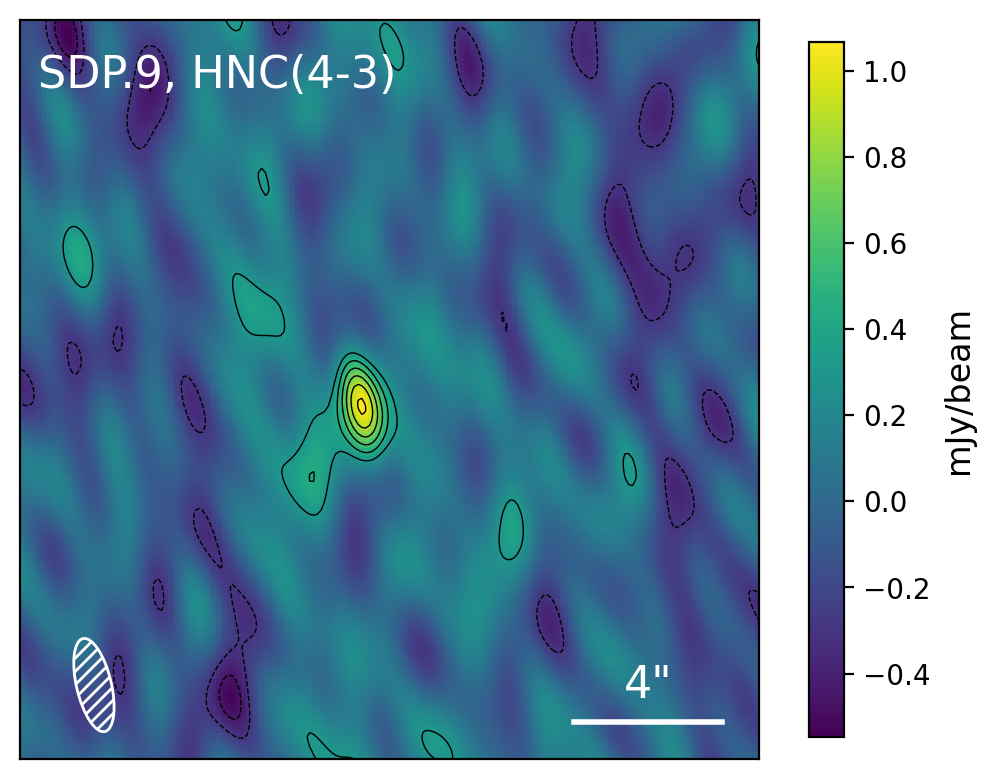}\\
    
    \caption{NOEMA narrow-band images of the HCN/HCO$^+$/HNC lines for SDP.9, which is detected in all three transitions. For the remaining sources, see Appendix. Contours start at $\pm$2$\sigma$ and increase in steps of 2$\sigma$. }
    \label{fig:noema_hcnhcohnc}
\end{figure*}

To derive the line fluxes, we created narrow-band images over $\pm$0.5~FWHM of the linewidth of the lowest-$J$ CO line available (Fig.~\ref{fig:noema_hcnhcohnc}). For a Gaussian profile, this range should include 76\% of the total line flux; we corrected the inferred total line luminosities accordingly. We preferred this approach to fitting parametric profiles to the spectra, as the lines might not follow exactly a Gaussian profile (see Section~\ref{subsec:linewidths}).

\section{Results}
\label{sec:results}

\subsection{Continuum emission}
\label{subsec:continuum}

We first examined the continuum images for individual sources (Fig.~\ref{fig:noema_continuum} and ~\ref{fig:alma_continuum_compact}), which corresponds to rest-frame wavelengths 845~$\mu$m or 1.1~mm.  All NOEMA targets are marginally resolved. In the ALMA imaging, G09v1.326 and NAv1.195 are resolved into two separate components; SDP.130 is marginally resolved, with a small extension to the south-west, which is arising from a source at a lower redshift \citep{Falgarone2017}. For the lensed sources, a significant fraction of the observed 2-mm or 3-mm continuum might arise from AGNs in the foreground lensing galaxies, such as seen in ALMA Band~3 observations of SDP.81 \citep{Rybak2023}.

\subsection{HCN, HCO$^+$ and HNC emission}
\label{subsec:results_hcn}

Figure~\ref{fig:noema_hcnhcohnc} shows the narrow-band images at the positions of HCN, HCO$^+$ and HNC lines. We detect HCN/HCO$^+$/HNC (3--2) emission lines in all targeted galaxies except SDP.130, and HCN/HCO$^+$/HNC (4--3) emission lines in seven out of nine targets. The two sources without any line detections are G09v1.326 and G15v2.235. Additionally, J1202 is not detected in the HNC(4--3) line, while SDP.130 is not detected in the HCO$^+$(4--3) line \citep{Rybak2023}. The inferred line luminosities (or 3$\sigma$ upper limits) are listed in Tab.~\ref{tab:results_lines}. {We provide details on the detections of CO isotopologues and CN(3--2) lines in Appendices~\ref{sec:co_isotopologues} and \ref{sec:cn}.}

\begin{table*}
\caption{Rest-frame frequencies and luminosities (in units of $10^8$ K km s$^{-1}$ pc$^2$) for  HCN, HCO$^+$, and HNC lines in individual galaxies. The reported values are not corrected for lensing magnifications. The upper limits are given as 3$\sigma$.}
    \centering
    \begin{tabular}{l|c|ccc|ccc}
    \hline
     Line& HCN(1--0) & HCN(3--2) & HCO$^+$(3--2) & HNC(3--2) & HCN(4--3) & HCO$^+$(4--3) & HNC(4--3)\\
 $f_0$ [GHz] & 88.632 & 256.886 & 267.558 & 271.981 & 354.505 & 356.734 & 362.630\\
     \hline 
   SDP.9 & 380$\pm$130$^\mathrm{O17}$ & 164$\pm$27$^\mathrm{O17}$ & 85$\pm$34$^\mathrm{O17}$ & 99$\pm$19$^\mathrm{O17}$ & 84$\pm$9 & 80$\pm$13 & 83$\pm$13\\
   SDP.11 & $\leq$540$^\mathrm{O17}$ & 170$\pm$25$^\mathrm{O17}$ & 73$\pm$6$^\mathrm{O17}$ & $\leq$20$^\mathrm{O17}$ & 94$\pm$24 & $\leq$54 & $\leq$35 \\
   G09v1.40 & --- & $94\pm19$ & $113\pm19$ & $80\pm18$ & --- & --- & ---\\
   SDP.17  & --- & --- & --- & --- & $126\pm16$ & $80\pm16$ & $125\pm17$ \\
   J1202 & 800$\pm$400$^\mathrm{R22}$ & 620$\pm$90 & 600$\pm$90 & 176$\pm$90 & 230$\pm$60 & 240$\pm$60 & $\leq$120 \\ 
   G15v2.235 & --- & --- & --- & ---  & $\leq74$ & $\leq75$ & $\leq78$\\
    J0209  & $\leq$1000$^\mathrm{R20}$ & 560$\pm$70 & 310$\pm$70 & 300$\pm$50 & 960$\pm$50$^\mathrm{NG}$&  480$\pm$30$^\mathrm{NG}$ & 570$\pm$30$^\mathrm{NG}$\\
   G09v1.326 & --- & --- & --- & --- & 52$\pm$13 & 61$\pm$15 & 78$\pm$13 \\
   SDP.130 & $\leq$140$^\mathrm{R22}$ & $\leq$130 & $\leq$110 & $\leq$80 & 52$\pm$23 & $\leq$45 & 31$\pm$24 \\
   NAv1.195  & --- & --- & --- & --- & 180$\pm$30 & 170$\pm$20 & $\leq$290 \\
   SDP.81$^\mathrm{R23}$ & $\leq$370 & --- & --- & --- & $\leq$72 & 160$\pm$43 & $\leq$71\\ 
   G12v2.43 & --- & --- & --- & --- & 210$\pm$15 & 146$\pm$14 & 210$\pm$15 \\
     
     \hline
     \multicolumn{8}{l}{References: {O17} - \citet{Oteo2017}; {R22} - \citet{Rybak2022a}; R23 - \citet{Rybak2023}; NG - N. Geesink,} \\
     \multicolumn{8}{l}{MSc Thesis, Leiden 2025} \\
    \end{tabular}
    \label{tab:results_lines}
\end{table*}

\subsubsection{Linewidths of dense-gas tracers vs CO}
\label{subsec:linewidths}

Do linewidths of dense-gas tracers differ significantly from those of low-J CO lines, which trace the bulk of molecular gas? We fit a one-dimensional Gaussian profile to the slices of spectra within $\pm$1000~km/s of the HCN, HCO$^+$, and HNC lines. Although some lines show more complex profiles, a one-dimensional Gaussian profile allows a direct comparison to low-J CO linewidths reported in the literature. The measured linewidths are listed in Table~\ref{tab:linewidths}.

\begin{figure}
    \centering
    \includegraphics[width=0.45\textwidth]{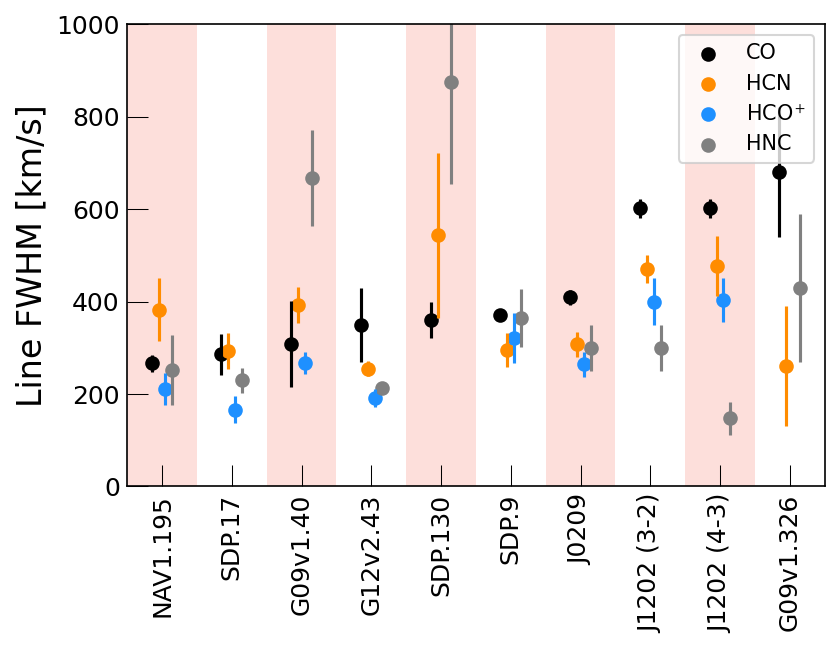}
    \caption{Comparison of CO, HCN, HCO$^+$ and HNC linewidths in individual galaxies, ordered by increasing CO linewidth. The linewidths of different tracers are generally consistent within 1$\sigma$ uncertainties. Note the large discrepancies between HCN/HCO$^+$ and HNC linewidths in G04v1.40 and J1202.}
    \label{fig:fwhm_fwhm}
\end{figure}

Figure~\ref{fig:fwhm_fwhm} compares the linewidths of HCN(3--2)/(4--3) and CO(1--0) lines. For J1202, we fit both the (3--2) and (4--3) lines; the HCN and HCO$^+$ linewidths are consistent, but HNC(4--3) is significantly narrower than HNC(3--2). 

On average, we find that the linewidths of dense-gas tracers are consistent with those of CO(1--0) emission. Namely, the mean linewidth (FWHM) ratios are as follows: HCN/CO: 0.95$\pm$0.20; HCO$^+$/CO: 0.70$\pm$0.13; HNC/CO: 1.00$\pm$0.27. However, several sources - particularly G09v1.40 and J1202 - show significant differences between different lines; we discuss them in detail in Section~\ref{sec:offsets}. 

\begin{table}[]
\caption{HCN, HCO$^+$ and HNC FWHM linewidths for individual sources, inferred from one-dimensional Gaussian fits to the observed (i.e., not de-lensed) spectra.}
    \label{tab:linewidths}
    \centering
    \begin{tabular}{lc|ccc}
    \hline
    Source & Lines & HCN & HCO$^+$ & HCN \\
         &  & [km/s] & [km/s] & [km/s] \\
         \hline
     SDP.9 & (3--2) & 300$\pm$40 & 320$\pm$50 & 370$\pm$60 \\
     G09.v140   & (3--2) & 390$\pm$40 & 270$\pm$ 20 & 670$\pm$100 \\
     SDP.17 & (4--3)& 290$\pm$40 & 170$\pm$30 & 230$\pm$40 \\
     J1202 & (3--2) & 470$\pm$30& 470$\pm$60 & 300$\pm$40 \\
     J1202 & (4--3) & 480$\pm$50& 400$\pm$50 & 150$\pm$40 \\
     J0209 & (3--2) & 310$\pm$30 & 260$\pm$30 & 300$\pm$50 \\
     G09v1.326 & (4--3) & 260$\pm$130 & --- & 430$\pm$160 \\
     SDP.130 & (4--3)& 810$\pm$50& --- & 870$\pm$220 \\
     NAv1.195 & (4--3) & 380$\pm$70 & 210$\pm$40 &  250$\pm$70 \\
     G12v2.43 & (4--3) & 260$\pm$20 & 190$\pm$20 & 210$\pm$10 \\
     
     \hline
    \end{tabular}
    
\end{table}

\section{Discussion}
\label{sec:discussion}

\subsection{Dense-gas tracer scaling relations}
\label{subsec:scaling_rels}

\begin{figure*}
\centering
\includegraphics[width = 0.49\textwidth]{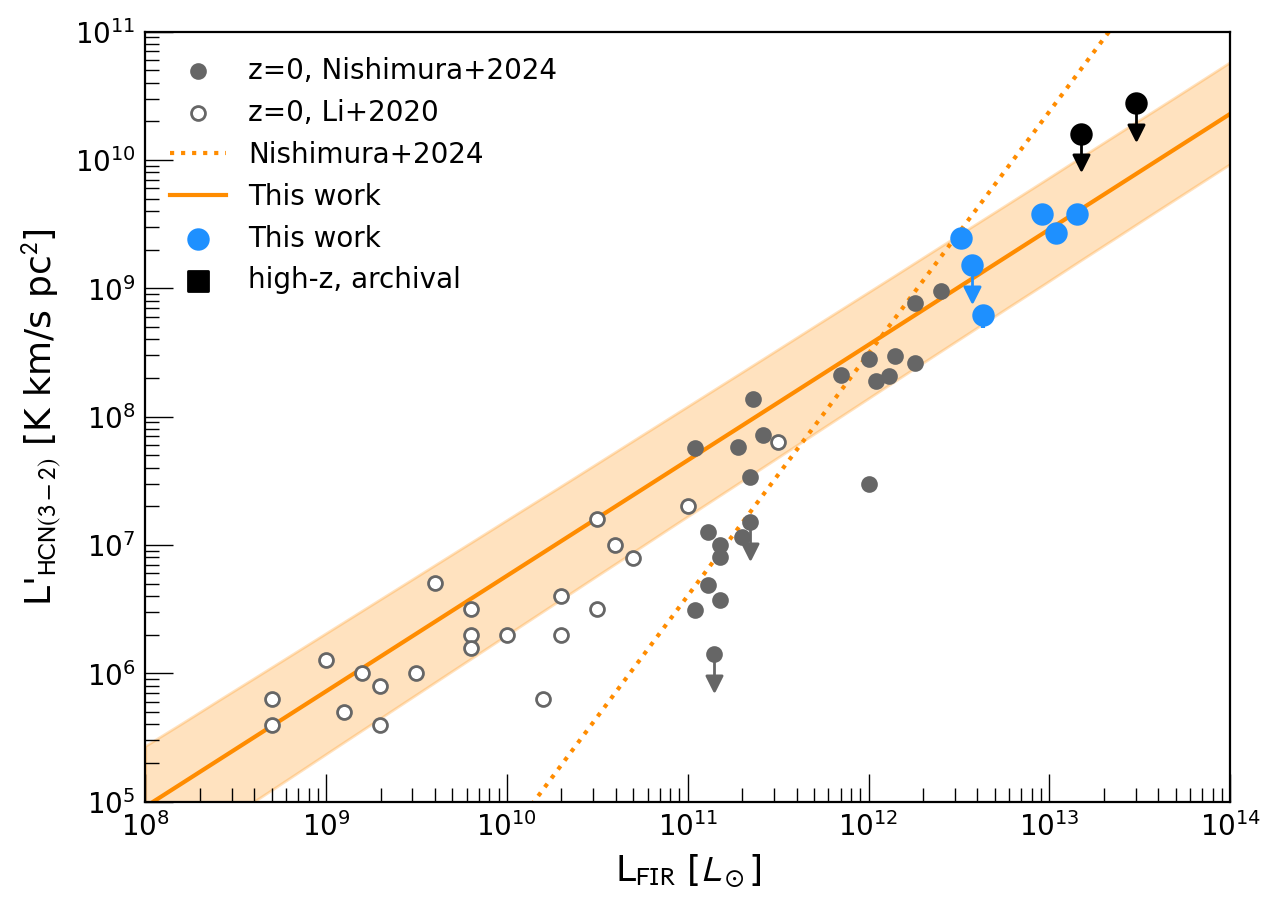}
    \includegraphics[width = 0.49\textwidth]{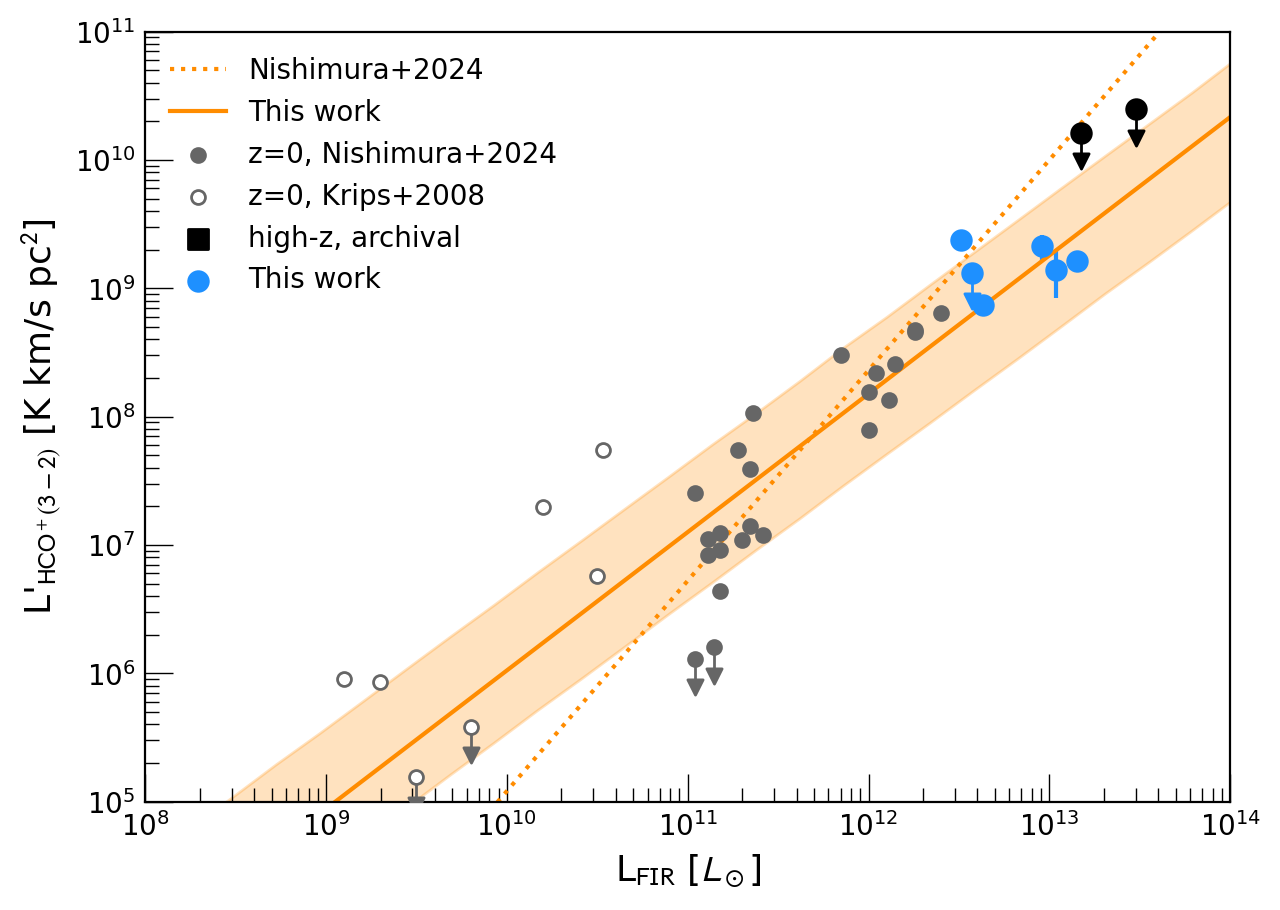} \\
    
    \includegraphics[width = 0.49\textwidth]{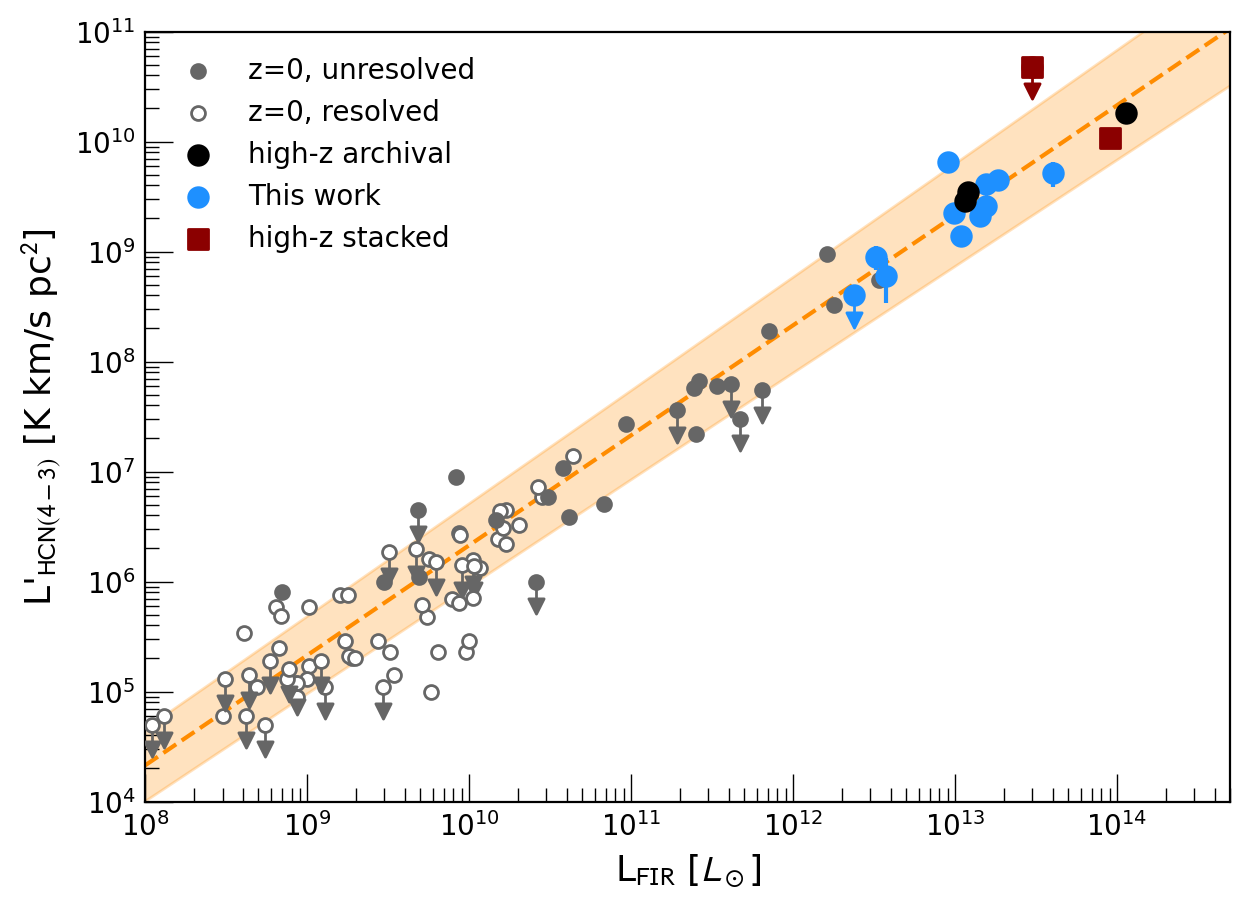}
    \includegraphics[width = 0.49\textwidth]{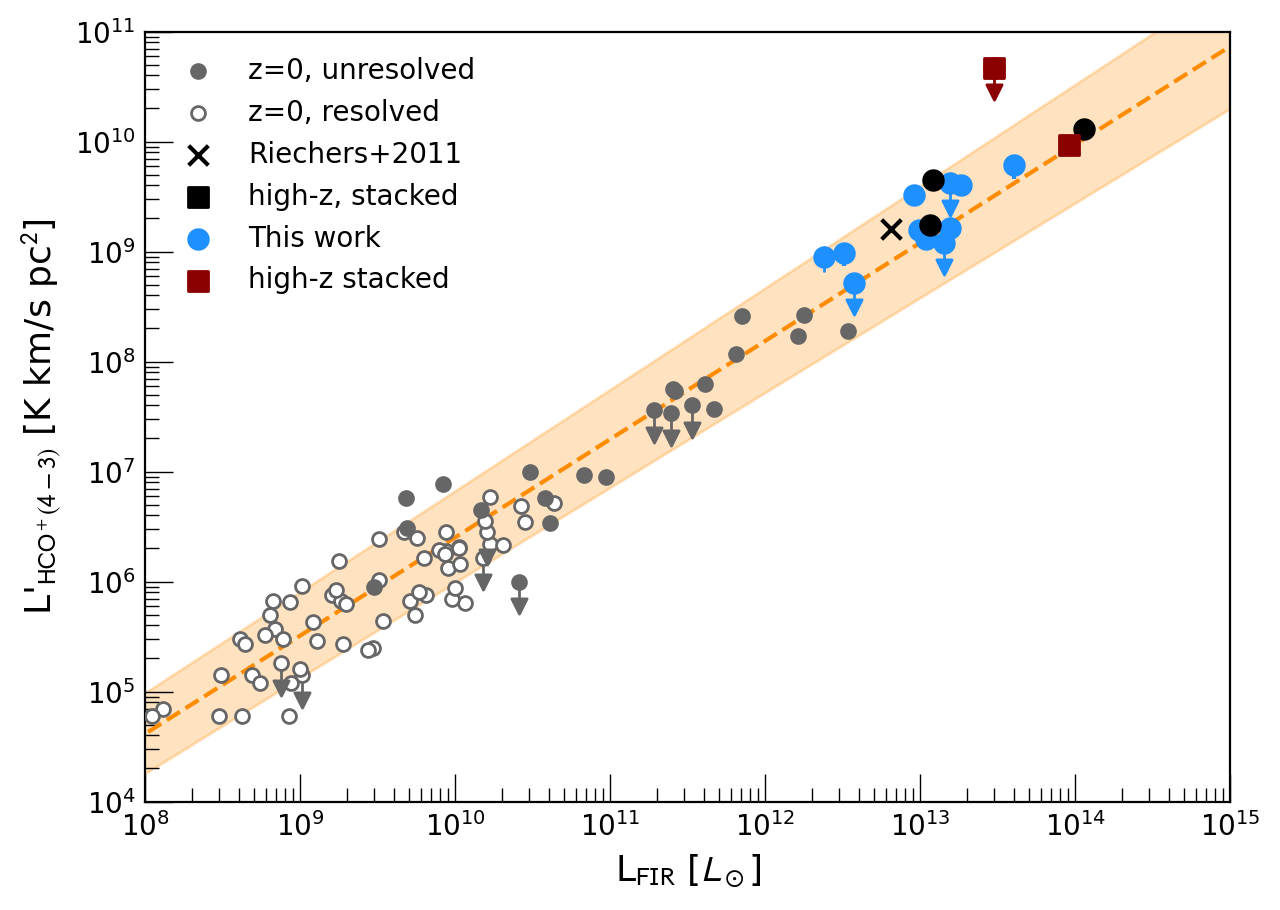} \\

    \caption{Correlation between far-IR luminosity and HCN/HCO$^+$(3--2) and (4--3) luminosities. Orange solid/dashed lines indicate the empirical trends from this work and \citet{Zhang2014} with $\pm$1$\sigma$ scatter; orange dotted lines indicate the empirical trend from \citet{Nishimura2024}. Individual data points show $z=0$ galaxy-averaged \citep{Greve2009, Papadopoulos2014, Zhang2014, Nishimura2024} and resolved observations \citep{Tan2018}, and high-$z$ detections in individual galaxies \citep[black, solid]{Riechers2011, Canameras2021} and this work (blue). We also show results from spectral stacks \citep{Reuter2022, Hagimoto2023}. Where appropriate, we correct for the lensing magnification. {We also plot errorbars on high-z measurements; these are typically smaller than symbol size.}} \label{fig:FIR_HCN_HCO}
\end{figure*}

We now compare our HCN, HCO$^+$ and HNC observations to previous observations of local and high-redshift galaxies. Figure~\ref{fig:FIR_HCN_HCO} presents our data in the context of $z=0$ resolved \citep{Tan2018, Li2020} and galaxy-averaged measurements \citep{Greve2009, Krips2008, Greve2009, Papadopoulos2014, Zhang2014, Nishimura2024}. Additionally, we compare our observations to the empirical power-law scaling relations of \citet{Nishimura2024} and \citet{Zhang2014}. 

However, the \citet{Nishimura2024} relations were derived over a narrow range of $L_\mathrm{FIR}= (1-15)\times10^{11}$~$L_\odot$ and do not provide a good fit to data at higher and lower FIR luminosities. We therefore re-fit\footnote{We fit the data in the log-log space using the \textsc{Linmix} package \citep{Kelly2007} which explicitly includes non-detections.} the HCN(3--2) and HCO$^+$(3--2) data using a power-law model, fitting data from \citet{Krips2008, Li2020, Nishimura2024} and our high-z measurements.
We obtain:
\begin{equation}
    \log L_\mathrm{FIR} = (0.91\pm0.05) \times  \log L'_\mathrm{HCN(3-2)} + (3.70\pm0.44)
\end{equation}

\begin{equation}
    \log L_\mathrm{FIR} = (1.04\pm0.09) \times  \log L'_\mathrm{HCO^+(3-2)} + (3.90\pm0.54)
\end{equation}

The slopes for HCN(3--2) and HCO$^+$(3--2) are close to unity (within 1--2$\sigma$). The almost-linear slopes contrast with predictions for sub-linear correlation from older theoretical work \citep{Krumholz2007, Narayanan2008} and surveys of local galaxies (e.g., \citealt{Bussmann2008} and \citealt{Juneau2009}, who found slopes of $\approx0.7$ for the log($L_\mathrm{FIR}$)--log ($L'_\mathrm{HCN(3-2)})$' \citealt{Juneau2009}).

For the $J_\mathrm{upp}=4$ lines, we consider the \citet{Zhang2014} relations\footnote{The original fit from \citet[p.\,4]{Zhang2014} contains a typo; we give the correct form (Z. Zhang, priv. comm.)}, which were
derived using spatially unresolved observations of nearby galaxies:

\begin{equation}
\log L_\mathrm{FIR} = (1.00\pm0.04) \times  \log L'_\mathrm{HCN(4-3)} + (3.67\pm0.28)
\end{equation}
and:
\begin{equation}
\log L_\mathrm{FIR} = (1.12\pm0.05) \times  \log L'_\mathrm{HCO^+(4-3)} + (2.83\pm0.34).    
\end{equation}

As shown in Fig.~\ref{fig:FIR_HCN_HCO}, {all our measurements are consistent with the \citet{Zhang2014} $L_\mathrm{FIR}$--HCN(4--3) trend within the $\pm$1$\sigma$, similar to the single detection from \citet{Canameras2021} and stacking results from \citet{Reuter2022}.} Similar to the HCN/HCO$^+$(3--2) lines, the linear slopes inferred from the data contrast with predictions from \citet{Narayanan2008} who predicted a slope of $\approx$0.6.

{While the HCN(4--3) luminosities in our DSFGs scatter equally above and below the \citet{Zhang2014} trend, they are slightly overluminous in HCO$^+$(4--3), although still within the 1$\sigma$ scatter. We provide a more detailed discussion of line ratios in Fig.~\ref{fig:line_ratios}}. We do not show the corresponding plots for the HNC lines due to the paucity of $z\approx0$ measurements.

\subsection{HCN, HCO$^+$, and HNC line ratios}
\label{subsec:line_ratios}

The ratios of HCN, HCO$^+$, and HNC line luminosities provide insights into the thermodynamics and energetics of high-density gas. In particular, HCN/HCO$^+$ enhancements have been proposed as an AGN signature (e.g., \citealt{Kohno2003, Gracia2006, Krips2008, Privon2015, Izumi2016}).
However, as demonstrated by \citet{Viti2017} and \citet{Privon2020}, the HCN enhancements show rapid time evolution (on $\approx$1-Myr timescales) which decouples them from the AGN activity. Moreover, recent studies of nearby galaxies indicate that  HCN/HCO$^+$ can be only used as a reliable AGN diagnostic at scales of $\leq$100~pc \citep{Butterworth2025}.

\begin{figure}
\centering
    \includegraphics[width = 0.4\textwidth]{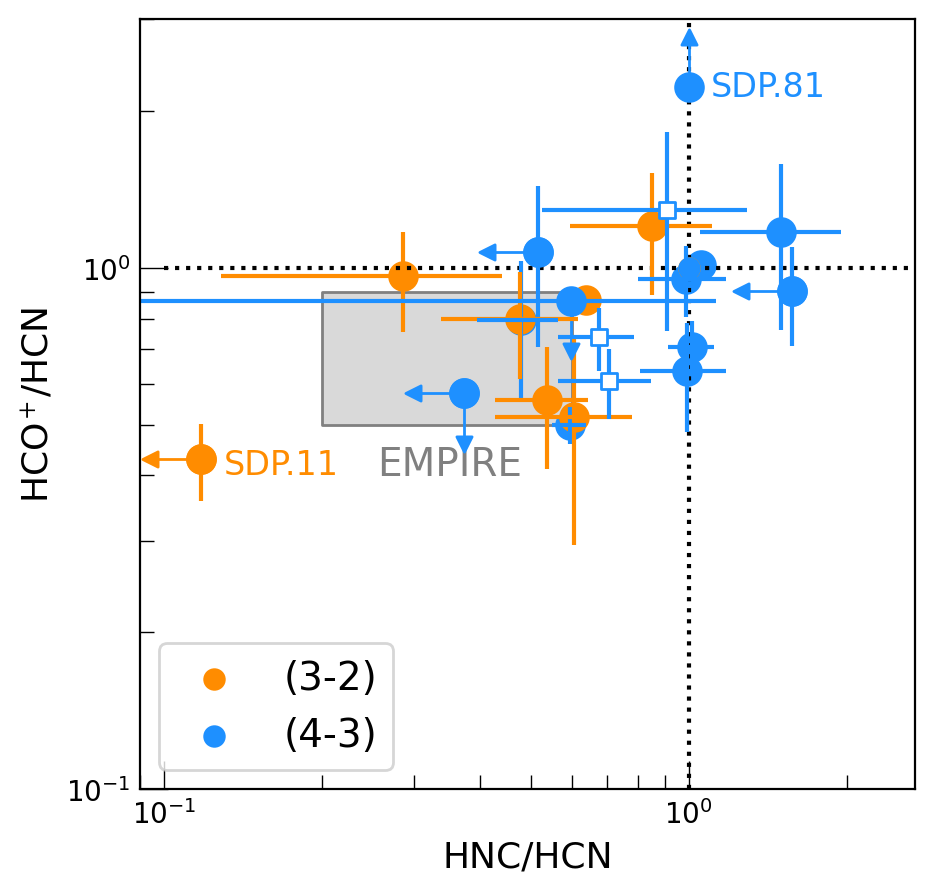}
    \caption{HCO$^+$/HCN and HNC/CO luminosity ratios in our sample (solid circles). We also include data from  \citet{Yang2023} for comparison (empty squares). The shaded regions indicate HCO$^+$/HCN and HNC/HNC ratios for the EMPIRE sample \citet{Jimenez2019}. HCN is the brightest line in most of our targets. SDP.81 shows strong HCO$^+$ enhancement, potentially due to low metallicity \citep{Rybak2023}. The HCN(3--2) emission line in SDP.11 is unusually bright, potentially due to AGN-driven chemistry.} 
    \label{fig:line_ratios}
\end{figure}

As shown in Fig.~\ref{fig:line_ratios}, the HCO$^+$/HCN ratios in our sample are consistent with $\geq1$ with the exception of G12v2.43 and SDP.81. The elevated HCO$^+$ luminosity in these sources might be due to sub-solar metallicity or time-evolution effects \citep{Rybak2023}. 

The HNC/HCN ratio has been proposed to be a sensitive diagnostic of the gas temperature. Specifically, while HCN and HNC form via the same pathways with a branching ratio of $\approx$1:1, their relative abundances are modified via the HNC + H $\rightarrow$ HCN + N reaction and HNC + O $\rightarrow$ NH + CO reations, with $\Delta T\approx200$~K (e.g., \citealt{Graninger2014, Jin2015, Hacar2020}). As a result, the HNC/HCN ratio is predicted to decrease with increasing gas kinetic temperature (e.g., \citealt{Hacar2020}).

We find HNC/HCN $\leq$1 in all sources except~G12v2.43 (HNC/HCN=$1.4 \pm 0.2$). Compared to the EMPIRE sample, our DSFGs have comparable HCO$^+$/HCN ratios, but have elevated HNC/HCN ratios. 
Using the \citet{Hacar2020} calibration $T_\mathrm{kin}=10\times [I_\mathrm{HCN}/I_\mathrm{HNC}]$, we find kinetic temperatures of 10 -- 35~K, similar to those in the Orion Nuclear Cluster \citep{Hacar2020}.

\subsection{Dense-gas tracer excitation}
\label{subsec:ladders}

Measuring the HCN/ HCO$^+$/ HNC spectral line energy distribution (SLED) in DSFGs is necessary for a proper comparison of observations of different rotational transitions.

\begin{figure*}
    \centering
    \includegraphics[width = 0.33\textwidth]{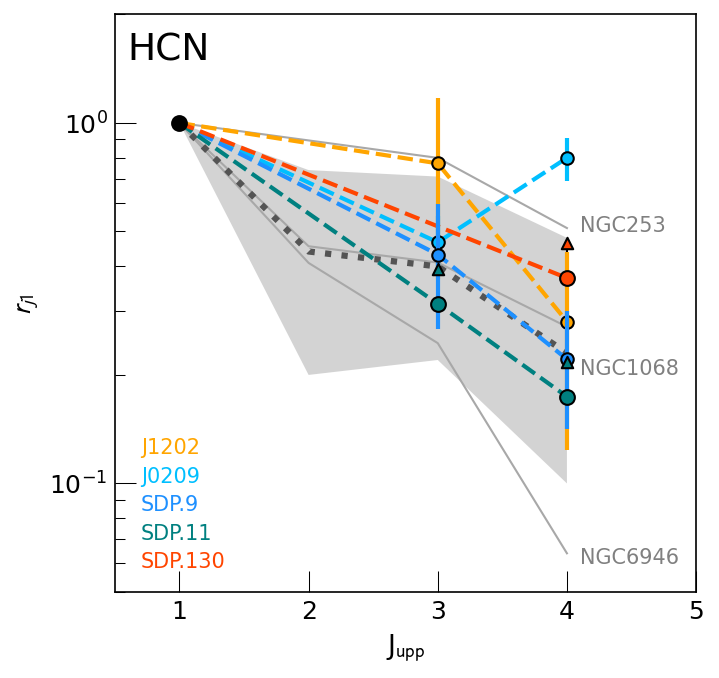}
     \includegraphics[width = 0.33\textwidth]{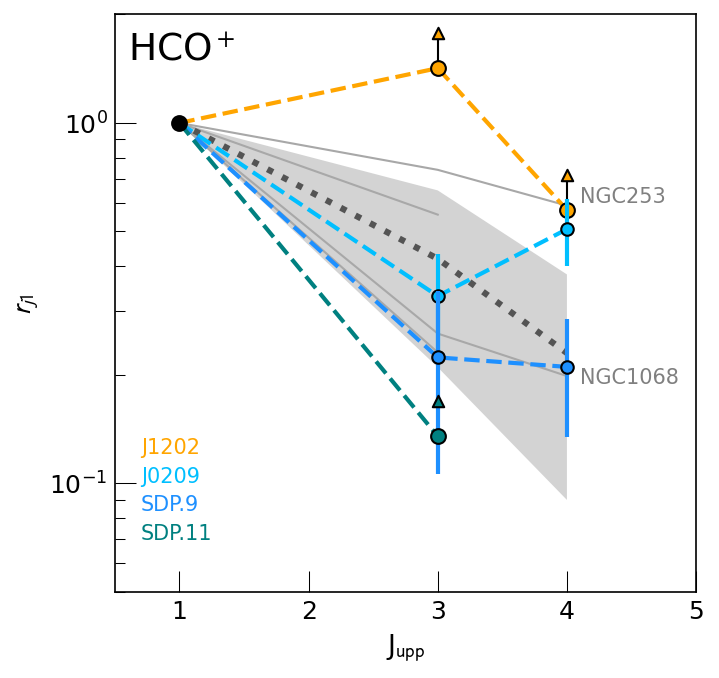}
     \includegraphics[width = 0.33\textwidth]{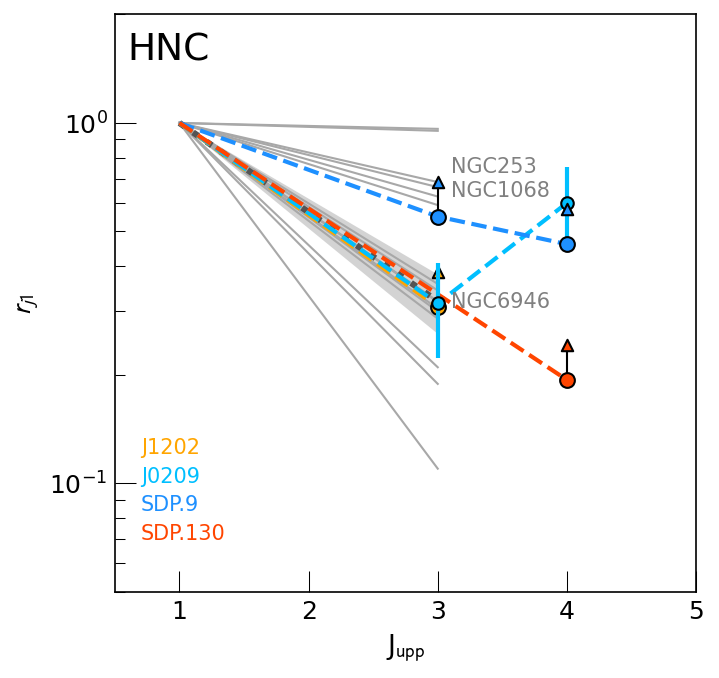}\\
    \caption{Excitation coefficients of HCN, HCO$^+$ and HNC in DSFGs from our sample (coloured), normalised to the (1--0) transition and compared to $z\sim0$ galaxies from the \citet{Israel2023} compilation (grey). The arrows indicate the 3$\sigma$ upper / lower limits. For comparison, we highlight SLEDs for nearby (U)LIRGs NGC253, NGC1068, and NGC6946. In J0209 and J1202, the HCN and HCO$^+$ SLEDs appear to be superthermalised. The sparseness of the HNC plot reflects the lack of detections at $z\approx0$.}
    \label{fig:excitation_hcn}
\end{figure*}

We observed four galaxies in multiple HCN transitions -- J0209, J1202, SDP.9, and SDP.130. To anchor our SLEDs, we use ground-state observations from \citet[][SDP.9]{Oteo2017}, \citet[][J1202, SDP.130]{Rybak2022a} and Riechers et al., in prep. (J0209). For J0209, we include source-integrated HCN/HCO$^+$/HNC(4--3) fluxes from high-resolution ALMA observations (N. Geesink, MSc thesis 2025). 
Figure~\ref{fig:excitation_hcn} shows the dense-gas SLEDs for our sample, compared to the HCN SLEDs in $z\sim0$ galaxies from \citet{Israel2023}. While most SLEDs are consistent with being sub-thermalised ($r_\mathrm{j1}\leq1$), J0209 and J1202 show superthermal ratios. {This might be a result of non-thermal excitation by X-rays (e.g., \citealt{Meijerink2007, Viti2017, Privon2020} or infrared pumping (e.g., \citealt{sakamoto2010, Rangwala2011, Martin2021}, or optical depth effects (e.g., if the ground-state transition has lower optical depth than the upper-state ones .}

Combining the detections and upper limits on $r_\mathrm{j1}$ in our sample, we calculate the mean excitation ratios $r_{31}$ and $r_{41}$, assuming all high-z DSFGs share a common $r_\mathrm{j1}$. The inferred mean $r_\mathrm{j1}$ are listed in Table~\ref{tab:rj1}.

The inferred median $r_{j1}$ are systematically higher than in most $z\sim0$ (U)LIRGs from \citet{Israel2023} (median $r_\mathrm{31}=0.33\pm0.18$, $r_\mathrm{41}=0.19\pm0.15$), although the low-redshift and high-redshift values are consistent within 1$\sigma$.
However, some local sources have HCN excitation similar to (e.g., NGC~1068) or even higher (e.g., NGC~253, NGC~6240) than our high-redshift targets.

Finally, we test whether $r_{j1}$ excitation ratio depends on the FIR luminosity. We find that the data (taken from this work, \citealt{Li2020, Israel2023}) is consistent with a null hypothesis of a constant $r_{j1}$. The lack of correlation of HCN excitation with $L_\mathrm{FIR}$ mirrors the results of \citet{Li2020}, who report no significant correlation of HCN(3--2)/HCN(1--0) luminosity with $L_\mathrm{FIR}$ over $L_\mathrm{FIR}=10^8-10^{12}$~$L_\odot$.

In this analysis, we remain agnostic about the physical mechanism driving the dense-gas excitation (e.g., heating by turbulence, cosmic rays or X-rays; mid-infrared pumping); our objective was simply to derive the median excitation coefficients to obtain the HCN(1--0) luminosities. We will explore  dense-gas thermodynamics in future work.
\begin{table}[]
\caption{Median excitation ratios for HCN, HCO$^+$, and HNC lines, derived from data in Fig.~\ref{fig:excitation_hcn} . The listed uncertainties correspond to the 16th and 84th percentile, respectively.}
    \centering
    \begin{tabular}{c|cc|cc}
    \hline
    Species & \multicolumn{2}{c}{This work} & \multicolumn{2}{c}{Israel (2023)}\\
     & $r_{31}$ & $r_{41}$ & $r_{31}$ & $r_{41}$\\
    \hline
        HCN & $0.59^{+0.17}_{-0.14}$ & $0.41^{+0.04}_{-0.04}$ & $0.33^{+0.18}_{-0.18}$ &
        $0.19^{+0.15}_{-0.15}$\\
        HCO$^+$ & $0.36^{+0.10}_{-0.10}$ & $0.27^{+0.05}_{-0.05}$ & $0.42^{+0.23}_{-0.21}$ & $0.22^{+0.16}_{-0.13}$ \\
        HNC & $\geq0.34$ & $\geq0.22$ & $0.33^{+0.34}_{-0.09}$& --- \\
        \hline
    \end{tabular}
    
    \label{tab:rj1}
\end{table}

\subsection{Star-forming efficiencies {and dense-gas fractions} in DSFGs}
\label{subsec:sfe}

{Armed with the constraints on the HCN excitation, we assess the dense-gas star-forming efficiency SFE$_\mathrm{dense}=M_\mathrm{dense}$/SFR and dense-gas fractions for our sample.} We first calculate SFE$_\mathrm{dense}$, assuming it is directly proportional to the ratio of HCN(1--0) luminosity and obscured star formation, SFR = $1.71\times10^{-10}\times L_\mathrm{FIR}$ for the {Salpeter} stellar initial mass function.

We derive the dense-gas mass $M_\mathrm{dense}$ for each galaxy as:
\begin{equation}
    M_\mathrm{dense}=r_{j1} \alpha_\mathrm{HCN(1-0)} L'_\mathrm{HCN(j-j-1)},
\label{eq:M_dense}
\end{equation} 

where and $\alpha_\mathrm{HCN}$ is the conversion factor between HCN(1--0) luminosity and the  dense-gas mass.

Specifically, we correct our HCN(4--3) and HCN(3--2) observations using $r_{41}$ and $r_{31}$ factors derived in Section~\ref{subsec:ladders}. For galaxies where both HCN(3--2) and HCN(4--3) are detected, we list values inferred from the lower-$J$ line. For $\alpha_\mathrm{HCN}$, we assume the canonical value of 10 $M_\odot$\,(K km s$^{-1}$)$^{-1}$ \citep{Gao2004b}; see Section~\ref{subsec:systematics} for a more detailed discussion of this choice.

\begin{figure}
    \centering
     \includegraphics[height = 7cm]{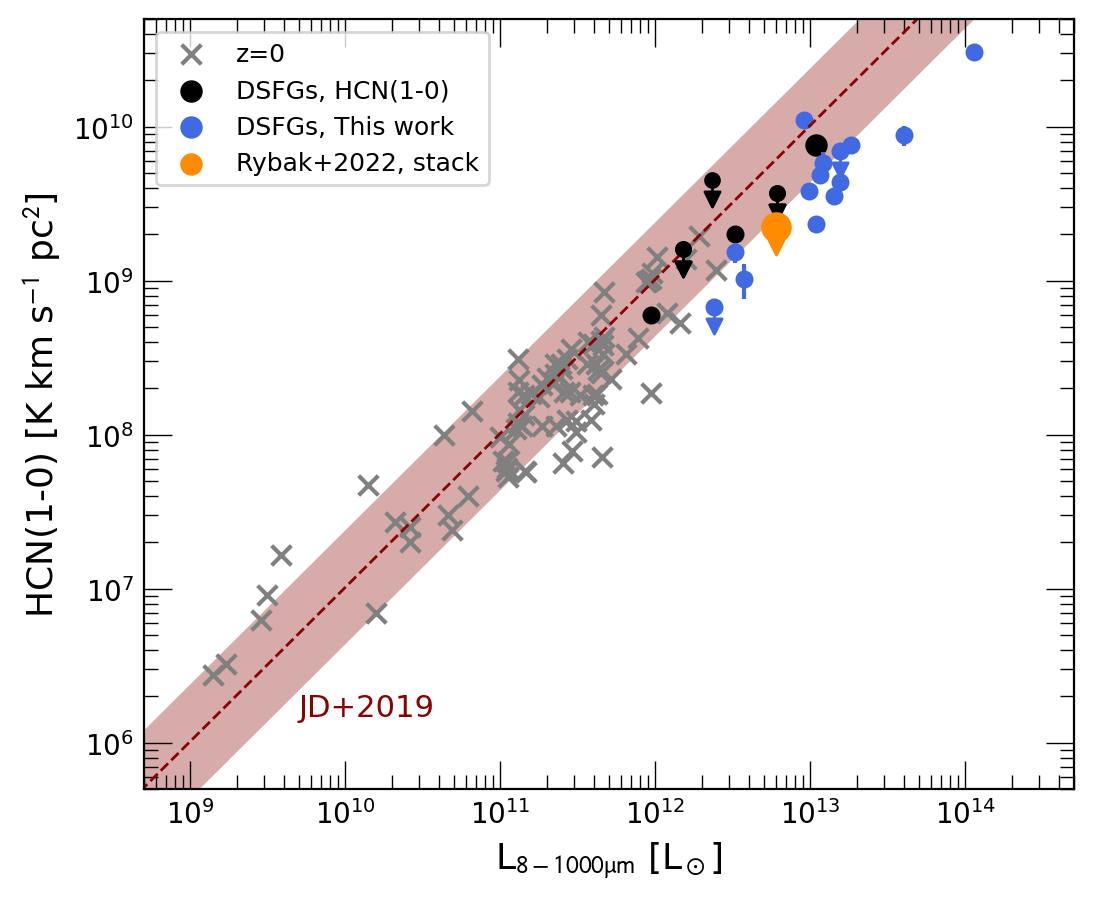}

    \caption{HCN(1--0) vs FIR luminosity (8-1000~$\mu$m) for our sample (blue), compared to other high-redshift (black) and selected low-redshift (grey) measurements for entire galaxies \citep{Gao2004a, Gracia2008, Krips2008, Garcia2012, Privon2015}, and the empirical linear trend from \citet[][red]{Jimenez2019}. Our measurements fall systematically below the linear HCN-FIR correlation.}
    \label{fig:GaoSolomon}
\end{figure}

As shown in Fig.~\ref{fig:GaoSolomon}, almost all of our data points fall below the mean HCN(1--0)/FIR trend in nearby galaxies (HCN(1--0)/FIR=$10^{-3}$, \citealt{Jimenez2019}, with 1$\sigma$ spread of 0.37~dex). Our sample has a median HCN(1--0)/FIR ratio of $4^{+11}_{-1}\times10^{-4}$, approximately 1$\sigma$ below the median for nearby galaxies ($1^{+10}_{-5}\times10^{-3}$, \citealt{Jimenez2019}). The joint probability of our HCN/FIR ratios being consistent with the \citet{Jimenez2019} value is $p\approx10^{-3}$; i.e., DSFGs as a population have significantly lower HCN/FIR ratios. As there are no $z\approx0$ galaxies with FIR luminosities comparable to our sample, we cannot determine whether this ``turnover'' in HCN/FIR ratio (i.e., a deviation from a linear HCN--FIR correlation) is related to the higher redshift or a higher FIR luminosity of DSFGs. We note that a ``turnover'' in HCN/FIR in intensely star-forming galaxies in has been predicted by theoretical models of  \citet{Krumholz2007}, where it is a consequence of median density exceeding the critical density for HCN emission.

\begin{figure*}
    \centering

    \includegraphics[height = 6cm]{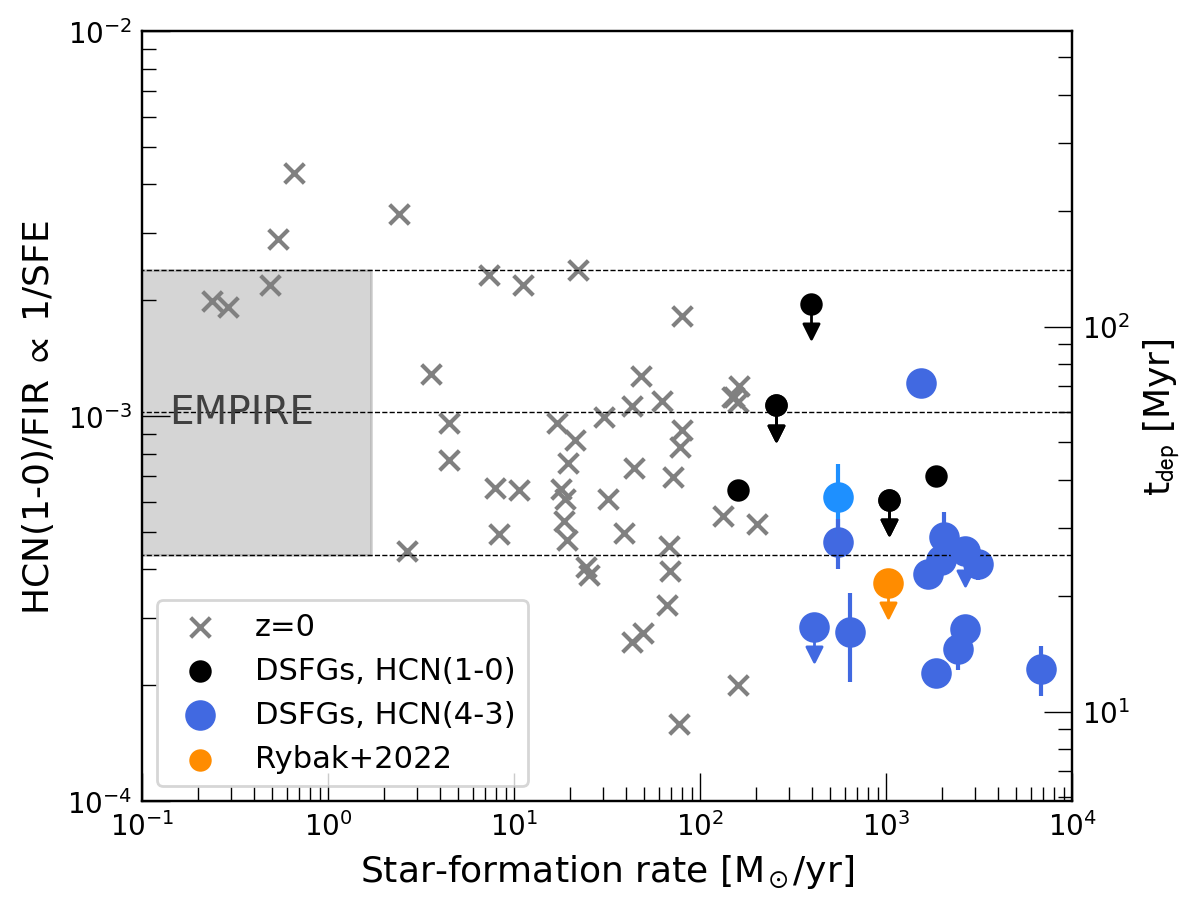}
    \includegraphics[height = 6cm]{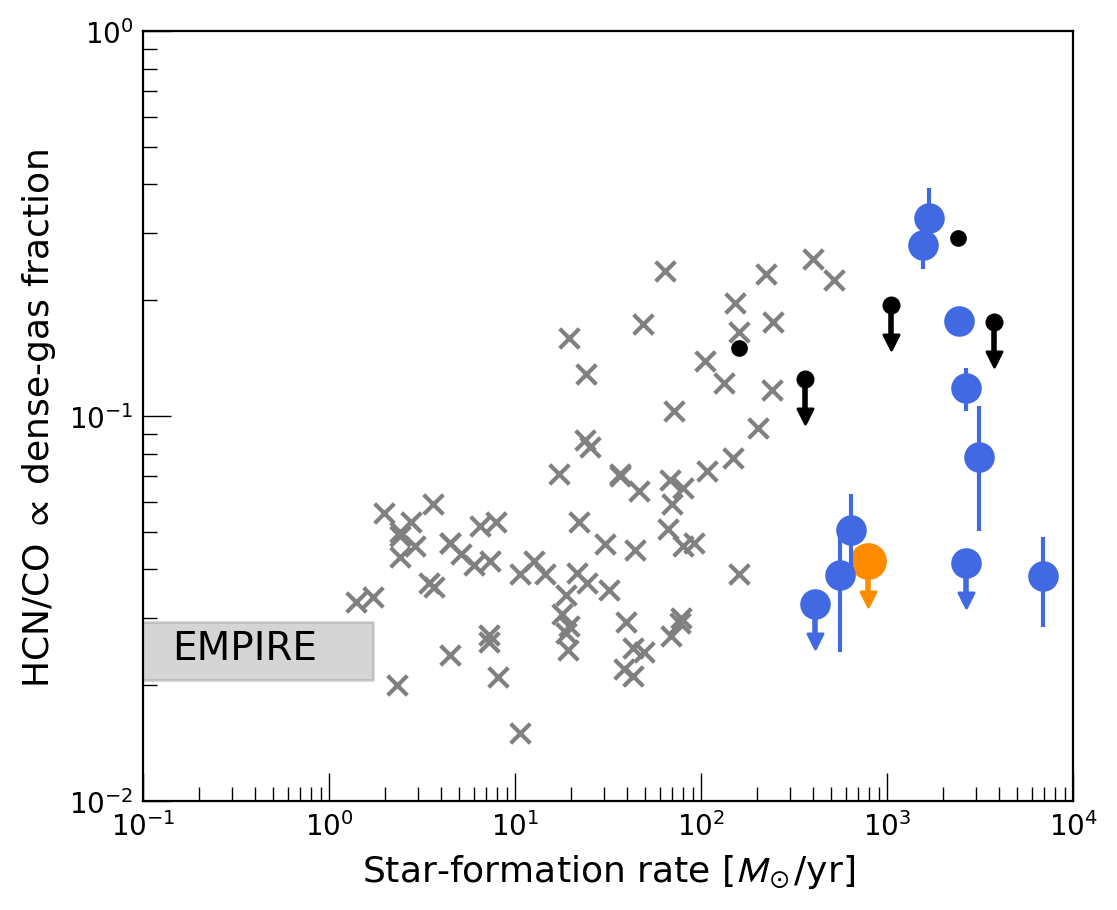}
    \caption{{Left:} Ratio of HCN(1--0) and FIR luminosities (a proxy for star-forming efficiency) as a function of star-formation rate. The dotted lines indicate the mean and 1$\sigma$ scatter for local galaxies from the EMPIRE sample \citep{Jimenez2019}. We infer HCN(1--0) luminosities for our DSFGs using the following excitation coefficients: $r_\mathrm{31}=0.41$  and $r_\mathrm{41}=0.59$. {Right:} Ratio of  HCN(1--0) and CO(1--0) luminosities (a proxy for the dense-gas fraction). The inferred HCN/CO ratios span $\approx$1~dex, suggesting that dense-gas fraction in high-z DSFGs vary significantly source-to-source. }
    \label{fig:SFE}
\end{figure*}

Figure~\ref{fig:SFE} shows the inferred HCN(1--0)/FIR (a proxy for dense-gas depletion time,  $\tau_\mathrm{dense}=M_\mathrm{dense}$/SFR) and HCN(1--0)/CO(1--0) (a proxy $f_\mathrm{dense}$). For $z=0$ galaxies, we plot the galaxy-integrated observations of \citet{Gao2004a, Gracia2008, Krips2008, Garcia2012, Privon2015}.

The HCN(1--0)/CO(1--0) luminosity ratios range between 0.01 (SDP.9) and 0.36 (SDP.11), with a median of $0.08^{+0.12}_{-0.03}$; about half of our sample has HCN/CO$\leq$0.05. While the HCN/CO ratios can be affected by differential magnification (Section~\ref{subsec:diff_magnification}), the wide spread of values likely refers real physical differences between individual sources. We do not find a significant correlation between HCN/FIR and HCN/CO.

Assuming that $\alpha_\mathrm{HCN}$ does not change between $z=0$ and $z=2.5$, this implies a significantly elevated SFE$_\mathrm{dense}$. We can estimate the dense-gas and total molecular gas mass -- and the corresponding depletion timescales -- using conversion factors $\alpha_\mathrm{HCN}=10$ and $\alpha_\mathrm{CO}=1$ (Tab.~\ref{tab:masses}). {Our choice of $\alpha_\mathrm{CO}=1$ is motivated by kinematic constraints on gas masses in DSFGs (e.g., \citealt{Calistro2018, Frias2022,Amvrosiadis2025}), although higher values have been proposed for some systems (e.g., \citealt{Dunne2022, Harrington2021}).}

We find a median dense-gas fraction $f_\mathrm{dense}=0.8^{+1.2}_{-0.3}$ and a median dense-gas depletion timescale of $\tau_\mathrm{dense}=24^{+22}_{-12}$~Myr. This is significantly shorter than the typical $\tau_\mathrm{dense}\simeq60$~Myr in nearby spiral galaxies from the EMPIRE survey \citep{Jimenez2019}. The increase in SFE would be further reinforced if DSFGs have lower $\alpha_\mathrm{HCN}$ than $z\approx0$ spiral galaxies (e.g., \citealt{Gracia2008,Jones2023, Vollmer2024}; see Section~\ref{subsec:alpha_HCN} for more detail).

\begin{table}[]
\caption{Dense-gas masses, fractions and dense-gas and molecular-gas depletion timescales for individual galaxies.}
    \label{tab:masses}
    \centering
    \begin{tabular}{l|ccccc}
    \hline
    Source & Line & $M_
    \mathrm{dense}$ & $f_
    \mathrm{dense}$ & $
    \tau_
    \mathrm{dense}$ & $\tau_
    \mathrm{mol}$ \\
         & & [$10^9 M_\odot$] & & [Myr] & [Myr] \\
         \hline
SDP.9 & (3--2) & 66$\pm$4 & 0.08$\pm$0.01 & 35$\pm$2 & 12$\pm$3 \\
SDP.11 & (3--2) & 92$\pm$6 & 3.19$\pm$0.19 & 38$\pm$2 & 12$\pm$2 \\
G09v1.40 & (3--2) & 15$\pm$1 & 0.76$\pm$0.06 & 20$\pm$2 & 27$\pm$12 \\
SDP.17 & (4--3) & 44$\pm$3 & 0.82$\pm$0.06 & 16$\pm$1 & 20$\pm$7 \\ 
J1202 & (3--2) & 60$\pm$4 & 1.06$\pm$0.07 & 109$\pm$7 & 103$\pm$36 \\
G15v2.235 & (4--3) & $\leq$70  & $\leq$0.26 & $\leq$26 & 91$\pm$25 \\
J0209 & (3--2) & 93$\pm$5 & 1.63$\pm$0.09 & 60$\pm$3 & 37$\pm$5 \\
G09v1.326 & (4--3) & 88$\pm$13 & 0.27$\pm$0.04 & 13$\pm$2 & 48$\pm$13 \\ 
SDP.130 & (4--3) & 10$\pm$3 & 0.35$\pm$0.09 & 16$\pm$4 & 46$\pm$8 \\
NAv1.195 & (4--3) & 76$\pm$7 & 0.54$\pm$0.05 & 24$\pm$2 & 44$\pm$20 \\ 
SDP.81 & (4--3) & $\leq$7  & $\leq$0.23 & $\leq$17 & 73$\pm$15 \\
G12v2.43 & (4--3) & 38$\pm$2 & 2.28$\pm$0.10 & 23$\pm$1 & 10$\pm$2 \\

     \hline
    \end{tabular}

\justify
Note: the second column indicates the line used to derive dense-gas mass. Excitation coefficients adopted from Tab.~\ref{tab:rj1}. {We assume $\alpha_\mathrm{CO}=1$ and $\alpha_\mathrm{HCN}$=10.}
\end{table}

\begin{figure}
    \centering
    \includegraphics[width=0.49\textwidth]{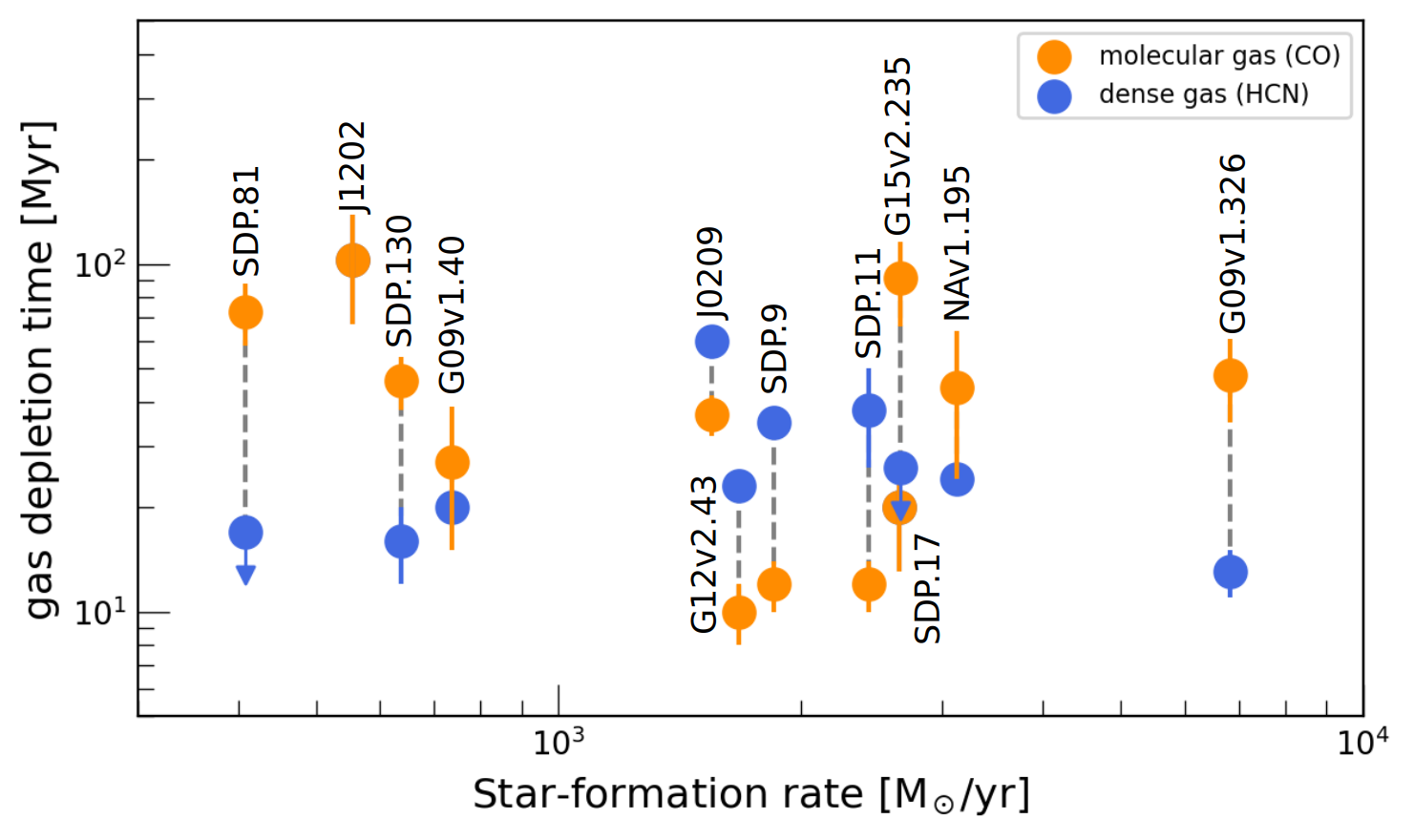}
    \caption{Depletion timescales for the total molecular gas $\tau_\mathrm{mol}$ (orange, based on CO(1--0)) and dense gas $\tau_\mathrm{dense}$ (blue, based on HCN) for individual DSFGs from our sample. The gas masses are derived using $\alpha_\mathrm{CO}$=1 and $\alpha_\mathrm{HCN}$=10. In several galaxies, $\tau_\mathrm{dense}\geq \tau_\mathrm{mol}$; this indicates that $\alpha_\mathrm{CO}$ and/or $\alpha_\mathrm{HCN}$ deviate from the assumed values. The median dense-gas depletion timescale is 24~Myr. }
    \label{fig:tdep}
\end{figure}

How do dense-gas depletion timescales compare to the depletion timescales for the total molecular gas? Figure~\ref{fig:tdep} compares the depletion times for the total molecular gas (traced by CO) and dense gas (traced by HCN). Assuming $\alpha_\mathrm{CO}=1$~$M_\odot$(K km s$^{-1}$ pc$^2$)$^{-1}$, the molecular gas depletion times $\tau_\mathrm{mol}$ range between 11~Myr (G12v2.43) and 110~Myr (SDP.9). The dense-gas depletion timescales $\tau_\mathrm{dense}$ range between 10 and 30~Myr, with a mean of 18$\pm$5~Myr. In almost all cases, $\tau_\mathrm{dense}$ is systematically lower than $\tau_\mathrm{mol}$. This discrepancy would increase further for $\alpha_\mathrm{CO}\geq1$ or $\alpha_\mathrm{HCN}\leq10$~$M_\odot$(K km s$^{-1}$ pc$^2$)$^{-1}$. 
In case of G12v2.43 and SDP.11, $\tau_\mathrm{dense}\geq \tau_\mathrm{mol}$, which is clearly unphysical. This is a clear indication that our assumption on $\alpha_\mathrm{CO}=1$ and $\alpha_\mathrm{HCN}=10$ are not applicable to all DSFGs.

We can get additional insights into the process of star-formation by considering the star-forming efficiency per free-fall time (e.g., \citealt{Krumholz2005}).
The free-fall timescale of a self-gravitating cloud with a density $\rho$ is $t_\mathrm{ff}=\sqrt{{3 \pi}/({32 G \rho})}$,
where $G$ is the gravitational constant and $\rho$ the mean gas density, $\rho=m_u n$. Setting $n=3\times10^4$~cm$^{-3}$, we obtain $t_\mathrm{ff}=0.25$~Myr which is comparable to the expected formation timescale for O-type stars ($\approx$0.5~Myr, e.g., \citealt{Sabatini2021}). The corresponding dense-gas star-forming efficiency per free-fall time $\epsilon_\mathrm{ff}=t_\mathrm{dep}/t_\mathrm{ff}$ \citep{Krumholz2005} is then 1.3$\pm$0.4\%, compared to $\approx$0.4\% for nearby galaxies \citep{Jimenez2019, Salim2020}. The higher value of $\epsilon_\mathrm{ff}$ in DSFGs is comparable with estimates for present-day (U)LIRGs \citep[1.4\%,][]{Usero2015}.

In reality, dense cores have a complex three-dimensional geometry and will be subject to various internal processes. For a more realistic comparison, we use the results of hydrodynamical simulations of individual star-forming clouds by \citet{Onus2018}. Using chemical and radiative transfer modelling, and allowing for an environment-dependent $\alpha_\mathrm{HCN}$, \citet{Onus2018} derived an empirical scaling relation between the FIR/HCN(1--0) luminosity:

\begin{equation}
  \mathrm{SFR} = 2.6\times10^{-7}\Big(\frac{\epsilon_\mathrm{ff}}{0.01}\Big)^{0.9} L'_\mathrm{HCN(1-0)}
\end{equation}

which yields $\epsilon_\mathrm{ff}=$2.3$\pm$0.6\%, similar to 1.3$\pm$0.4\% derived for a self-gravitating cloud.

\subsection{Redshift evolution of dense-gas mass density}
\label{subsec:z_evolution}

We now use our eleven detections to estimate the evolution of dense-gas mass density $\rho_\mathrm{dense}$ between the redshift range probed by our sample ($z=1.5-3.2$) and the present-day. This ``Madau-Dickinson'' plot complements the evolution of the cosmic star-formation density (e.g., \citealt{Madau2014}) and molecular gas density (e.g., \citealt{Riechers2019, Tacconi2020, Walter2020}). {Such trends can serve as a powerful benchmark for the latest cosmological simulations that now directly include cold, dense gas (e.g., COLIBRE, \citealt{Schaye2025}).}

We make the following assumptions:
\begin{enumerate}
\item[(i)] HCN(1--0) luminosity is directly proportional to the dense-gas mass, $M_\mathrm{dense}=\alpha_\mathrm{HCN}\times L'_\mathrm{HCN(1-0)}$.
\item[(ii)] A universal $\alpha_\mathrm{HCN}$ that does not depend on galaxy properties or redshift.
\item[(iii)] The HCN excitation ($r_\mathrm{j1}$) does not depend on galaxy properties, but varies with redshift.
\item[(iv)] The $L'_\mathrm{HCN(1-0)}$/$L_\mathrm{FIR}$ ratio (i.e., dense-gas star-forming efficiency) varies with redshift, but not with galaxy properties.
\end{enumerate}

The cosmic dense-gas mass density then can be expressed as:
\begin{equation}
\phi_\mathrm{dense}(z)=\alpha_\mathrm{HCN} \frac{L'_\mathrm{HCN(1-0)}}{L_\mathrm{FIR}}\Bigg\rvert_z \int \phi(L_\mathrm{FIR},z)\, \mathrm{d} L_\mathrm{FIR}
\end{equation}

The assumptions above are, by necessity, simplifying. However, there is a strong case for a redshift evolution of dense-gas excitation and dense-gas star-forming efficiency. This is because the morphology and properties of ``typical'' star-forming galaxies, as well as (U)LIRGS, change dramatically between present day and the Cosmic Noon.

First, high-redshift galaxies have both higher star-formation rates and star-forming efficiency, even at the same stellar mass as $z\approx0$ ones (e.g., \citealt{Tacconi2020}). Second, high-redshift DSFGs are morphologically distinct from present-day ultraluminous infrared galaxies (ULIRGs) with comparable $L_\mathrm{FIR}$: while star formation in ULIRGs is typically concentrated into a very compact region (just a few hundred pc across, e.g., \citealt{Lutz2016, Barcos2017}), in high-redshift galaxies, it is spread over a region few kiloparsecs across \citep[e.g.,][]{Hodge2016, Gullberg2019}. Third, DSFGs have significantly higher gas densities, far-UV irradiation (e.g., \citealt{Wardlow2017,Rybak2019}), and are potentially more turbulent (e.g., \citealt{Dessauges2019, Harrington2021}). Finally, as we will discuss in Appendix~\ref{sec:co_isotopologues}, high-redshift DSFGs might have a top-heavy stellar IMF. We discuss the potential impact of all these factors below.

We match our galaxies to the far-infrared luminosity functions (FIR LFs), rather than the stellar mass functions\footnote{While the stellar mass functions at high redshift are better observationally constrained, the stellar masses of dusty DSFGs are highly uncertain. Moreover, for lensed sources, the light from the background galaxy is blended with that from the foreground lens.}. As our fiducial model, we adopt FIR LFs derived by  \citet{Casey2018} and \citet{Zavala2021}. For comparison, we repeat our calculations using FIR LFs of \citet{Gruppioni2013} for $z=0$ and \citet{Gruppioni2020} for high redshift; while the latter overpredicts the star-formation rate density at $z\geq$4 (e.g., \citealt{Zavala2021,vanderVlugt2022}), it still provides a good fit to the observational data over the redshift range considered here. 

\begin{figure}
    \centering
    \includegraphics[width=0.4\textwidth]{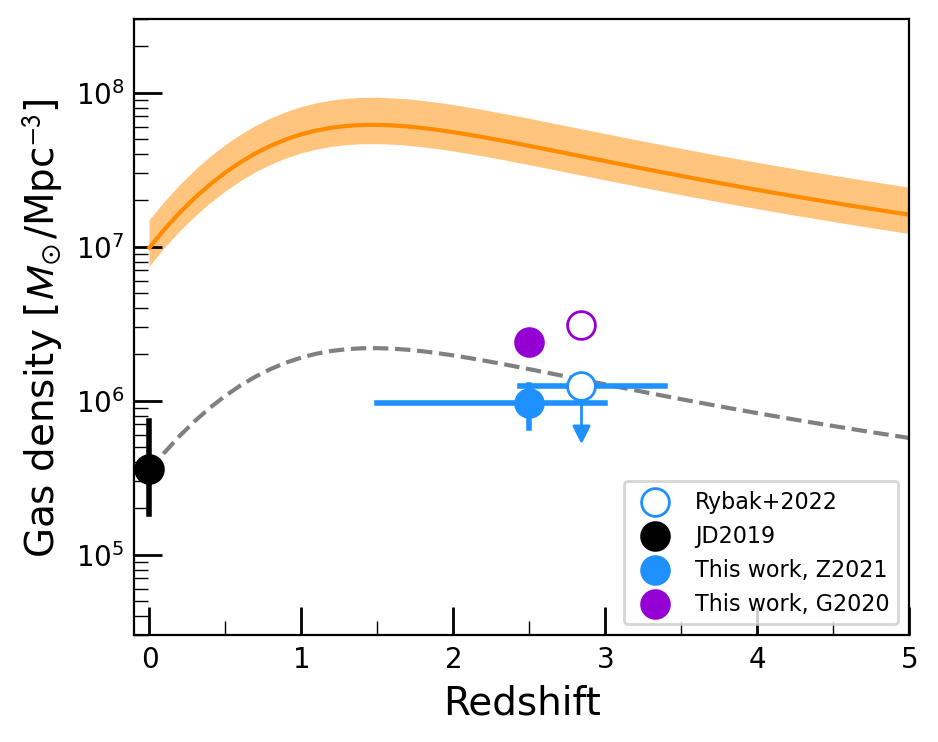}
    \caption{Redshift evolution of dense-gas mass density between the present-day (black point). The blue and purple points show dense-gas mass density inferred using the far-IR LFs from \citet[][blue]{Zavala2021} and \citet[][purple]{Gruppioni2020}, respectively.  The $z\approx0$ point is inferred using HCN/FIR ratio from \citet{Jimenez2019}. We also show the upper limit from \citet{Rybak2022a}. Orange  line points shows the \emph{total} molecular gas density evolution from  \citet[][orange solid line]{Walter2020}. Grey dashed line shows the \citet{Walter2020}  trend normalised to match $\rho_\mathrm{dense}$ at $z=0$. The $\rho_\mathrm{dense}$ evolution is consistent with the redshift evolution of total molecular gas with a constant dense-gas fraction of $\approx$3.5\%.}
    \label{fig:madau}
\end{figure}

Figure~\ref{fig:madau} shows the resulting constraints on the redshift evolution of  $\rho_\mathrm{dense}$. For the \citet{Casey2018} and \citet{Zavala2021} LFs, $\rho_\mathrm{dense}$ increases from $3.6^{+3.4}_{-1.9}\times10^5$~$M_\odot$ Mpc$^{-3}$ at $z=0$ to $(9.6\pm3)\times10^6$~$M_\odot$ Mpc$^{-3}$ at $z=2.5$. This corresponds to an increase by factor of $\approx$3, but the $z=0$ and $z=2.5$ values are essentially consistent within 1.3$\sigma$, {due to the considerable uncertainties on both measurements}. For the Gruppioni et al. LFs, $\rho_\mathrm{dense}$ increases from $(3.6^{+3.4}_{-1.8})\times10^5$~$M_\odot$ Mpc$^{-3}$ at $z=0$ to $(24\pm8) \times10^5$~$M_\odot$ Mpc$^{-3}$ at $z=2.5$. This translates to an increase by a factor of $\approx$7 (1.8$\sigma$ significance). In summary, we see a tentative increase in $\rho_\mathrm{dense}$ towards high redshift. 

We complement our constraint on $\rho_\mathrm{dense}$ at $z=2.5$ by the upper limit on HCN/FIR ratio from \citet{Rybak2022a} ($\geq3.6\times10^{-4}$), which was derived directly from HCN(1--0) observations and is not affected by uncertainties on $r_\mathrm{j1}$. The \citet{Rybak2022a} upper limit puts a nominally stronger constraint on $\rho_\mathrm{dense}$ at $z=2.5-3.5$ (increase by a factor of $\leq3.0$ compared to $z=0$), but is derived from a smaller sample of galaxies.

{The tentative increase in $\rho_\mathrm{dense}$ is qualitatively in line with the higher molecular gas mass density $\rho_\mathrm{mol}$ at high redshift (e.g., \citealt{Tacconi2020, Walter2020}. For a more quantitative comparison, Fig.~\ref{fig:madau} shows the constraints on the total molecular gas mass from the \citet{Walter2020} compilation (orange), and the Walter et al., trend normalised to $\rho_\mathrm{dense}$ at $z$=0 (grey line).
Our constraints on the redshift evolution of dense gas are consistent with the $\rho_\mathrm{mol}$ trend, with a mean dense-gas fraction of $\approx$3.5\%. This fraction is considerably lower than the dense-gas fractions derived for individual DSFGs (Tab.~\ref{tab:masses} ); however, the latter might suffer from a considerable uncertainty due to conversion factors or differential magnification. The constant dense-gas fraction indicate that the increase of SFR density at high redshift is driven by enhanced star-forming efficiencies rather than increased availability of dense gas - a direct  }

\subsection{Systematic uncertainties}
\label{subsec:systematics}

\subsubsection{Spatial offsets and differences in line profiles}
\label{sec:offsets}

Throughout this study, given the low angular (and spectral) resolution of our data, we assume that HCN, HCO$^+$, and HNC emission trace the same gas. We now examine this assumption by looking for: (1)
differences between the HCN/HCO$^+$/HNC line profiles; (2) spatial offsets in moment-0 maps.

As shown in Fig.~\ref{fig:fwhm_fwhm}, line profiles -- parametrised by their Gaussian FWHM -- are generally consistent within $2\sigma$ uncertainties. A similar correspondence was found for $J_\mathrm{upp}=4,5$ lines in three galaxies high-z by \citet{Canameras2021}. However, noticeable differences are seen in several sources with high S/N: G09v1.40, J0209, and J1202. First, in G09v1.40, the HNC(3--2) line appears to be double-peaked, with a prominent blue-shifted emission offset by 350$\pm$50 km/s from the systemic redshift. In contrast, the HCN(3--2) and HCO$^+$(3--2) lines are consistent with a Gaussian profile. We note that G09v1.40 has a massive molecular outflow, detected in the CH$^+$ \citep{Falgarone2017} and OH$^+$ emission with an outflow velocity of $\approx$250 km/s \citep{Butler2021}. The HNC emission might be associated with this molecular outflow.

In J0209, the HCN, HCO$^+$, and HNC lines have comparable linewidths, but are systematically ($\approx$30\%) narrower than the CO emission. This is consistent with a picture where the bulk of low-$J$ CO emission arises from extended gas reservoir (e.g., \citealt{Koenig2018, rybak2025a}), as has been confirmed by high-resolution ALMA observations (N. Geesink, MSc thesis).

Finally, in J1202, the HNC(3--2) and HCN(4--3) are significantly narrower than the corresponding HCN and HCO$^+$ transitions. Compared to the HCN and HCO$^+$(4--3) lines, the peak of HNC emission is shifted 160$\pm$15~km/s towards red, with absorption in the blue wing of the line -- a P-Cygni profile, indicative of outflowing gas. We therefore hypothesise that in G09v1.40 and J1202, a significant fraction of the HNC (but not HNC or HCO$^+$) emission arises from a molecular outflow.

\begin{figure}[h]
    \centering
    \includegraphics[width = 0.4\textwidth]{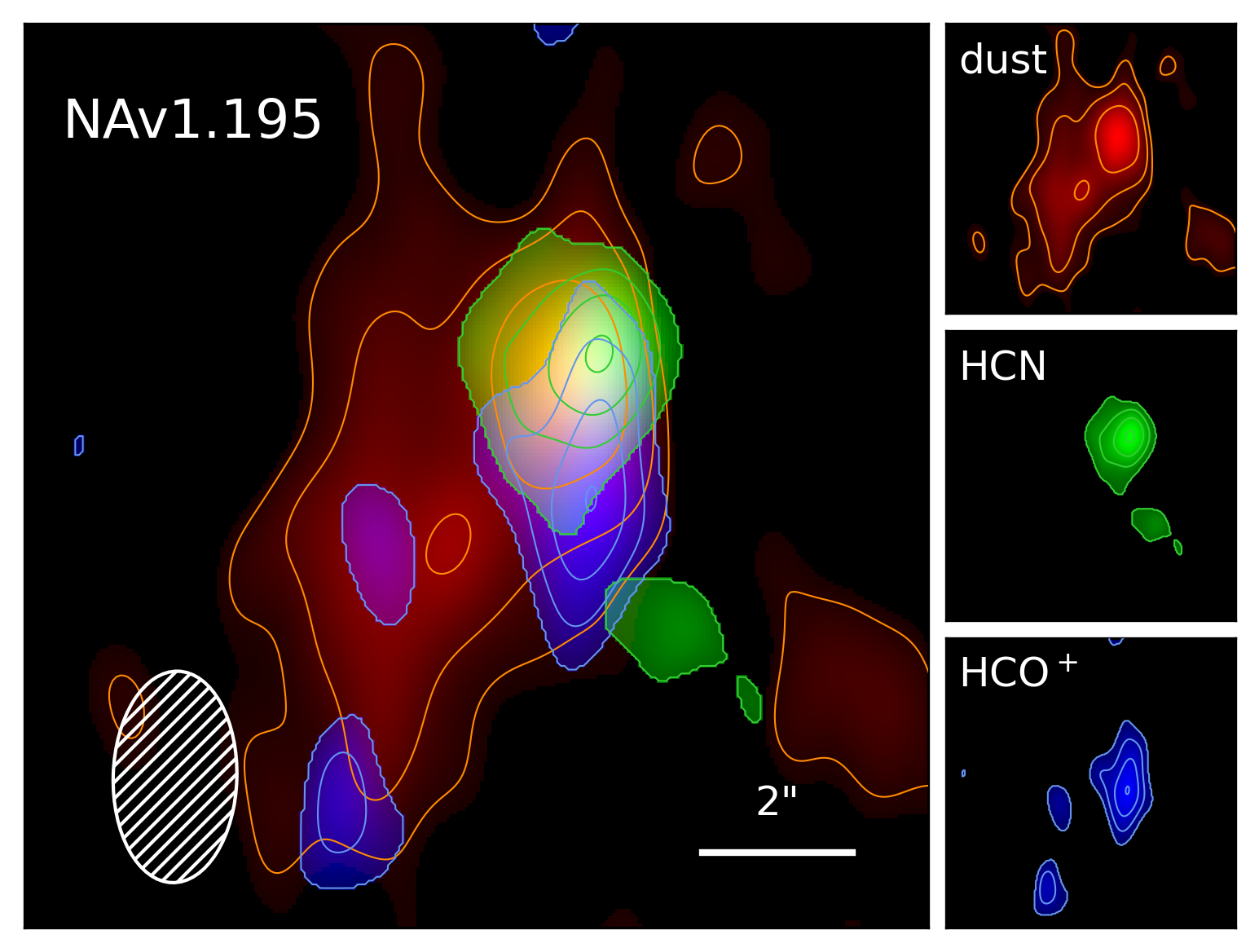}
    \caption{Spatial offsets between HCN (green) HCO$^+$ (blue) and dust (red) emission in NAv1.195. The contours are drawn at $\pm$(2, 4, 6, ...)$\sigma$ for the continuum, and $\pm$(2, 3, 4,...)$\sigma$ for the lines. The HCO$^+$(4--3) emission peaks $\approx$2.5" to the south of HCN and dust continuum. These offsets suggest that different dense-gas tracers do not necessarily trace the same gas.}
    \label{fig:nav1.195}
\end{figure}

Looking at the moment-0 maps, the dust, HCN, HCO$^+$, and HNC emission can be considered co-spatial for almost all the sources -- with the exception of NAv1.195. As shown in Fig.~\ref{fig:nav1.195}, in this source, the dust (peak S/N$\simeq9$) and HCN(4--3) surface brightness distribution (peak S/N$\simeq$6) peak the same position. However,  HCO$^+$(4--3) (peak S/N$\approx$5) is offset to the south by $\approx$2.5''. This offset is comparable to the synthesised beam FWHM (2.8"$\times$1.6") and thus likely to be physical. The HCN/HCO$^+$ enhancement in the north of the source could be caused, for example, by a buried AGN (see Section~\ref{subsec:line_ratios}).

\subsubsection{Differential magnification}
\label{subsec:diff_magnification}

In gravitational lensing, different spatial distributions of different tracers will cause them to be magnified by a different factor -- the so-called differential magnification (e.g., \citealt{Serjeant2012}). Namely, if the CO(1--0) emission is much more extended than HCN or FIR continuum and thus extends to low-magnification regions, the total magnification factor for CO will be smaller than for the more compact HCN or FIR. This would artificially inflate the observed HCN/CO ratios, while HCN/FIR will be less affected, as dense gas and obscured star-formation are likely almost co-spatial (and we can thus apply magnifications derived for FIR continuum to dense-gas tracers).

For example, in SDP.81, the difference between FIR and CO(3--2) magnifications is $\leq$10\% \citep{Rybak2020b}. Similarly, \citet{Canameras2018} found that CO and FIR magnifications differ by $\leq$20\% in a sample of nine lensed DSFGs from the \textit{Planck} sample. In extreme case of SDP.9 -- which, as a very compact source, is particularly sensitive to differential magnification --  \citet{Massardi2018} find a 50\% difference between the magnification of the FIR continuum and the stellar component. Typical differences in CO and FIR continuum magnifications are therefore likely $\leq50$\%. Overall, the high HCN/CO values are likely physical, and not driven by the differential magnification. 

\subsubsection{The HCN-to-dense-gas conversion factor}
\label{subsec:alpha_HCN}

The most significant systematic uncertainty for the interpretation of our results is the conversion factor between the HCN luminosity and dense-gas mass (Eq.~\ref{eq:M_dense}).

In particular, there is a considerable uncertainty regarding the ``effective'' gas density traced by the HCN(1--0) emission. Specifically, spatially resolved studies of Galactic molecular clouds have shown that on (sub-)parsec scales, a significant fraction of HCN(1--0) emission is associated with densities as low as $10^3$~cm$^{-3}$ \citep{Kauffmann2017, Evans2020}. Similarly, magneto-hydrodynamic simulations of individual molecular clouds indicate that HCN(1--0) traces gas with a mean density of $\approx3\times10^3$~cm$^{-3}$ \citep{Jones2023}. However, recent observations by \citet{Jimenez2023} have shown that in nearby galaxies, HCN(1--0) directly correlates with N$_2$H$^+$ emission on kiloparsec scales. As N$_2$H$^+$ is considered an unambiguous tracer of dense gas in molecular clouds (N$_2$H$^+$ traces gas with characteristic density $\approx4\times10^3$ cm$^{-3}$, \citealt{Kauffmann2017}), this implies that on galactic scales, HCN remains a reliable tracer dense gas.

\citet{Gao2004b} estimated $\alpha_\mathrm{HCN}$=10 $M_\odot$/(K km s$^{-1}$ pc$^{2}$) using two approaches: (1) large velocity-gradient modelling; and (2) considering optically thick emission from a virialised cloud\footnote{
In the inner regions of nearby galaxies, the HCN and HCO$^+$ lines are usually moderately optically thick, with an optical depth of 1--2 (e.g., \citealt{Jimenez2017}).}. 
However, this conversion factor is far from universal. As already noted by \citet{Gao2004b}, $\alpha_\mathrm{HCN}$  has to differ significantly between individual galaxies. For example, many $z\approx 0$ ULIRGs have HCN(1--0)/CO(1--0) luminosity ratios $\geq0.1$ (see Fig.~\ref{fig:SFE}); assuming $\alpha_\mathrm{HCN}=10$ and $\alpha_\mathrm{CO}$=0.8 would yield dense-gas fractions in the excess of 100\% - a clearly unphysical result. This was confirmed by \citet{Gracia2008} who found that $\alpha_\mathrm{HCN}$ decreases by a factor of $\approx$2.5 over the $L_\mathrm{FIR}=10^{11}-10^{12}$~$L_\odot$ range, based on LVG modelling of HCN/HCO$^+$ (1--0) and (3--2) lines. Finally, spatially resolved ALMA observations of HCN and CO emission in J0209 imply $\alpha_\mathrm{HCN}\leq2$ (Rybak, Geesink et al., in prep.).

Conversely, several studies of nearby galaxies report $\alpha_\mathrm{HCN}\geq10$. For example, \citealt{Papadopoulos2014} derived $\alpha_\mathrm{HCN}=20-60$ for Arp~193 and NGC~6240; for the latter galaxy, \citet{Tunnard2015} report $\alpha_\mathrm{HCN}=32_{-13}^{+89}$. More recently, very elevated $\alpha_\mathrm{HCN}\approx100$ has been reported for individual clouds in the Milky Way \citep{Dame2023} and Andromeda \citep{Forbrich2023}. However, as these would imply $f_\mathrm{dense}\geq100$\%, these are clearly not applicable to all DSFGs, although the actual $\alpha_\mathrm{HCN}$ value might vary significantly between individual sources.

Simulations of individual star-forming clouds and galaxies also support $\alpha_\mathrm{HCN}\leq10$. On cloud scales, \citet{Onus2018} found $\alpha_\mathrm{HCN}=16\pm4$; while \citet{Jones2023} found $\alpha_\mathrm{HCN}=6.8\pm3.8$ in simulations of colliding clouds, with high-velocity collisions showing  $\alpha_\mathrm{HCN}\leq3$.
On galactic scales, simulations by \citet{Vollmer2024} predict a mean $\alpha_\mathrm{HCN}=11\pm4$ and $12\pm7$ for high-redshift starburst and main-sequence galaxies, respectively, with $\alpha_\mathrm{HCN}\approx6$ for $z=0-0.5$ LIRGs.

A variable $\alpha_\mathrm{HCN}$ which decreases at high SFR would in fact reinforce our main findings. Specifically, our inferred $M_\mathrm{dense}=\alpha_\mathrm{HCN}L'_\mathrm{HCN(1-0)}$ would become even smaller, further increasing the dense-gas star-forming efficiency. In other words, in the HCN/FIR -- SFR plane (Fig.~\ref{fig:SFE}, left), our datapoints would shift \emph{downwards}, towards very short depletion times ($\leq 10$~Myr). Similarly, in the HCN/CO -- SFR plane (Fig.~\ref{fig:SFE}, right),  our data points would shift downwards, towards even lower dense-gas fraction.

Finally, it is possible that $\alpha_\mathrm{CO}$ is $\geq1$ in (some) high-redshift DSFGs (e.g., \citealt{Harrington2021}). In such a case, the inferred dense-gas fractions would decrease even further. On the other hand, the inferred $\tau_\mathrm{dense}$ -- which is independent of $L'_\mathrm{CO(1-0)}$ -- would remain unchanged.

\section{Conclusions}
\label{sec:conclusions}

We have presented results of the NOEMA and ALMA survey of dense-gas tracers -- HCN, HCO$^+$, and HNC -- in eleven $z=1.6 - 3.1$ dusty star-forming galaxies. This is the largest study of dense-gas tracers in $z\geq1$ galaxies to date. Our main results are:

\begin{itemize}
    \item We detect dense-gas tracers in 10 out of 11 galaxies. In total, we detect 34 transitions of HCN, HCO$^+$, and HNC, increasing the number of $z\geq1$ detections by more than a factor of four. Only one galaxy (G15v2.235) is not detected in any spectral line. Additionally, we also detect several transitions of $^{13}$CO, C$^{18}$O, and CN.
    \item The linewidths of HCN, HCO$^+$, and HNC lines are generally consistent within 1$\sigma$ uncertainty. However, in G09v1.40 and J1202, the HNC lines show structures indicative of molecular outflows. In one source -- NAv1.195 -- the HCN and HCO$^+$ emission appear to be spatially offset from each other. 
    \item We derive the excitation coefficients for the HCN, HCO$^+$, and HNC $J_\mathrm{upp}=$3 and 4 lines, and find that DSFGs have more excited dense-gas ladders than $z\approx0$ (U)LIRGs.
    \item High-redshift DSFGs have systematically lower HCN(1--0)/FIR ratios compared to present-day star-forming galaxies. This trend agrees qualitatively with a break in the HCN-FIR correlation predicted by some theoretical models (e.g., \citealt{Krumholz2007}). 
    \item Assuming the ``canonical'' $\alpha_\mathrm{HCN}=10$, we find a median dense-gas depletion timescale of $\approx$24~Myr. These results indicate that high-redshift DSFGs had higher star-forming efficiency than present-day galaxies.
    \item High-redshift DSFGs have a wide range of dense-gas fractions, with HCN/CO ratios ranging from $\approx0.01$ to $\approx0.15$. This likely reflects variations in the amount and extent of molecular gas between individual galaxies.
    \item We put the first constraints on the redshift evolution of the Cosmic dense-gas mass density, which tentatively increases by a factor of $7.4\pm6.0$ between $z=0$ and $z\approx2.5$. This trend is consistent with the evolution of the total molecular gas mass density if $\approx$3.5\% of the total molecular gas is in the ``high-density'' state.
\end{itemize}

This work opens a new chapter in studies of dense gas and star formation in early galaxies. As underlined by the large number of detections, systematic surveys of dense gas in early galaxies with ALMA and NOEMA -- aided by gravitational lensing -- are now feasible. {Systematic studies of dense gas across cosmic time can provide a powerful benchmark for upcoming cosmological hydrodynamical simulations that directly include cold, dense gas (e.g., COLIBRE, \citealt{Schaye2025}) } Future observations will allow us to extend these studies to larger samples and down to sub-galactic scales.

\begin{acknowledgements}
The authors thank Z.~Zhang for sharing the data from \citet{Zhang2014} and providing valuable insights into the local dense-gas observations.\\

This work is based on observations carried out under project numbers S21CB and S23CB with the IRAM NOEMA Interferometer. IRAM is supported by INSU/CNRS (France), MPG (Germany) and IGN (Spain).
The research leading to these results has received funding from the European Union’s Horizon 2020 research and innovation programme under grant agreement No 101004719 [ORP].\\

This paper makes use of the following ALMA data: \#2016.1.00663.S, \#2017.1.01694.S, \#2018.1.00747.S, and \#2023.1.00432.S.
ALMA is a partnership of ESO (representing its member states),
NSF (USA) and NINS (Japan), together with NRC (Canada), MOST and ASIAA (Taiwan), and KASI (Republic of Korea), in cooperation with the Republic of Chile. The Joint ALMA Observatory is operated by ESO, AUI/NRAO and NAOJ.\\

The authors acknowledge assistance from Allegro, the European ALMA Regional Center node in the Netherlands.\\

M.R. is supported by the NWO Veni project "\textit{Under the lens}" (VI.Veni.202.225). J.A.H. acknowledges support from the ERC
Consolidator Grant 101088676 (VOYAJ). D.R. gratefully acknowledges support from the Collaborative Research Center 1601 (SFB 1601 sub-projects C1, C2, C3, and C6) funded by the Deutsche Forschungsgemeinschaft (DFG) – 500700252.
T.R.G. is grateful for support from the Carlsberg Foundation via grant No. CF20-0534. S.V. acknowledges support from the European Research Council (ERC) grant MOPPEX ERC-833460. 
The Cosmic Dawn Center (DAWN) is funded by the Danish National Research Foundation under grant No.\,140.

\end{acknowledgements}

\bibliographystyle{aa} 
\bibliography{references}

\onecolumn

\begin{appendix}

\section{Details of NOEMA and ALMA observations}
\label{app:noema_obs}

Table~\ref{tab:obs_summary} summarises the details of NOEMA and ALMA observations.

\begin{table*}[h]
\caption{Observations summary. Individual columns list: target ID, dates of observations, HCN transition targeted, on-source time, beam FWHM and position angle, {and rms noise over a 100 km\,s$^{-1}$ bandwidth at the position of the HCN line}. \label{tab:obs_summary} }
\begin{center}
 \begin{tabular}{@{}lccccc @{}}
 \hline \hline

Target & Transitions & Date &  $t_\mathrm{on}$ &   Beam FWHM \& PA & $\sigma_\mathrm{100 km/s}$\\
& & & [h] & [arcsec$^2$, deg] & [mJy/beam]\\
\hline
\multicolumn{5}{c}{NOEMA}\\
\hline
J1202 & (3--2) & 2021 August 26 &  2.6   & 5.6$\times$4.8 (56) & 0.32\\
J1202 & (4--3) & 2021 August 29 &  3.4 &  5.1$\times$4.0 (88) & 0.35 \\
SDP.130 & (3--2) & 2021 Sep 20, Dec 23; 2022 Jan 16 &  1.9 &  4.7$\times$2.5 (68) & 0.26  \\
 J0209 & (3--2) & 2021 August 13 &  5.3 &  7.5$\times$5.0 (35) & 0.42  \\
SDP.9 & (4--3) & 2023 October 6 & 3.4  & 2.6$\times$0.9 (13) & 0.39  \\
SDP.11 & (4--3) & 2023 October 8 & 4.5  & 3.5$\times$1.1 (11) & 0.43 \\
\hline
\multicolumn{5}{c}{ALMA}\\
\hline
G09v1.40 & (3--2) & 2018 April 9 & 0.86 & 2.7$\times$2.1 (154) & 0.14\\
SDP.17 & (4--3) & 2018 April 26, May 1, May 2 & 1.03 & 2.1$\times$1.6 (23) & 0.21\\
SDP.130 & (4--3) & 2018 April 9, April 24 & 0.89 &  2.4$\times$1.8 (153) & 0.11\\
G09v1.326 & (4--3) & 2018 April 22 & 0.91 & 2.1$\times$2.0 (143) & 0.14\\
G12v2.43 & (4--3) & 2018 April 3 & 0.84 & 2.4$\times$2.2 (22) & 0.13\\
NAv1.195 & (4--3) & 2018 March 30, April 26 & 1.69 & 2.8$\times$1.7 (86) & 0.20\\
G15v2.235 & (4--3) & 2018 April 6 & 0.47 & 2.3$\times$1.9 (29) & 0.18 \\
\hline
 
 \hline
 \end{tabular}
\end{center}
\end{table*}

\section{Imaging and spectra}
\label{app:spectra}

\begin{figure*}
\centering
    \includegraphics[height=3.3cm]{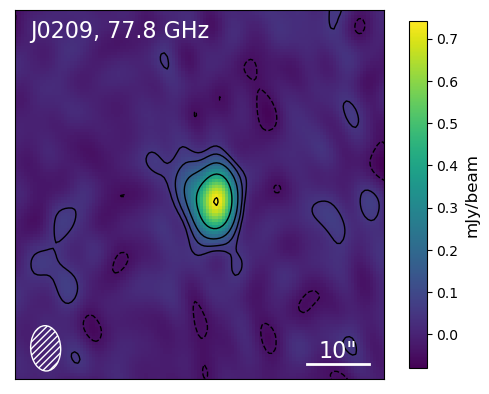}
    \includegraphics[height=3.3cm]{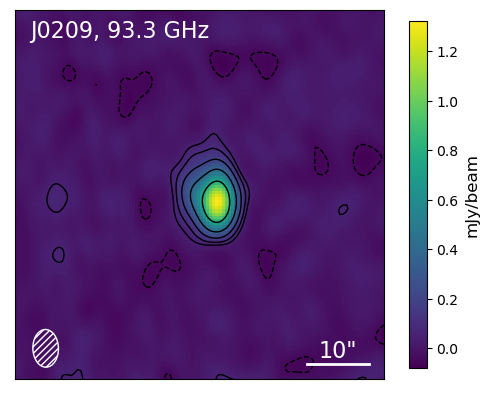}
    \includegraphics[height=3.3cm]{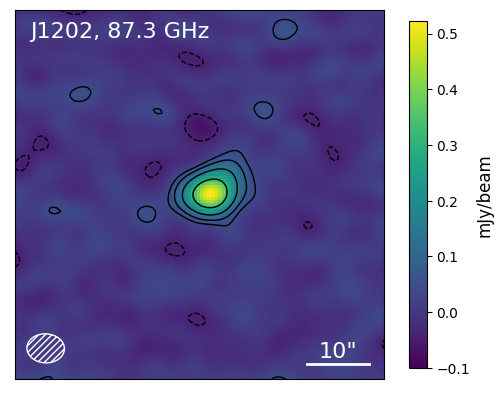}
    \includegraphics[height=3.3cm]
    {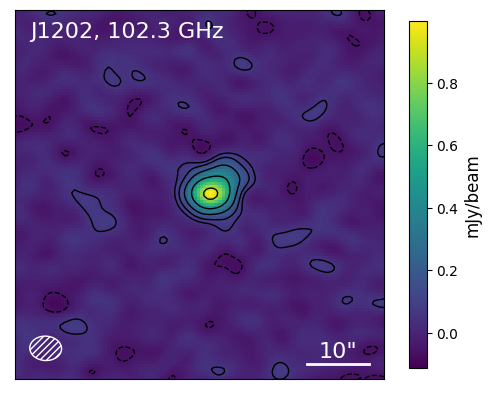} \\
     \includegraphics[height=3.3cm]{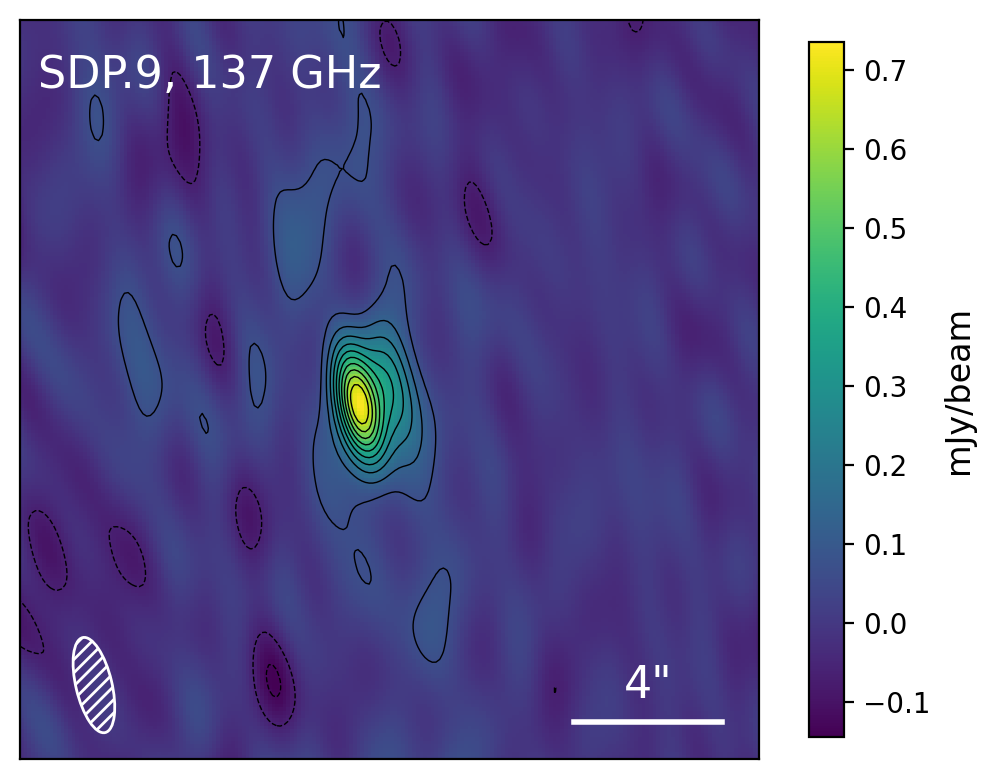}
    \includegraphics[height=3.3cm]{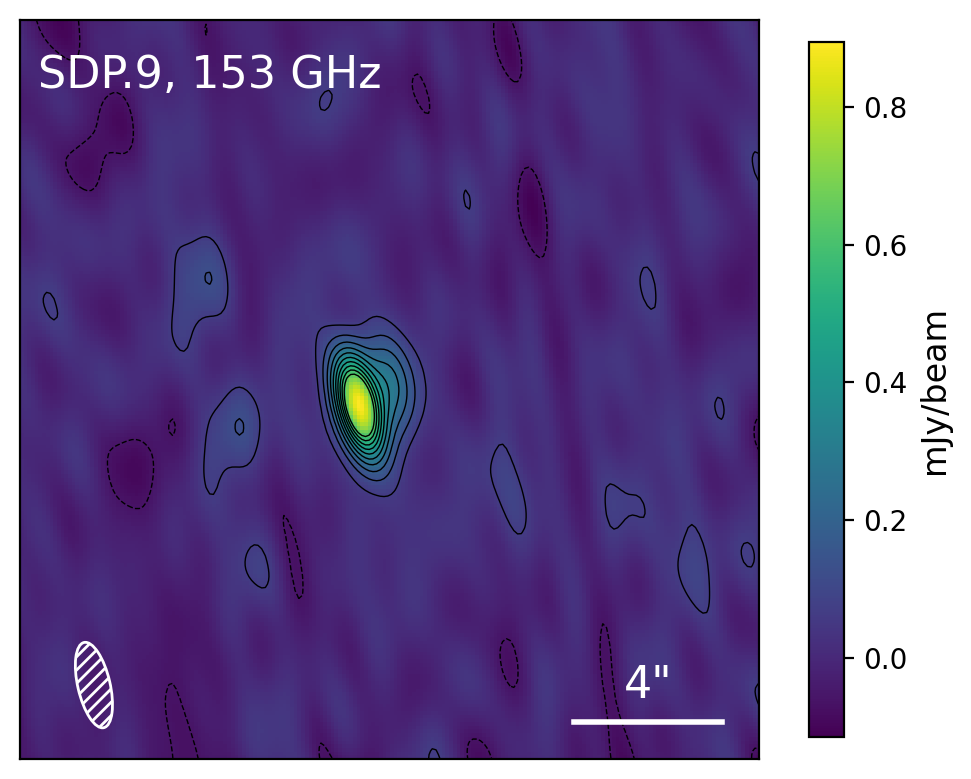} 
    \includegraphics[height=3.3cm]{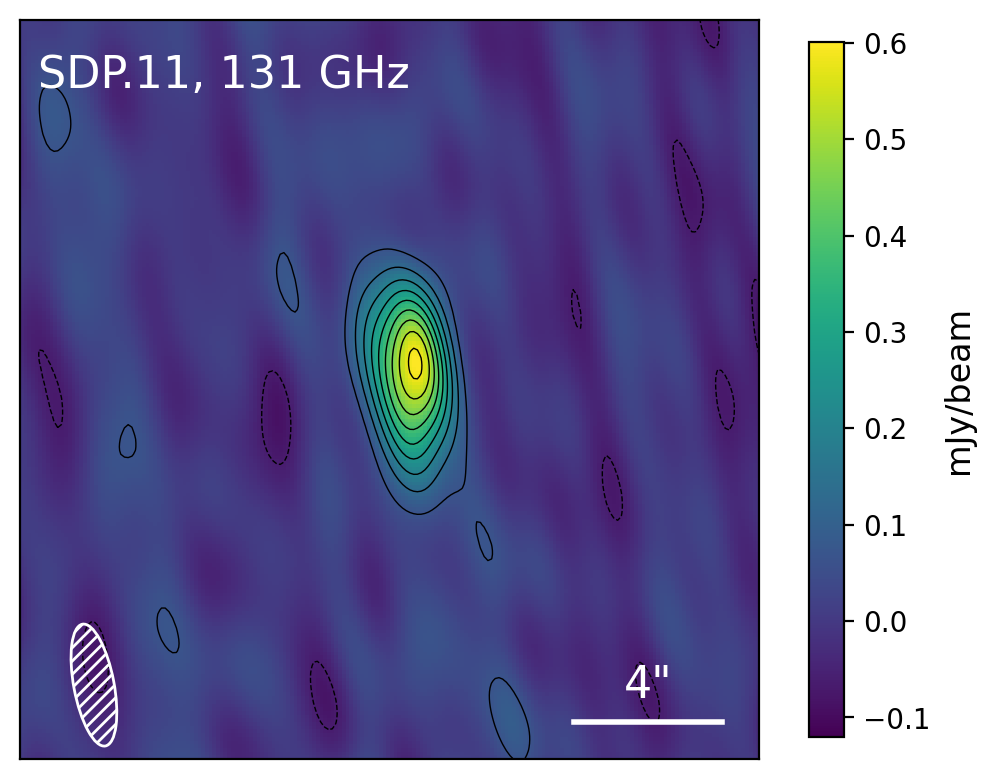}
    \includegraphics[height=3.3cm]{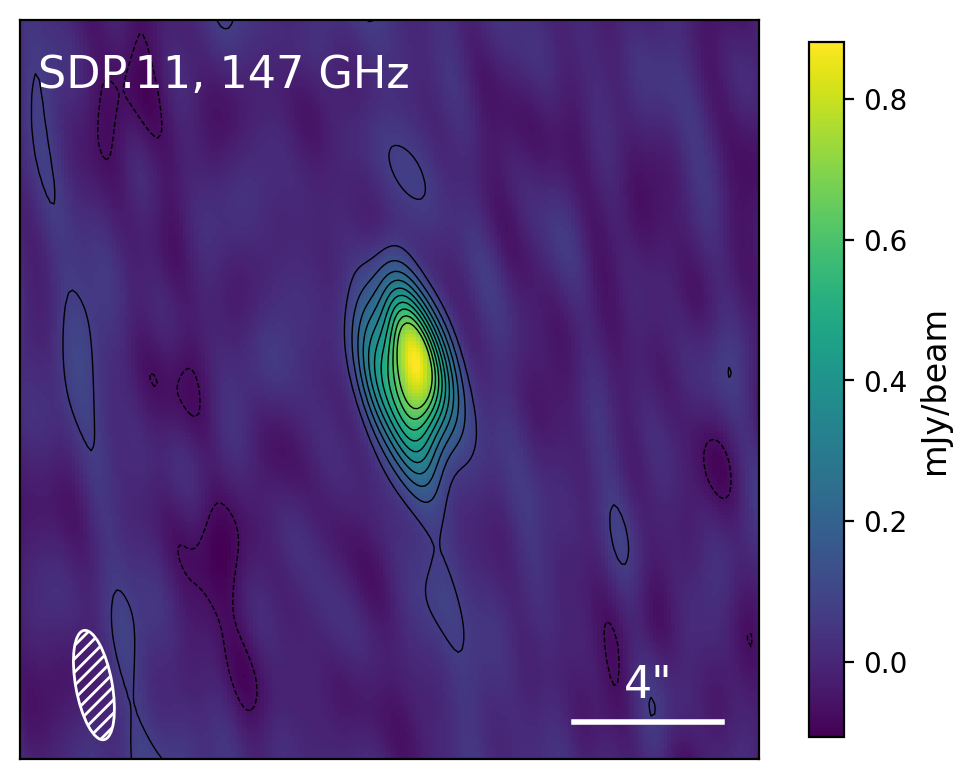} \\

    \includegraphics[height=3.3cm]{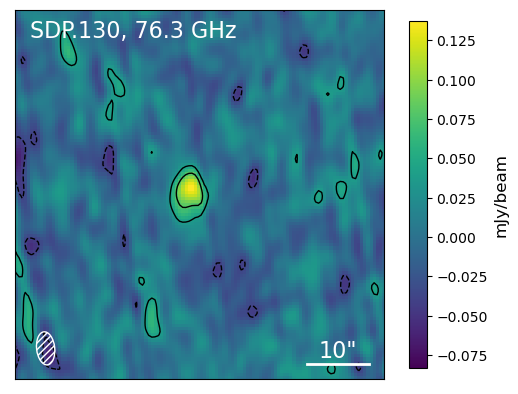}
    \includegraphics[height=3.3cm]{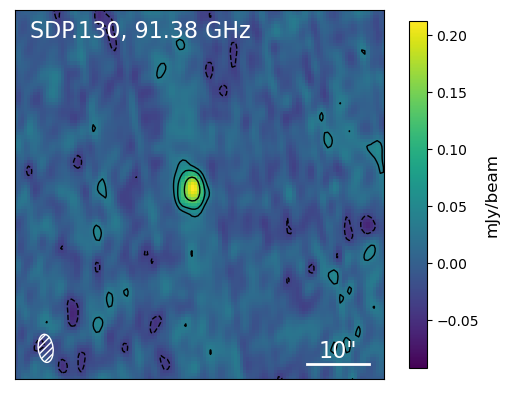}

    \caption{NOEMA Band 1/2 continuum imaging of individual galaxies. The observed-frame frequencies are given in each panel. For each target, we show separate images for the lower and upper sidebands. All images were produced using natural weighting. Contours start at $\pm2\sigma$ and terminate at 20$\sigma$, with a 2$\sigma$ increment. All sources are detected in the continuum emission and are marginally resolved.}
    \label{fig:noema_continuum}
\end{figure*}

\begin{figure*}
\centering
    \includegraphics[height=3.3cm]{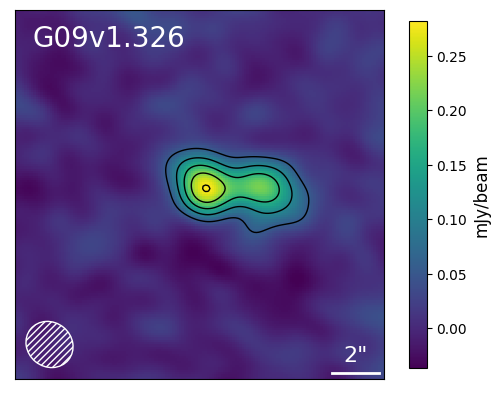}
    \includegraphics[height=3.3cm]{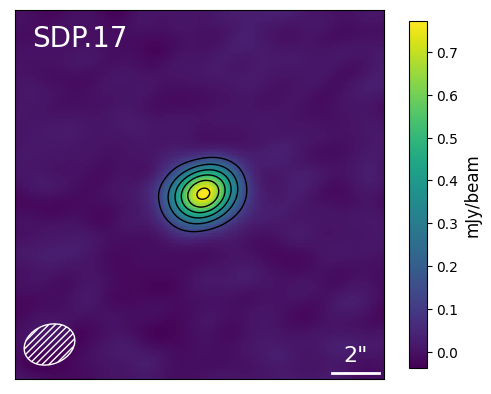}
    \includegraphics[height=3.3cm]{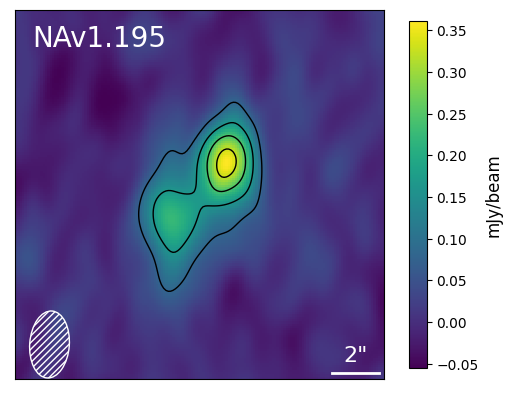}
    \includegraphics[height=3.3cm]{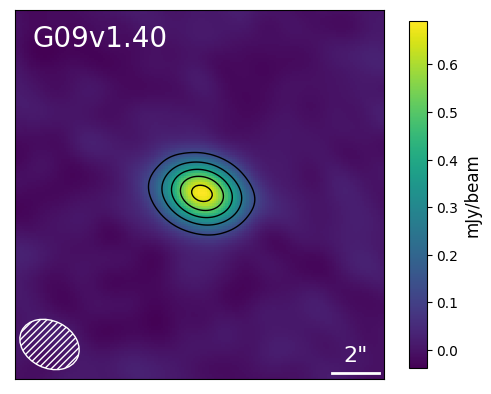}\\
    \includegraphics[height=3.3cm]{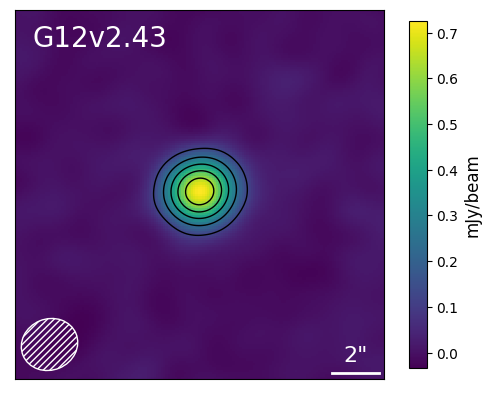}
    \includegraphics[height=3.3cm]{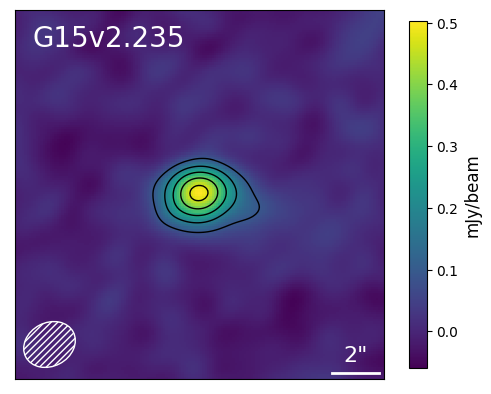}
    \includegraphics[height=3.3cm]{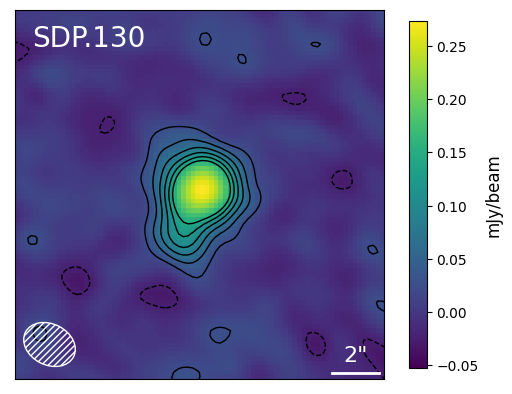} \\

    \caption{ALMA Band~3/4 continuum imaging of our targets. All images were produced using natural weighting. Contours start at $\pm2\sigma$, with a 2$\sigma$ increment. All sources are clearly detected in the continuum emission. G09v1.326 and NAv1.195 are resolved into two point-like sources due to the lensing morphology. SDP.130 is blended with a fainter galaxy (at a different redshift) to the south-west.}
    \label{fig:alma_continuum_compact}
\end{figure*}

Figures~\ref{fig:noema_continuum} and \ref{fig:alma_continuum_compact} present synthesised images of the continuum for individual sources, obtained with NOEMA and ALMA, respectively. For NOEMA observations, we show images for the upper and lower sideband separately.

\begin{figure*}[p]
\centering

\includegraphics[height=3.3cm]{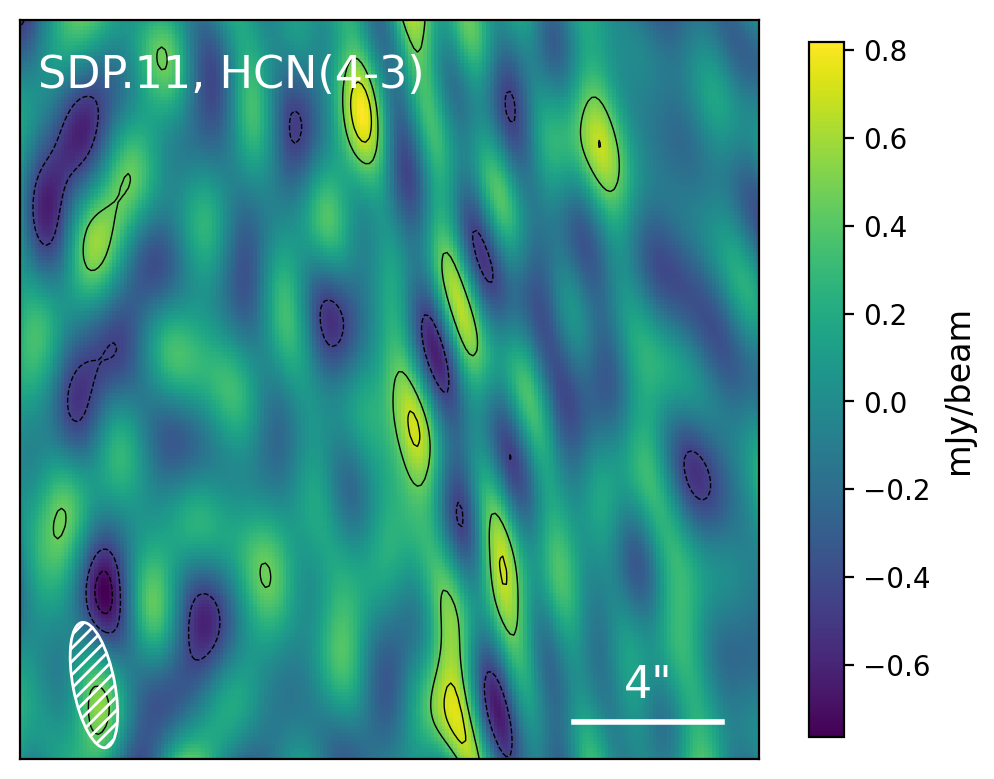}
    \includegraphics[height=3.3cm]{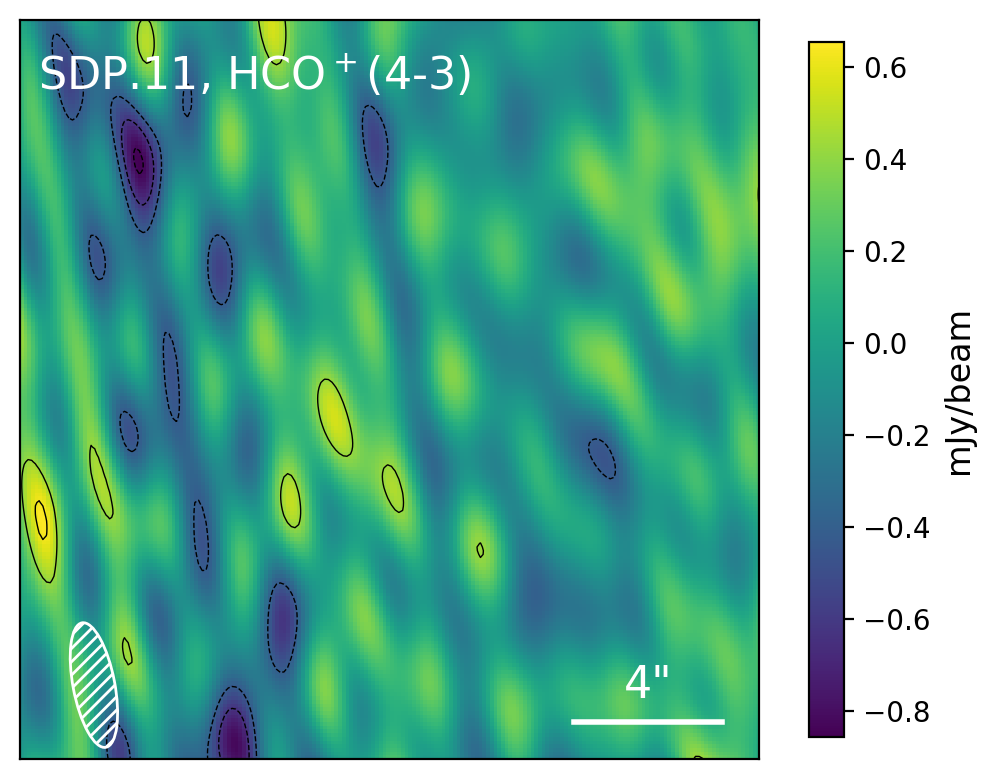}
    \includegraphics[height=3.3cm]{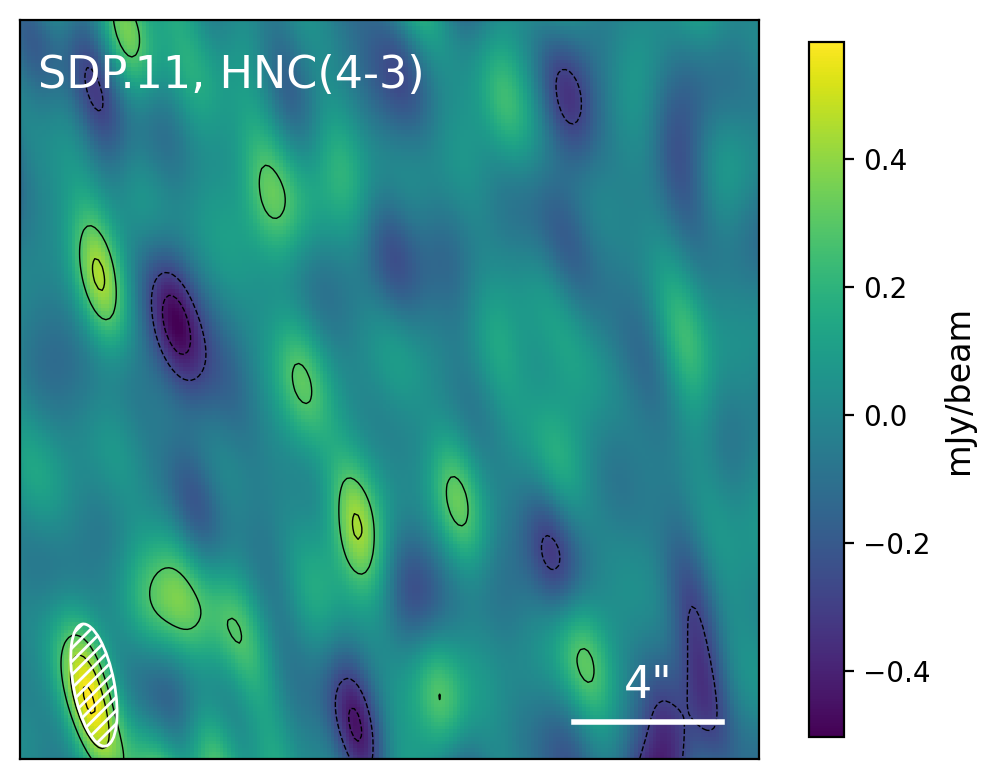}\\

     \includegraphics[height=3.3cm]{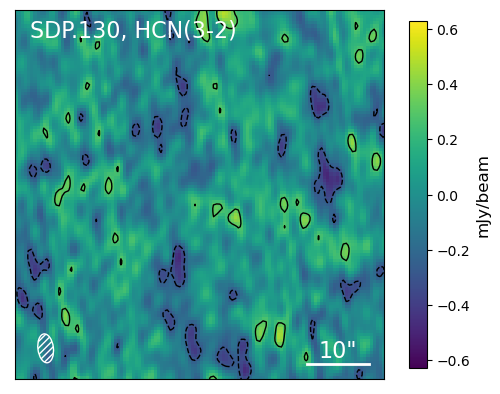}
    \includegraphics[height=3.3cm]{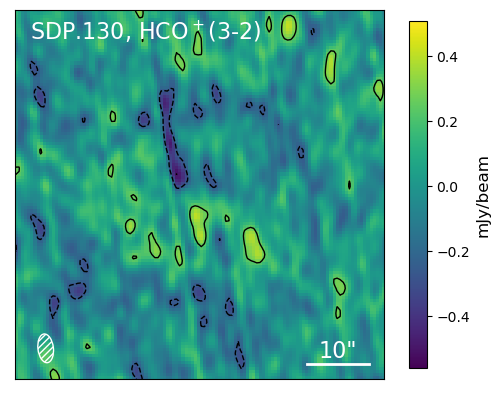}
    \includegraphics[height=3.3cm]{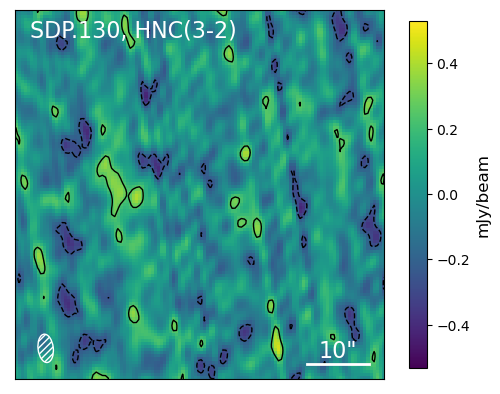}\\

    \includegraphics[height=3.3cm]{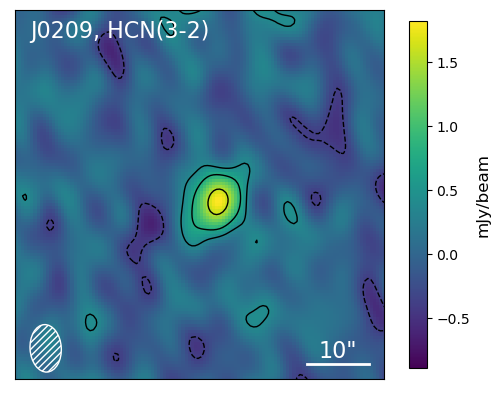}
    \includegraphics[height=3.3cm]{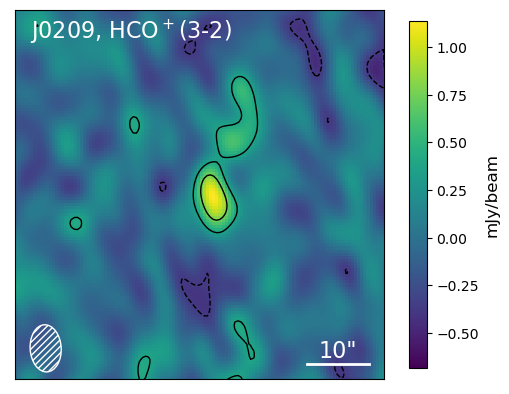}
    \includegraphics[height=3.3cm]{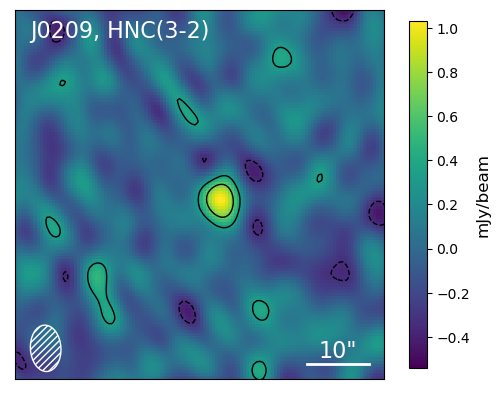}\\

    \includegraphics[height=3.3cm]{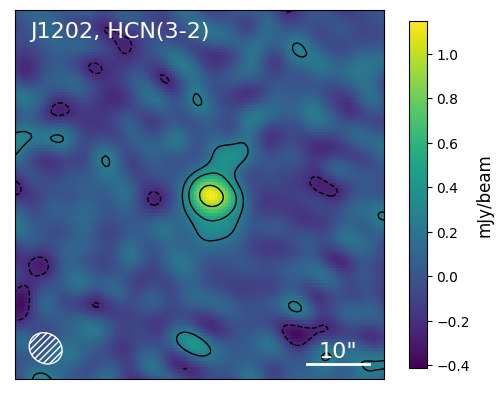}
    \includegraphics[height=3.3cm]{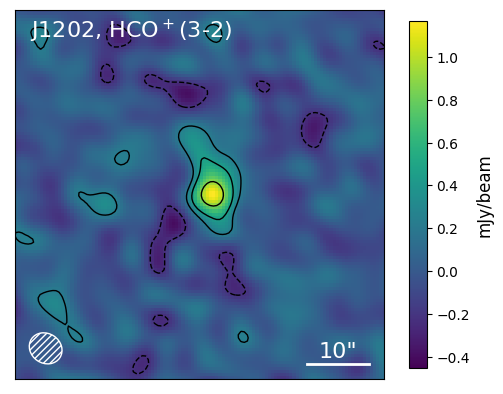}
    \includegraphics[height=3.3cm]{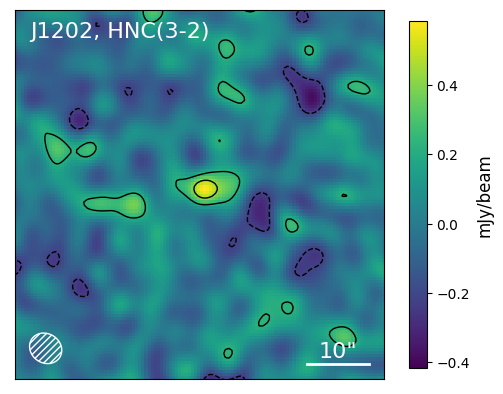}\\
    
    \includegraphics[height=3.3cm]{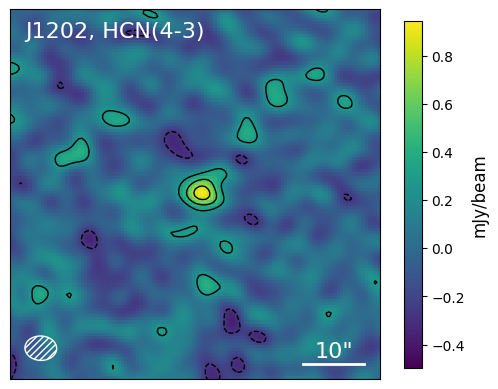}
    \includegraphics[height=3.3cm]{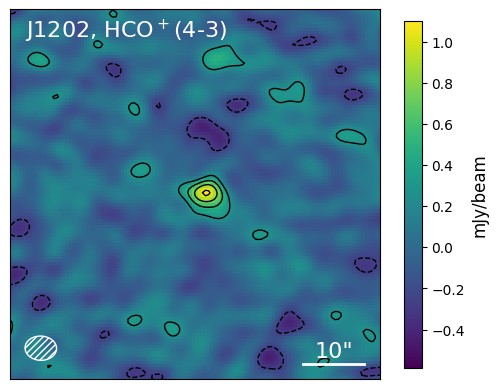}
    \includegraphics[height=3.3cm]{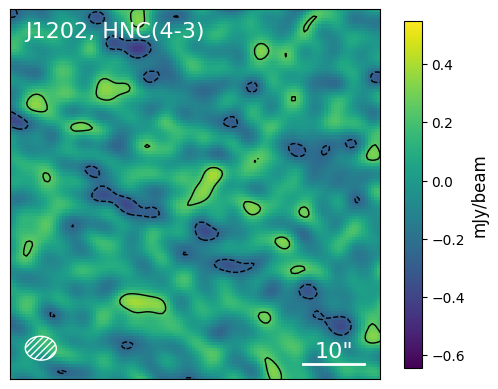}

    \caption{Figure~\ref{fig:noema_hcnhcohnc} continued: NOEMA narrow-band images of the HCN/HCO$^+$/HNC lines in individual sources. The field of view is 20''$
    times$20''; contours start at $\pm$2$\sigma$ and increase in steps of 2$\sigma$. J0209 is clearly detected in all lines except HCN(4--3), J1202 is detected in all transitions. SDP.130 is not detected in any transition.}
    \label{fig:app_noema_hcnhcohnc}
\end{figure*}

\begin{figure*}[p]
\centering

    \includegraphics[height=3.3cm]{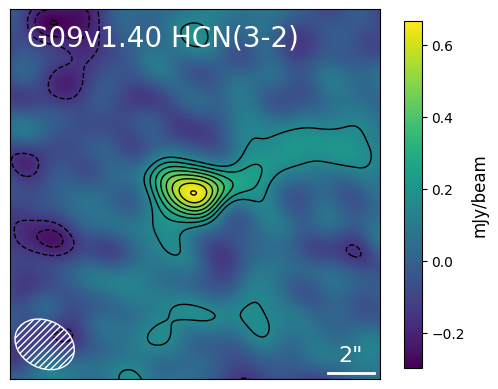}
    \includegraphics[height=3.3cm]{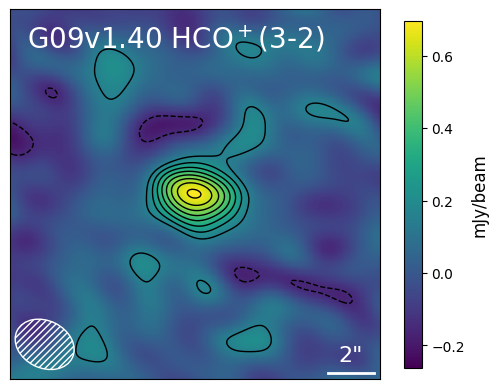}
    \includegraphics[height=3.3cm]{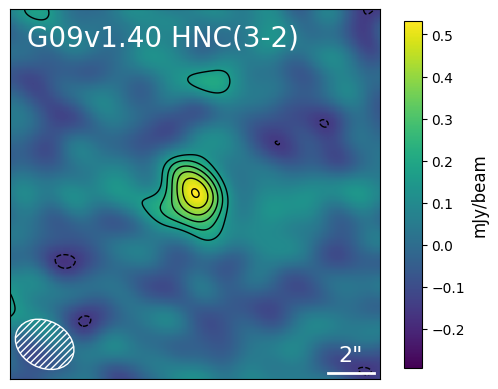}
    \\

    \includegraphics[height=3.3cm]{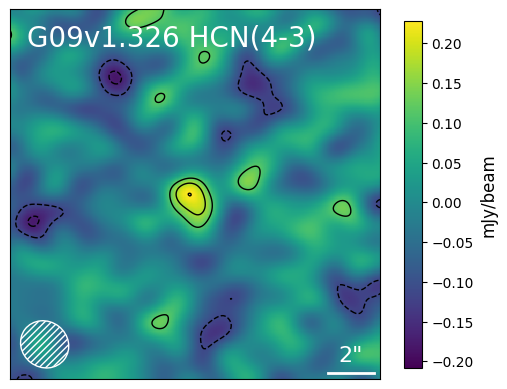}
    \includegraphics[height=3.3cm]{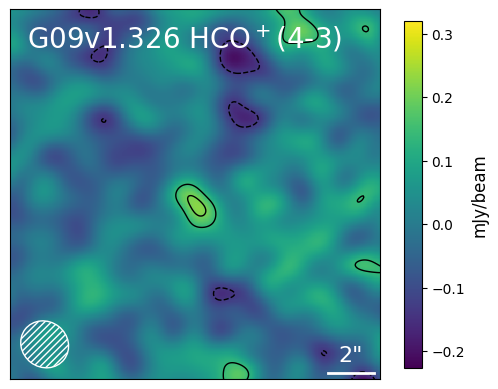}
    \includegraphics[height=3.3cm]{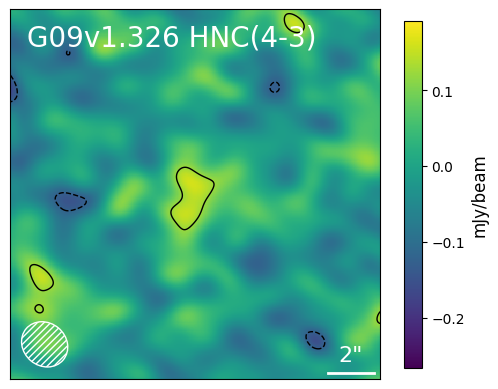}
    \\

    \includegraphics[height=3.3cm]{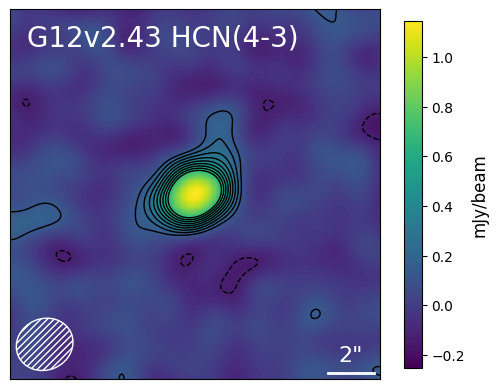}
    \includegraphics[height=3.3cm]{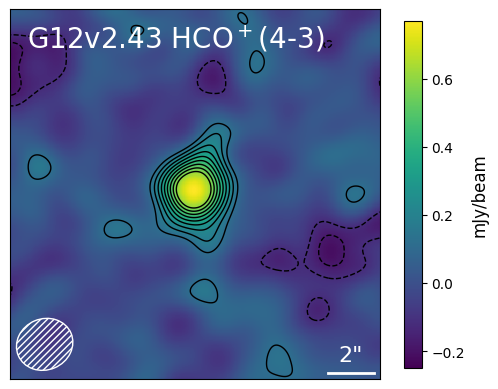}
    \includegraphics[height=3.3cm]{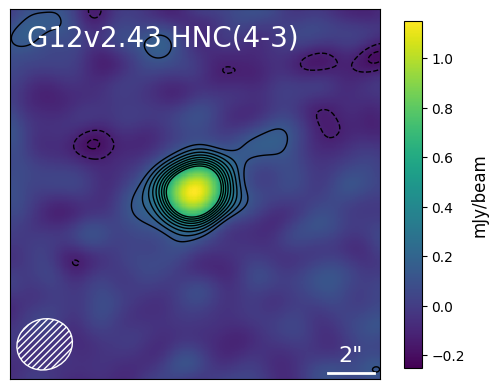}
    \\

      \includegraphics[height=3.3cm]{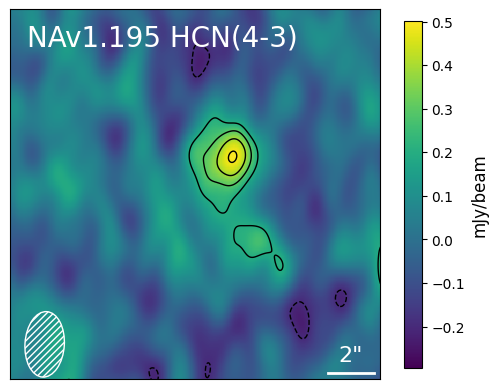}
    \includegraphics[height=3.3cm]{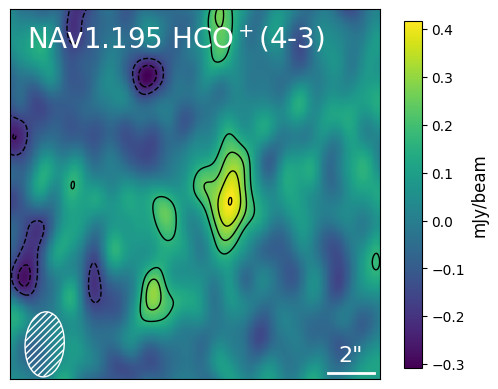}
    \includegraphics[height=3.3cm]{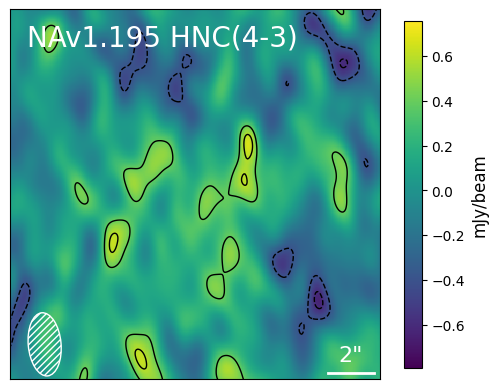}
    \\

      \includegraphics[height=3.3cm]{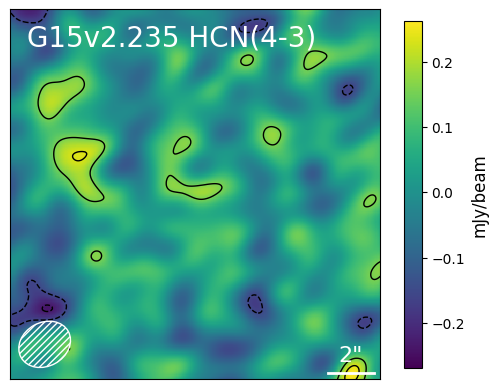}
    \includegraphics[height=3.3cm]{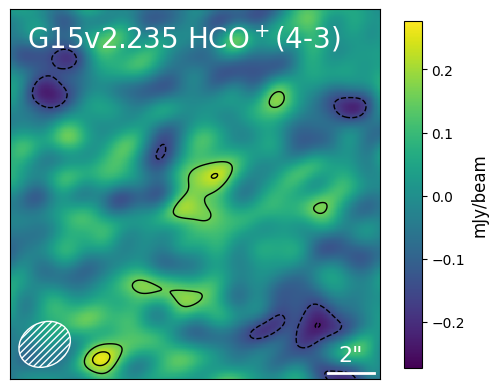}
    \includegraphics[height=3.3cm]{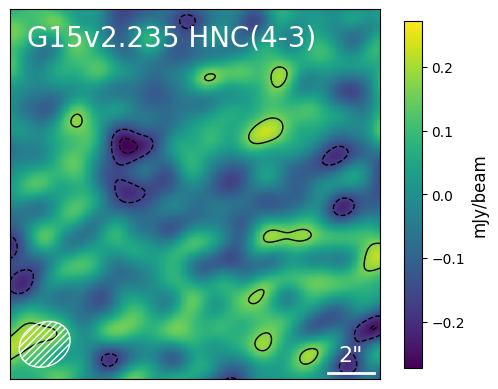}
    \\

   \includegraphics[height=3.3cm]{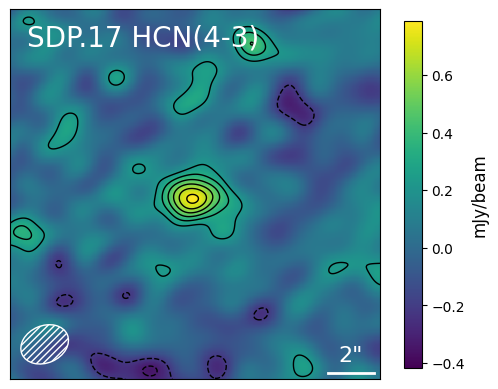}
    \includegraphics[height=3.3cm]{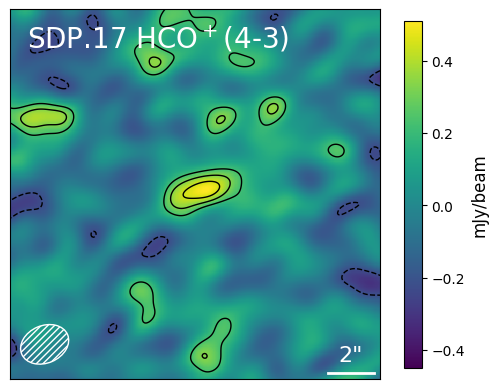}
    \includegraphics[height=3.3cm]{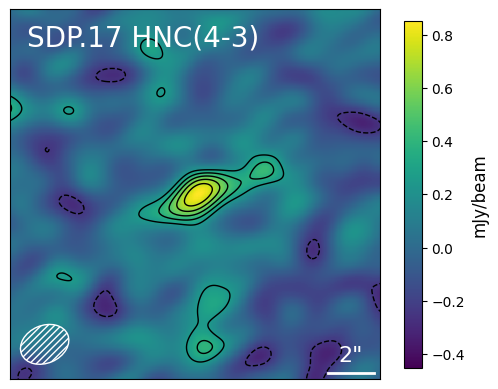}
    \\

     \includegraphics[height=3.3cm]{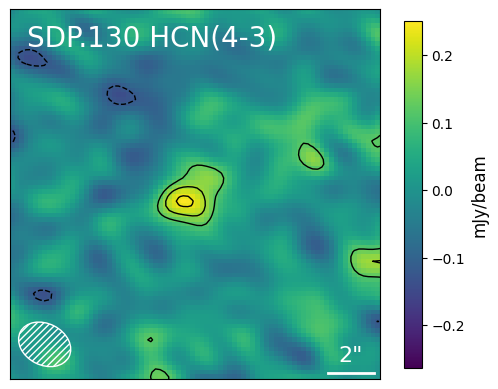}
    \includegraphics[height=3.3cm]{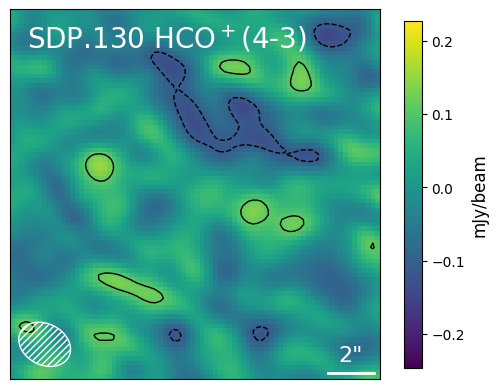}
    \includegraphics[height=3.3cm]{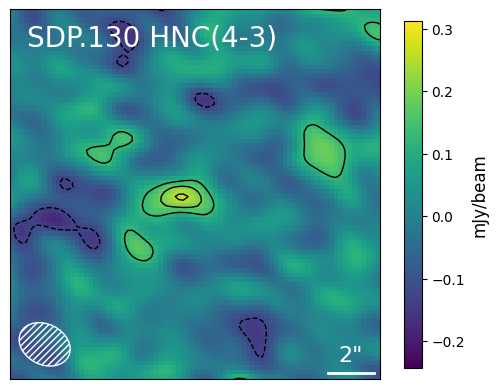}
    \\
    
    \caption{Figure~\ref{fig:noema_hcnhcohnc} continued: ALMA narrow-band images of the HCN/HCO$^+$/HNC lines for individual sources. Contours are drawn at $\pm$(2,3,...10)$\sigma$. }
     \label{fig:app_alma_hcnhcohnc}
    
\end{figure*}

Figures~\ref{fig:app_noema_hcnhcohnc} and \ref{fig:app_alma_hcnhcohnc} present the narrow-band images for the HCN, HCO$^+$, and HNC lines for individual sources. Finally, Fig.~\ref{fig:noema_spectra_2} and ~\ref{fig:alma_spectra} present the spectra extracted for individual sources. Table~\ref{tab:ALMA_fluxes} lists the fluxes for individual tracers inferred from the mom-0 maps.

\begin{table*}[h]
\caption{Line flux measurements from moment-0 images, not corrected for the lensing magnification.}
    \centering
    \begin{tabular}{l|ccc}
    \hline
     
     Source & HCN(4--3) & HCO$^+$(4--3) & HNC(4--3)\\
      & [Jy km/s] & [Jy km/s] & [Jy km/s] \\ 
     \hline
   SDP.130 & 0.15$\pm$0.05 & $\leq$0.13 & 0.09$\pm$0.05 \\ 
   SDP.17 &  0.45$\pm$0.09 & 0.29$\pm$0.09 & 0.47$\pm$0.10\\
   SDP.9 & 1.62$\pm$0.18 & 1.56$\pm$0.18 & 1.25$\pm$0.20 \\
     SDP.11 & 0.54$\pm$0.22 & $\leq$0.62 & $\leq$0.42 \\
     
   G09v1.326 & 0.15$\pm$0.11 & 0.18$\pm$0.13 & 0.24$\pm$0.12\\
   G12v2.43  & 0.44$\pm$0.04 & 0.32$\pm$0.04 & 0.47$\pm$0.04 \\
   G15v2.235  & $\leq$0.24 & $\leq$0.24 & $\leq$0.27 \\
   NAv1.195  & 0.43$\pm$0.12 & 0.39$\pm$0.10 & $\leq$0.7\\
   J1202 & 0.93$\pm$0.23 & 1.01$\pm$0.24 & $\leq$0.66 \\
   \hline
     Source & HCN(3--2) & HCO$^+$(3--2) & HNC(3--2) \\ 
     & [Jy km/s] & [Jy km/s] & [Jy km/s] \\ 
      \hline
     G09v1.140 & 0.24$\pm$0.04 & 0.28$\pm$0.04 & 0.20$\pm$0.04\\
     SDP.130 & $\leq$0.43 & $\leq$0.38  & $\leq$0.38 \\
     J0209 & 1.77$\pm$0.23 & 1.00$\pm$0.23 & 1.30$\pm$0.20 \\
     J1202 & 1.44$\pm$0.21 & 1.40$\pm$0.21 & 0.56$\pm$0.21\\
     \hline
     
     \hline
    \end{tabular}
    \label{tab:ALMA_fluxes}
\end{table*}

\vfill
\newpage

\begin{figure*}[ht]
\centering
    \includegraphics[height=4.25cm]{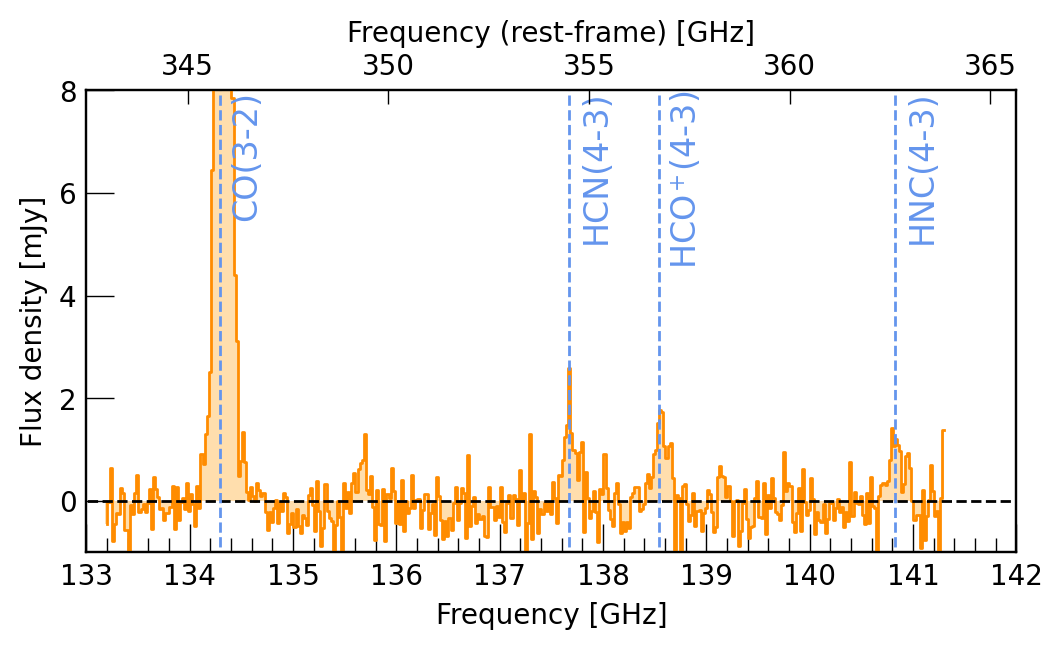}
    \includegraphics[height=4.25cm]{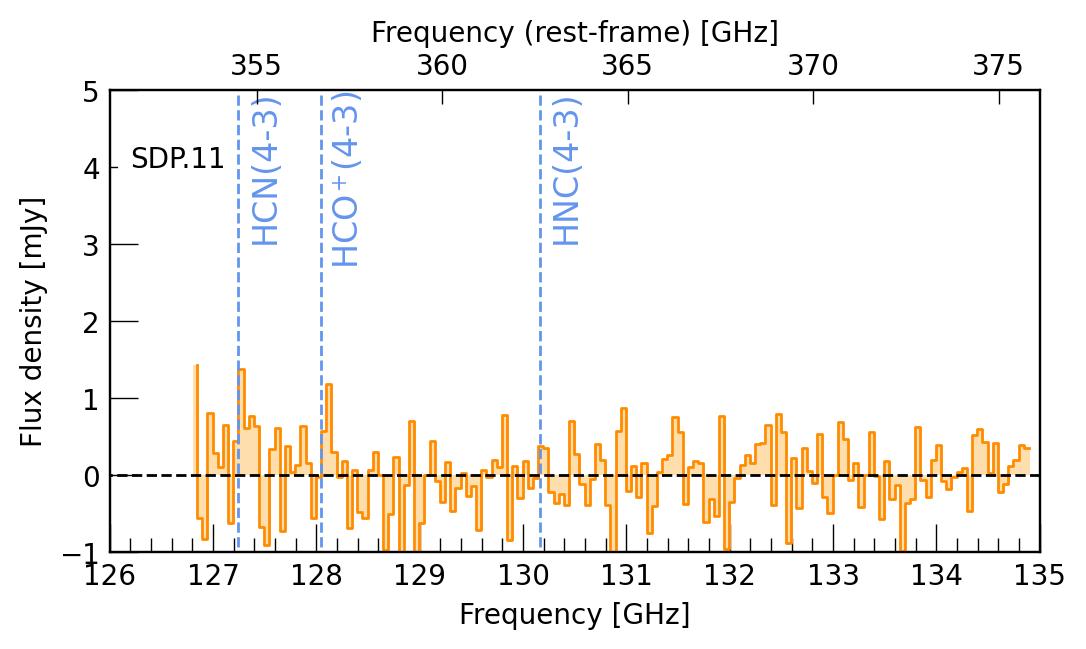}\\
  
    \includegraphics[height=4.25cm]{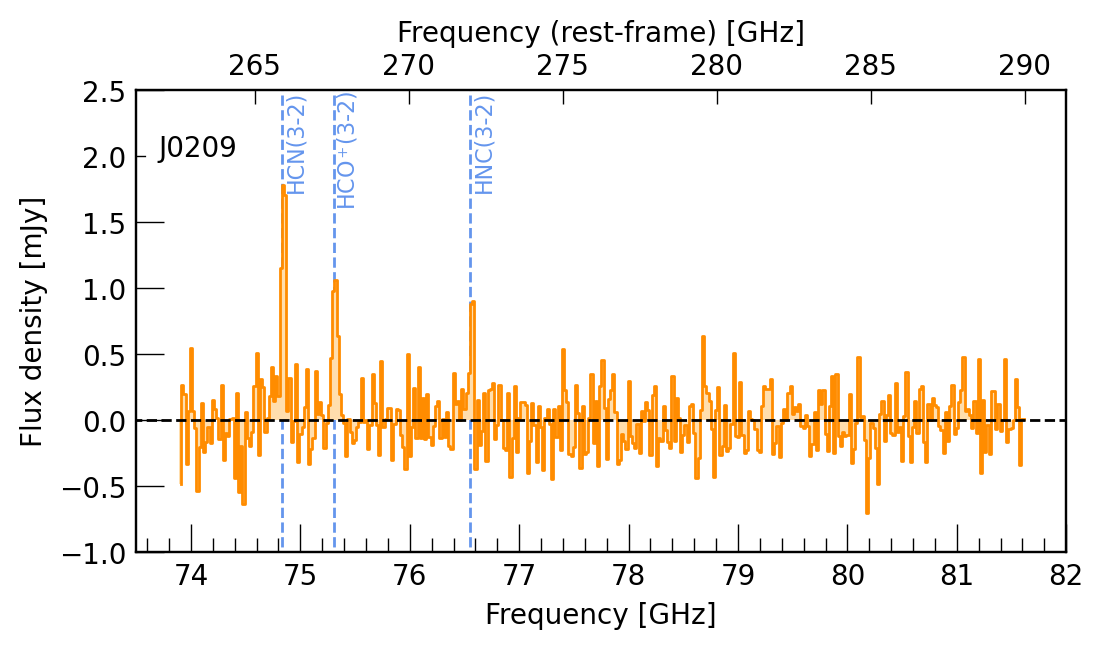}
    \includegraphics[height=4.25cm]{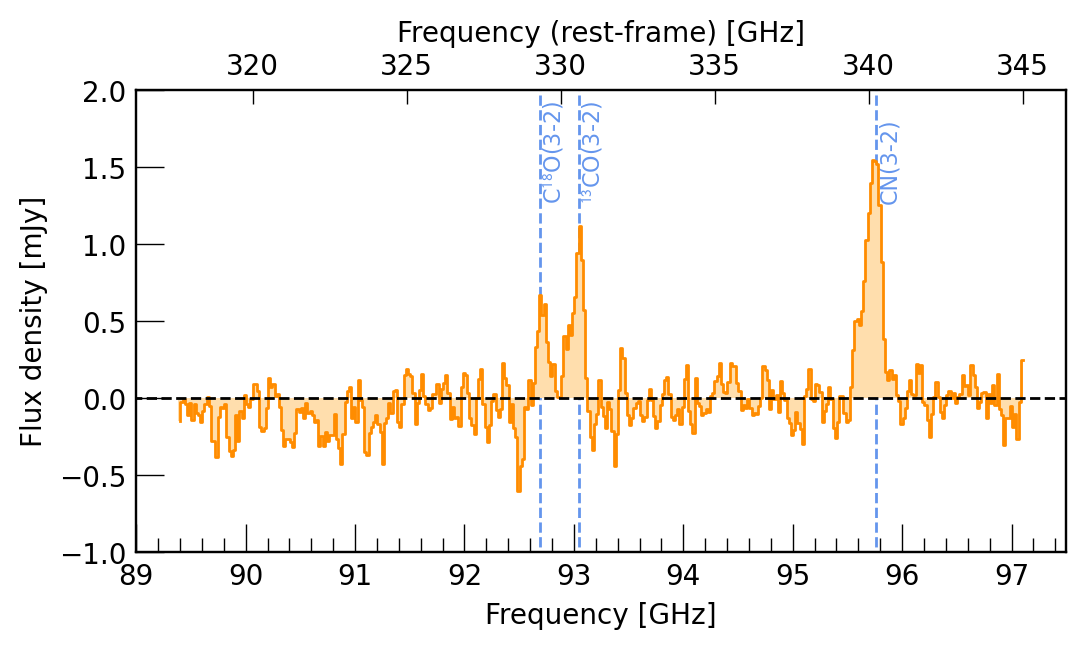}\\
    
    \includegraphics[height=4.25cm]{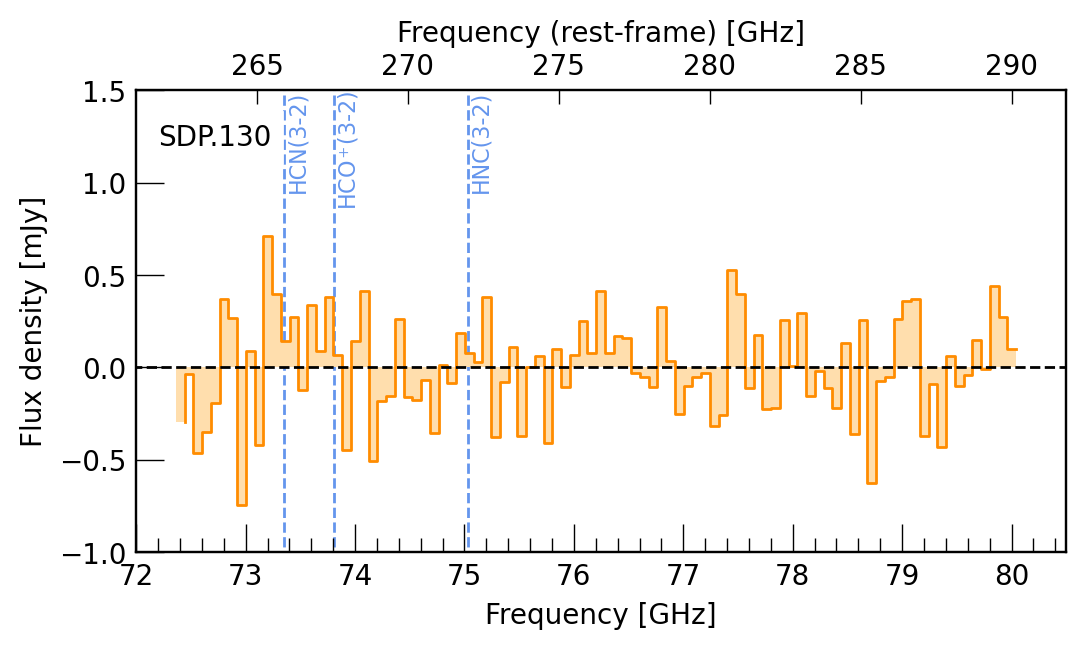}
    \includegraphics[height=4.25cm]{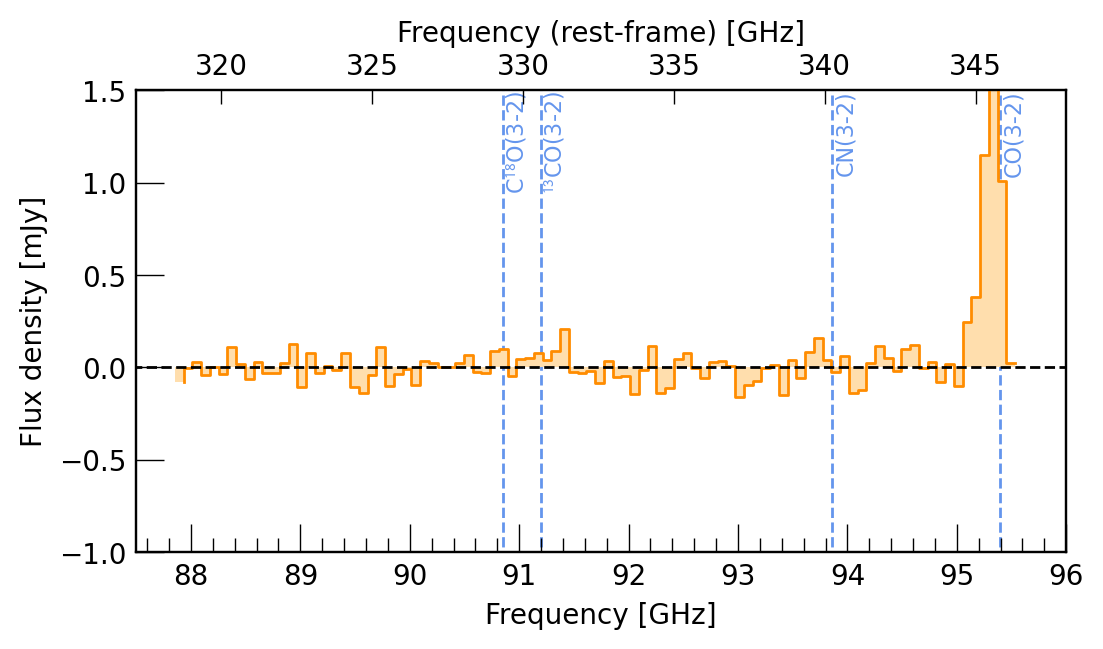}\\
    
    \includegraphics[height=4.25cm]{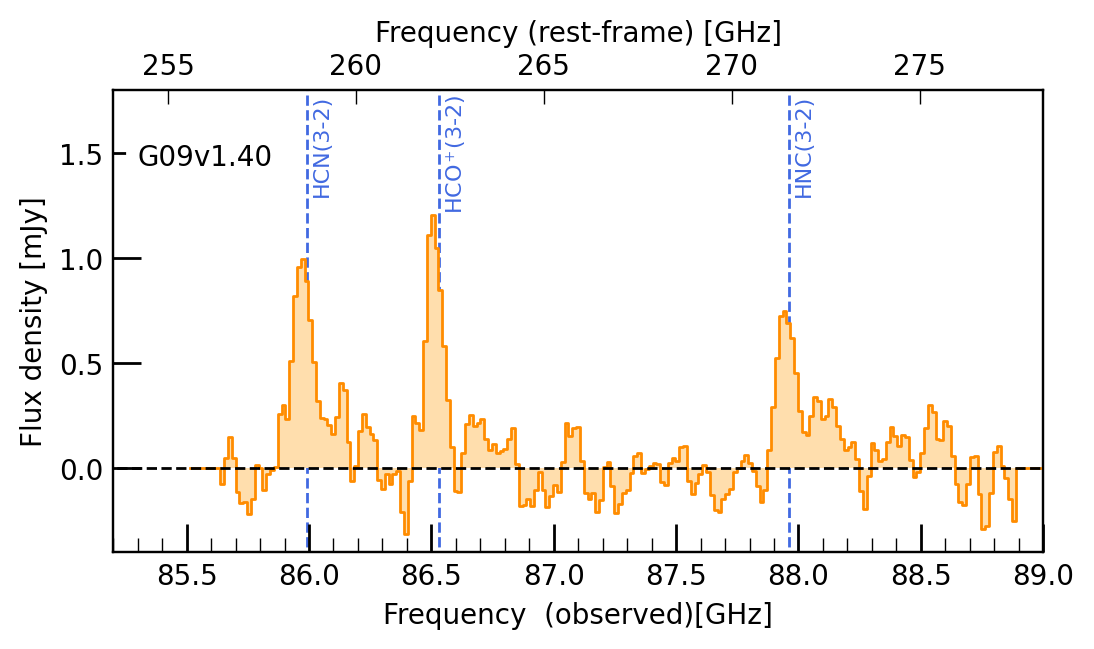}
    \includegraphics[height=4.25cm]{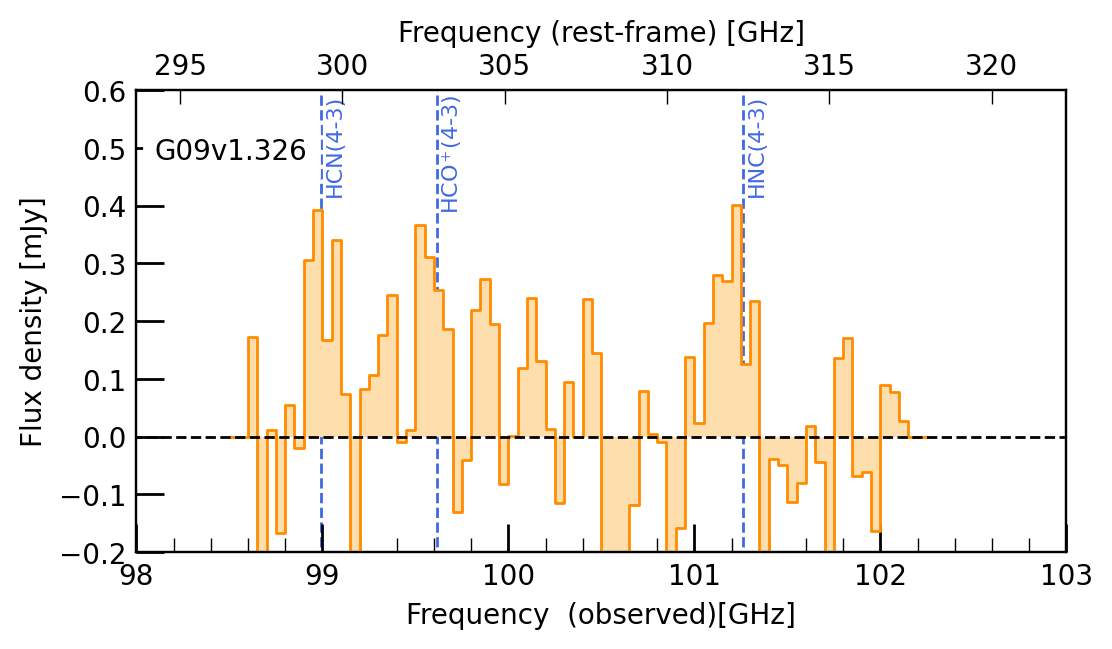} \\
    \includegraphics[height=4.25cm]{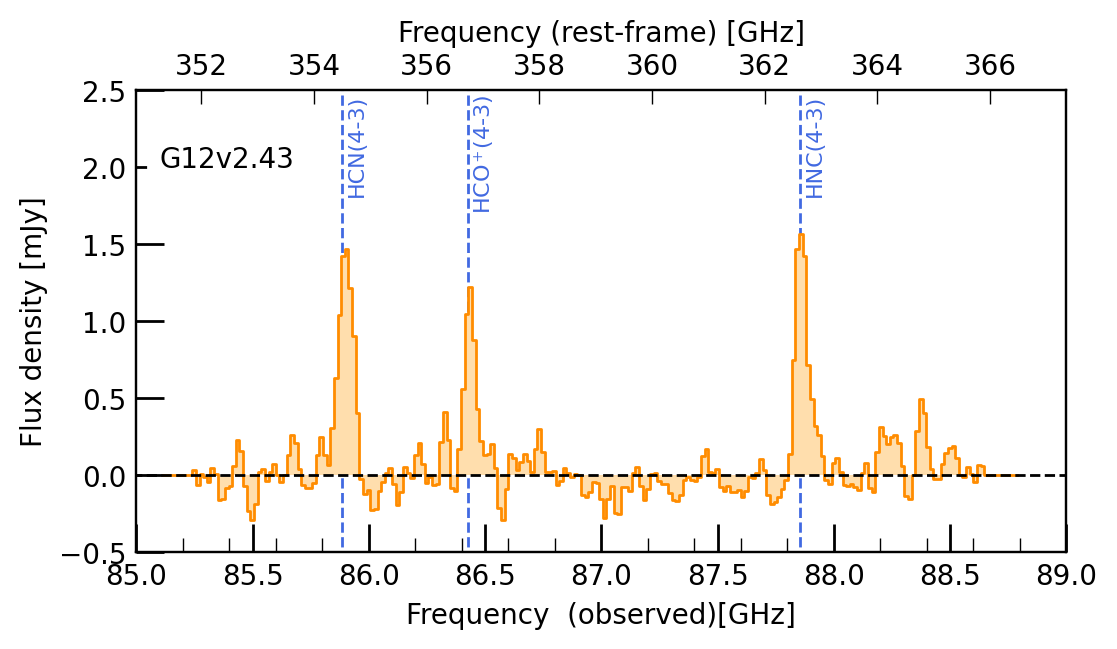} 
    \includegraphics[height=4.25cm]{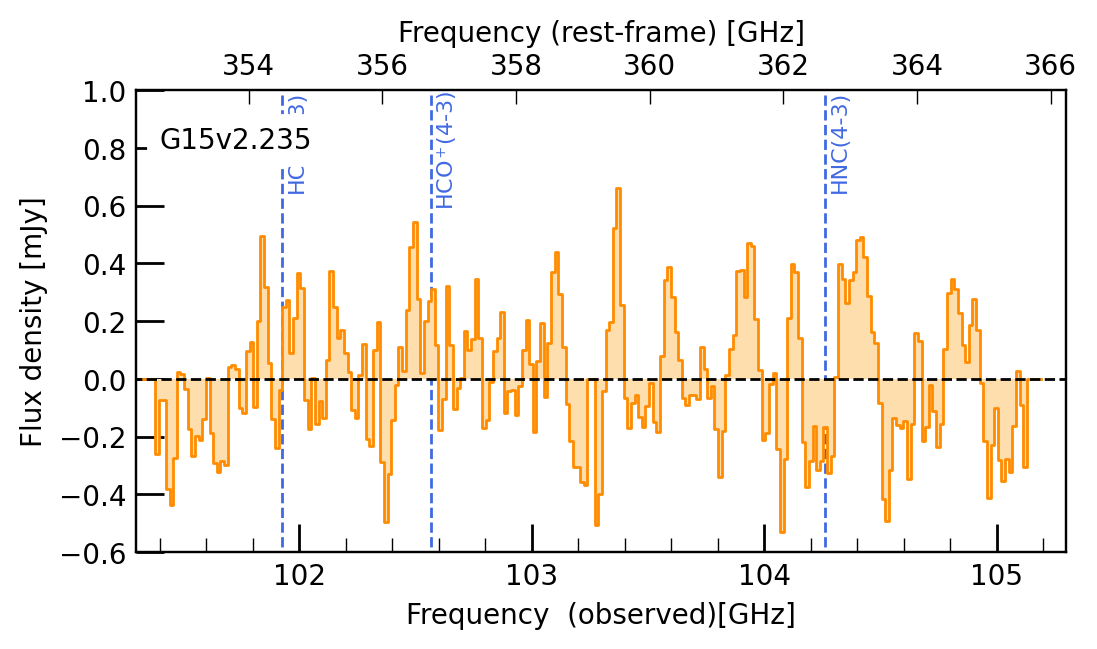} \\

    \caption{Spectra for individual targets, after continuum subtraction. In SDP.9, we detect HCN, HCO$^+$ and HNC(4--3) lines. For SDP.11, no lines are detected.  In J0209, besides HCN, HCO$^+$ and HNC, we also detect the $^{13}$CO, C$^{18}$O and CN emission. We do not detect any dense-gas tracers in SDP.130. Note the blue-shifted HCN(3--2) emission in G09v1.40, which might be coming from a molecular outflow.}
    \label{fig:noema_spectra_2}
\end{figure*}

\begin{figure*}[h]
\ContinuedFloat

    \centering
    \includegraphics[height=4.25cm]{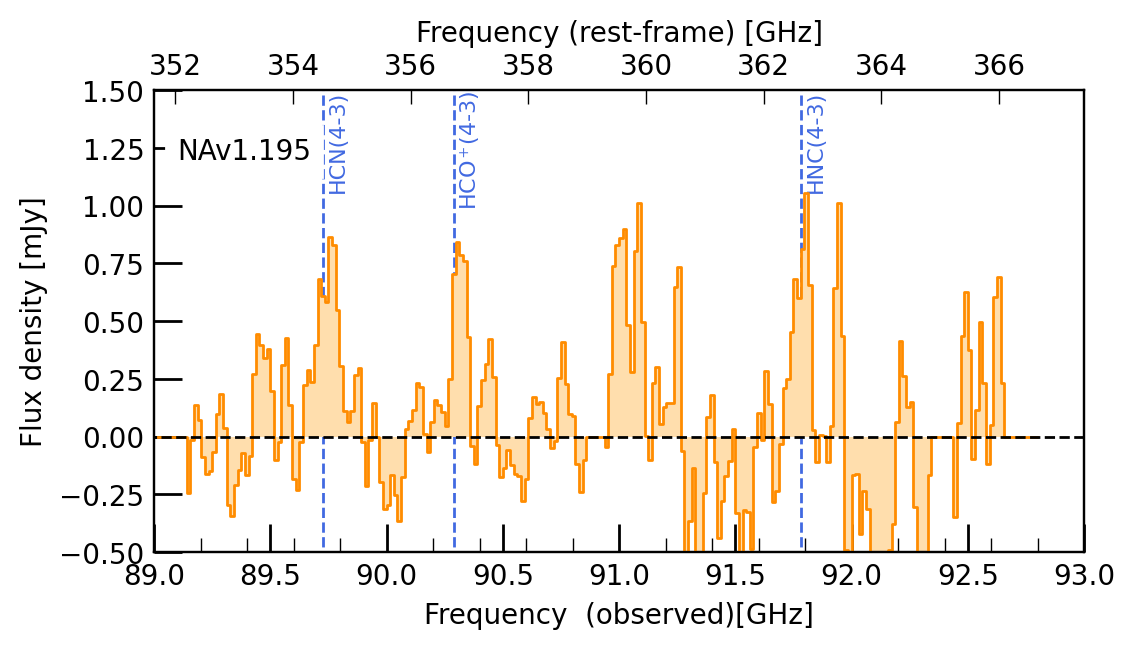} 
    \includegraphics[height=4.25cm]{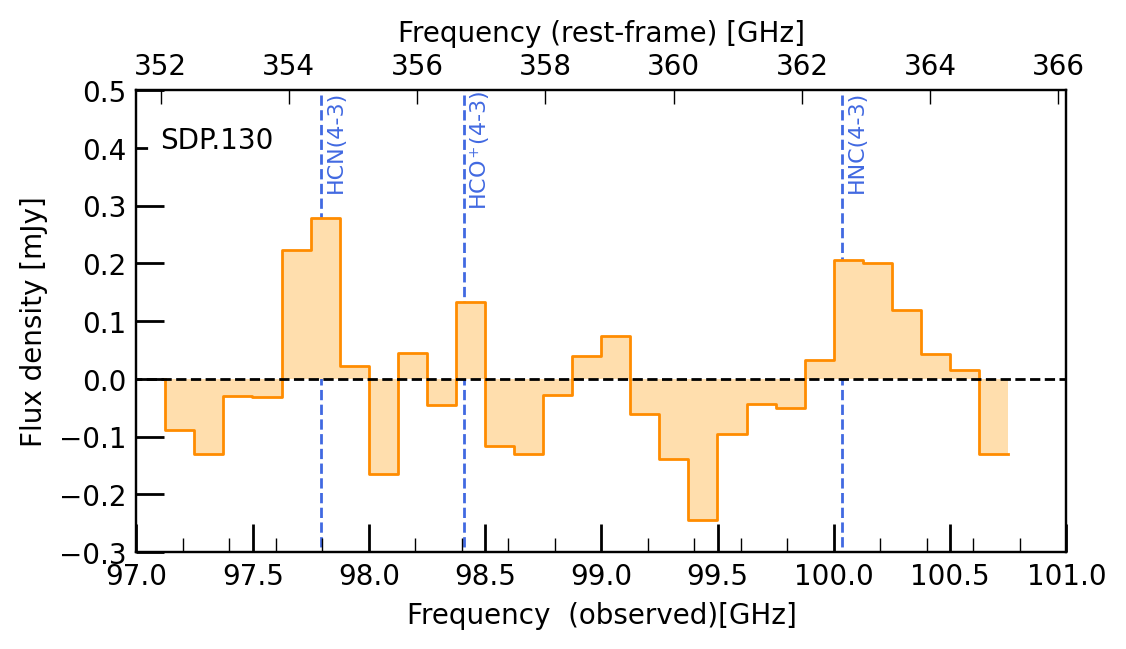} \\
    \includegraphics[height=4.25cm]{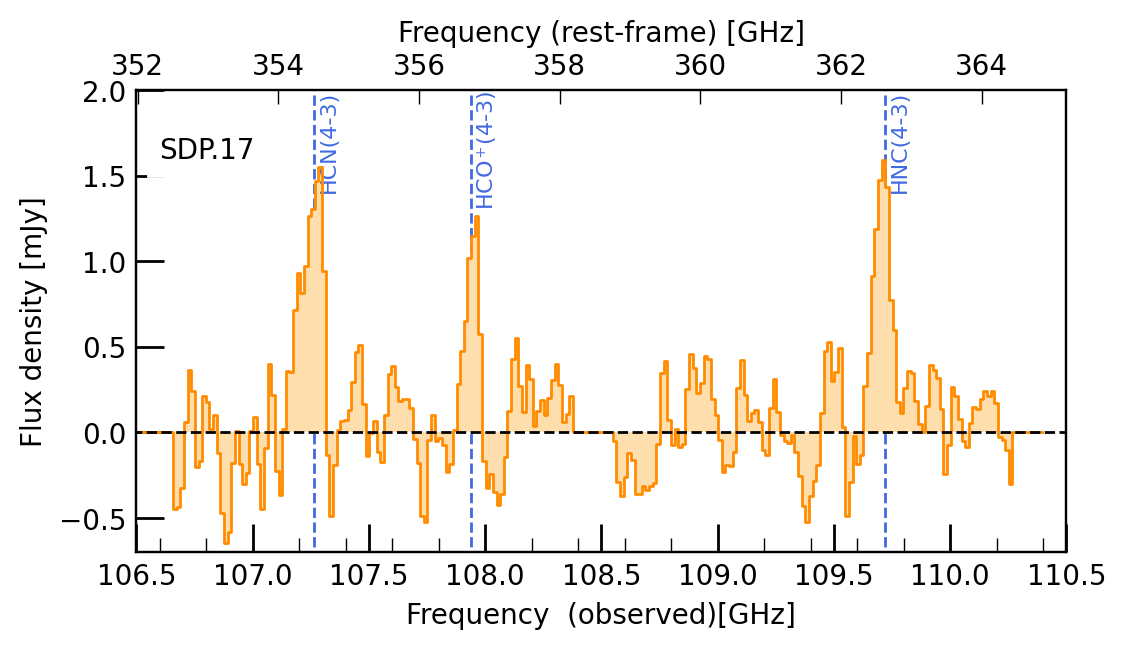}

    \caption{continued.}
    \label{fig:alma_spectra}
\end{figure*}

\section{Detections of CO isotopologues: $^{13}$CO and C$^{18}$O}
\label{sec:co_isotopologues}

\begin{figure*}[h]
\centering
    \includegraphics[height =3.5cm]{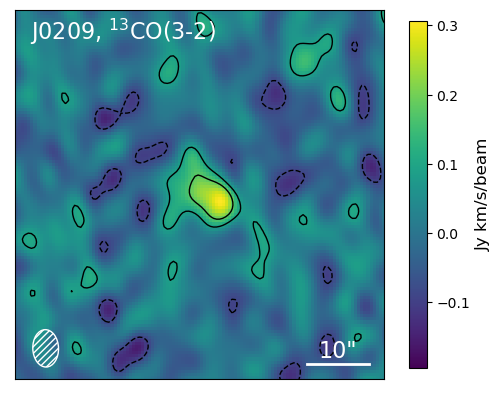}
    \includegraphics[height = 3.5cm]{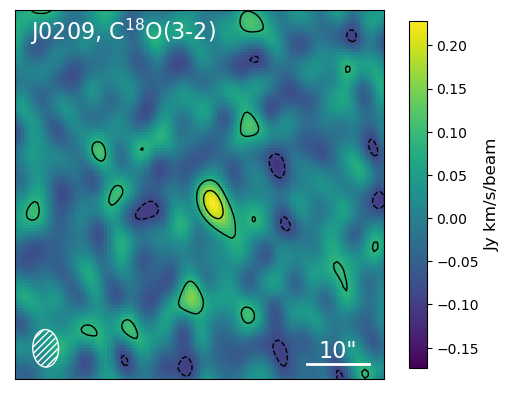}
    \includegraphics[height = 3.5cm]{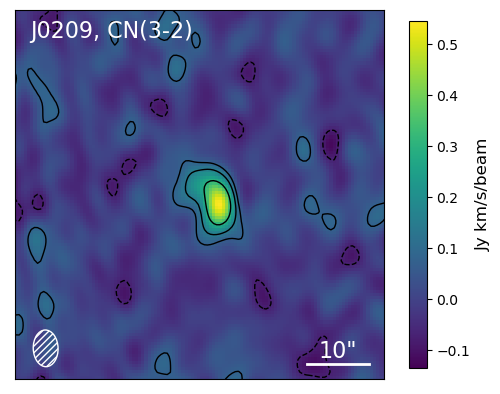}\\
    \includegraphics[height = 3.5cm]{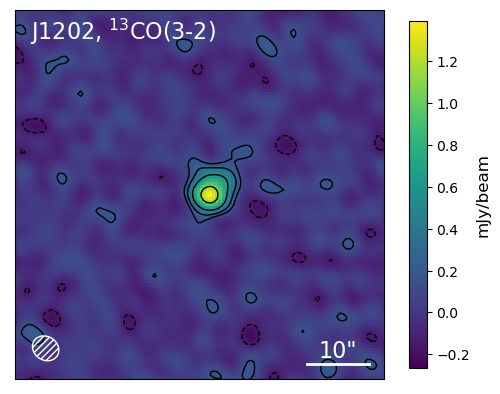}
    \includegraphics[height = 3.5cm]{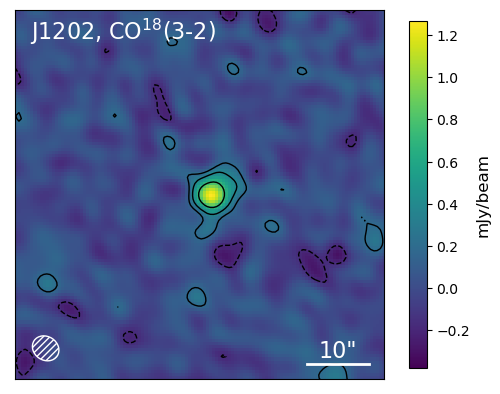}
    \includegraphics[height = 3.5cm]{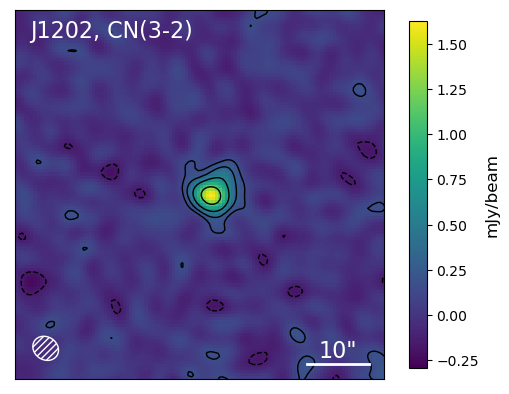}
       
\caption{Detections of CO isotopologues and CN lines in J0209 and J1202. Contours start at $\pm$2$\sigma$ and increase in steps of 2$\sigma$.}
\label{fig:co_cn}
\end{figure*}

In J0209 and J1202, our NOEMA spectral setup covers the CO(3--2) isotopologues $^{13}$CO(3--2) ($f_0=330.588$~GHz) and C$^{18}$O(3--2) ($f_0=329.331$~GHz). As shown in Fig.~\ref{fig:co_cn}, we detect both transitions in both sources; the measured fluxes and luminosities are reported in Tab.~\ref{tab:co_isotopologues}.
The $^{13}$CO/C$^{18}$O isotopologues luminosity ratios in J0209 and J1202 are $L'_
\mathrm{13CO}/L'_\mathrm{C18O}$=1.8$\pm$0.6 and 1.1$\pm$0.1, respectively. These are consistent with previous results for high-redshift DSFGs by \citep{Zhang2018,Yang2023}, and with the lower limit from the \citet{Spilker2014} spectral stack ($^{13}$CO/C$^{18}$O(3--2)$\geq$1.1).

The $^{13}$CO/C$^{18}$O ratio has been proposed as a potentially sensitive tracer of the IMF in obscured galaxies \citep[e.g.][]{Sage1991, Henkel1993, Romano2003, Romano2017}. This is because $^{13}$C and $^{18}$O are produced and dispersed via different routes. Namely $^{13}$CO is primarily produced by low/intermediate-mass stars and novae ($M_\star\leq8$~$M_\odot$), whereas C$^{18}$O production is dominated by high-mass stars. In the Milky Way, $^{13}$CO/C$^{18}$O luminosity ratio is $\approx$7  \citep{Romano2017} ; lower $^{13}$CO/C$^{18}$O ratios would thus indicate an excess of massive stars - a top-heavy IMF. Indeed, previous studies of CO isotopologues in DSFGs by \citet{Zhang2018, Yang2023} have found ratios close to unity. {Taken at the face value, CO isotopologue ratios in J0209 and J1202 are consistent with a top-heavy IMF. However, there are considerable uncertainties in linking CO isotopologue ratios to the IMF slope, particularly when stellar rotation is taken into account (C. Kobayashi, priv. comm.)}

\begin{table}[]
    \centering
    \caption{Flux densities of CO isotopologues and CN transitioms in J0209 and J1202, in units of Jy km/s. The reported fluxes are not corrected for lensing magnification.}
    \begin{tabular}{c|ccc}
    \hline
        Source & $^{13}$CO(3--2) & C$^{18}$O(3--2) & CN(3--2)\\
        \hline
        J0209 & 0.77$\pm$0.14 & 0.43$\pm$0.13 & 2.9$\pm$0.3\\
        J1202 & 1.77$\pm$0.11 & 1.61$\pm$0.14 & 3.7$\pm$0.2\\
         \hline
    \end{tabular}
    
    \label{tab:co_isotopologues}
\end{table}

\section{Detections of CN}
\label{sec:cn}

{In J0209 and J1202, we obtain strong detection of the cyanide radical CN($N=3-2$) transition (with different fine- and hyperfine structure transition spanning $f_0=339.447 - 340.279$~ GHz). CN arises primarily from regions exposed to strong ultraviolet (UV) or X-ray radiation (e.g., \citealt{Boger2005, Meijerink2007}) The total flux of the CN(3--2) transitions is 2.9$\pm$0.3 Jy km/s in J1202 and 3.7$\pm$0.2 Jy km/s in J1202 (Tab.~\ref{tab:co_isotopologues}), which corresponds to $L'_\mathrm{CN(3-2)}=(1.0\pm0.1)\times10^{11}$~and $L'_\mathrm{CN(3-2)}=(1.2\pm0.1)\times10^{11}$~K km s$^{-1}$ pc$^2$, respectively.}

{Adopting the CO(3--2) luminosities from \citet{Harrington2021}, we find CN(3--2)/CO(3--2) luminosity ratios of 0.15$\pm$0.04 and 0.17$\pm$0.05, respectively. These are comparable to measurements for the Cloverleaf quasar at $z=2.5$ (CN(3--2)/CO(3--2) ratio of 0.11$\pm$0.02 \citep{Riechers2007_CN}, but significantly higher than values reported for NCv1.143 (CN/CO=0.06, \citealt{Yang2023}) and for the spectral stack of SPT-selected galaxies (0.04, \citealt{Reuter2022}). Curiously, the CN(3--2) luminosities in J0209 and J1202 are about twice that of HCN(3--2) line (CN/HCN =1.8$\pm$0.3 and 2.0$\pm$0.4, respectively), in contrast to nearby star-forming galaxies, which show an almost-constant CN/HCN ratio of $0.86\pm0.27$ \citep{Wilson2023}.}

\end{appendix}
   
\end{document}